\documentclass[pdflatex,sn-mathphys]{sn-jnl}

\usepackage{siunitx}
\usepackage{subcaption}

\jyear{2023}%

\theoremstyle{thmstyleone}%
\theoremstyle{thmstyletwo}%

\theoremstyle{thmstylethree}%

\begin{document}

\title[Wake Tail Plane Interactions for a Tandem Wing Configuration]{Wake Tail Plane Interactions for a Tandem Wing Configuration in High-Speed Stall Conditions}


\author*[1]{\fnm{Johannes} \sur{Kleinert}}\email{johannes.kleinert@iag.uni-stuttgart.de}

\author[1]{\fnm{Maximilian} \sur{Ehrle}}\email{maximilian.ehrle@iag.uni-stuttgart.de}

\author[1]{\fnm{Andreas} \sur{Waldmann}}\email{andreas.waldmann@iag.uni-stuttgart.de}

\author[1]{\fnm{Thorsten} \sur{Lutz}}\email{thorsten.lutz@iag.uni-stuttgart.de}

\affil*[1]{\orgname{University of Stuttgart}, \orgdiv{Institute of Aerodynamics and Gas Dynamics (IAG)}, \orgaddress{\street{Pfaffenwaldring 21}, \city{ 70569 Stuttgart}, \state{Germany}}}


\abstract{
In this work, wake-tail plane interactions are investigated for a tandem wing configuration in buffet conditions, consisting of two untapered and unswept wing segments, using hybrid Reynolds-Averaged Navier-Stokes / Large Eddy Simulations (RANS/LES) with the Automated Zonal Detached Eddy Simulation (AZDES) method.
The buffet on the front wing and the development of its turbulent wake are characterized, including a spectral analysis of the fluctuations in the wake and a modal analysis of the flow.
The impact of the wake on the aerodynamics and loads of the rear wing is then studied, with a spectral analysis of its lift and surface pressure oscillations.
Finally, the influence of the position and the incidence angle of the rear wing is investigated.
For the considered flow conditions, 2D buffet is present on the front wing.
During the downstream movement of the shock, the amount of separation reaches its minimum and small vortices are present in the wake. 
During the upstream movement of the shock, the amount of separation is at its maximum and large turbulent structures are present accompanied by high fluctuation levels.
A distinct peak in the corresponding spectra can be associated with vortex shedding behind the wing.
The impingement of the wake leads to a strong variation of the loading of the rear wing. A low-frequent oscillation of the lift, attributed to the change of the intensity of the downwash generated by the front segment, can be distinguished from high-frequent fluctuations that are caused by the impingement of the wake's turbulent structures. 
}




\maketitle

\section{Introduction}\label{chp:introduction}

The reliable prediction and control of aircraft flight characteristics and flow conditions at the edges of the flight envelope is essential for continued improvement of commercial aviation and for the reduction of fuel consumption. The envelope boundaries demarcate the limits of safe and economic flight regimes of a given aircraft, and are typically associated with Mach number and angle of attack values that should not be exceeded.

One typical phenomenon that limits the flight envelope at transonic Mach numbers is transonic buffet. This refers to the occurrence of a coupled periodic oscillation of a shock and the corresponding shock-induced boundary layer separation over a lifting surface, possibly causing vibrations, which endanger the integrity of the structure.

Two-dimensional buffet has been widely studied and descriptions of the phenomenon on airfoils have been published based on computational~\cite{deck2005,grossi2014} and experimental~\cite{hartmann2013,jacquin2016} investigations. Lee at al.~\cite{lee1990} presented a comprehensive model relating the shock motion to disturbances from the shock propagating downstream and scattering at the airfoil trailing edge. The roundtrip duration of the downstream propagation and the scattered upstream propagation corresponds well to the buffet period, indicating a feedback loop. Jacquin et al.~\cite{jacquin2009} experimentally observed buffet periods consistent with this. Hartmann et al.~\cite{hartmann2013} and Feldhusen et al.~\cite{feldhusen2018} proposed an additional interaction mechanism involving propagation along the pressure side surface.

In the context of a conventional aircraft, buffet-induced separation may cause a wake flow incorporating a large bandwidth of turbulent scales which propagate downstream. These structures can impinge on the tail plane and cause undesirable unsteady loads. General insight of such wake-tail plane interactions is provided by Tan et al.~\cite{tan2018}. More specifically, Waldmann et al.~\cite{waldmann2020} observed strong load fluctuation of the horizontal tail plane (HTP) due to turbulent wake impingement via Computational Fluid Dynamics (CFD). The magnitude of turbulent kinetic energy and the extent of the wake strongly impact the imparted loads. There is a strong variation of the loads due to angle of attack. For a transonic cruise condition, Illi et al.~\cite{illi2013} showed that the fundamental frequency of the tail plane load oscillation correlated to that of the main wing shock movement. 

The wake of a wing or airfoil that exhibits transonic buffet or buffeting includes large-scale fluctuations caused by the periodic shock motion and small-scale turbulence due to the decay of vortices arising from flow separation and associated shear layers. Due to this broadband nature of the wake flow, the impingement of turbulent structures on an airfoil or wing immersed in a wake is comparable to the situation of the same airfoil or wing being subjected to an inflow with broadband atmospheric disturbances. M\"uller et al.~\cite{mueller2020} analyzed the impact of broadband turbulence on aircraft surfaces at flight-relevant subsonic conditions at $Ma_{\infty} = 0.25$ and $Re_{\infty} = 11.6 \cdot 10^6$. They observed that the leading edge and suction peak areas of airfoils and swept wings are most sensitive to unsteady pressure fluctuations due to oscillating inflow. Local surface pressure spectra in these areas correspond to the spectrum of the inflow velocity fluctuation, correlating inflow turbulence characteristics with load oscillations over a broad frequency range.

\section{Motivation and Background}\label{chp:tandem_intro}

The present results are part of the efforts carried out in sub-project 4 of the research unit 2895 funded by Deutsche Forschungsgemeinschaft (DFG), whose goal is the study of aerodynamic phenomena and interactions occurring in high-speed stall conditions of transport aircraft up to flight Reynolds numbers~\cite{lutz2022}. The research unit consists of multiple sub-projects researching specific aspects of high-speed stall, including transonic buffet on the main wing's upper surface, the impact of an Ultra-high Bypass (UHBR) nacelle on buffet, shock oscillations on the lower wing surface between fuselage and nacelle, the interaction of the wing's wake with the tail plane as well as the development of methods for flow analysis and reduced order modeling. Sub-project 4 involving the authors' working group is focused on the analysis of the development of the wake of the wing at buffet conditions, and on its interaction with the tail plane.

Several measurement campaigns carried out in the European Transonic Windtunnel (ETW) support the various computational studies. These involve a realistic transport aircraft configuration suited to high Reynolds number testing, the XRF-1 model ("eXternal Research Forum") designed by Airbus~\cite{mann2019}. These experiments include time-resolved pressure-sensitive paint (PSP) of the surfaces and particle image velocimetry (PIV) measurements of the wake flow, and enable a combined analysis of the phenomena with as well as a validation of the numerical simulations. While testing in cryogenic conditions is necessary to attain flight-relevant Reynolds numbers, it proves rather costly, which limits the number of flow conditions and test points which can be realized. Moreover, high Reynolds numbers increase the difficulty of obtaining detailed optical measurements, i.e.~due to very thin boundary layers. For these reasons, the investigations of the XRF-1 configuration are complemented by studies using a generic tandem wing configuration in high-speed stall conditions which creates flow phenomena representative of a wing-wake-tail plane configuration. This configuration is analyzed experimentally in the Trisonic Wind Tunnel at the RWTH Aachen University (Rheinisch-Westf{\"a}lische Technische Hochschule Aachen) in the context of sub-project 6, at smaller Reynolds numbers of around two million.
In addition to the easier measurements, the smaller Reynolds numbers also allow for validation and further development of wall-modeled large eddy simulation (WM-LES) methods developed in the research unit, enabling them to tackle the high Reynolds number cases of the XRF-1.

The measurement campaign consists of two different phases. In the first phase, inflow conditions and angle of attack are chosen so that the front wing segment exhibits buffet, while the rear wing segment experiences steady flow conditions, which allows for an unhindered study of the impact of the front wing's wake, denoted here as configuration (A). For the second phase of the measurements, the angle of incidence of the rear wing segment is changed to increase its loading, provoking buffet also on the latter (configuration (C)). Therefore, the influence of the turbulent wake inflow on the buffet characteristics can be studied. Both cases are investigated numerically in this work by means of hybrid RANS/LES simulations (Reynolds-Averaged Navier-Stokes / Large Eddy Simulations), presented in section~\ref{sec5}. 

Thus, a suitable airfoil has to be selected for the rear wing segment in order to ensure steady flow at the inflow conditions that cause buffet on the front wing. Therefore, a preliminary computational analysis is performed using different freely accessible airfoil geometries. This preliminary study involves simulation methods of varying fidelity, with its results and the corresponding airfoil selection criteria described in section~\ref{chp:airfoil}. 

The present investigation is centered on the interaction of the turbulent wake created by the front wing in buffet conditions with the rear wing segment. The study is conducted using a total of four hybrid RANS/LES simulations of the tandem configuration with two different vertical positions of the rear wing, and two different settings of its angle of incidence, as explained in section~\ref{sec4}. These incidence settings are selected to create buffet conditions on the rear wing in the second case, whereas the lower incidence in the first case leads to smooth flow without buffet.

\section{Airfoil Selection for the Rear Wing Segment of the Tandem Wing Configuration}\label{chp:airfoil}

In order to enable a study of its interaction with the front wing's wake, a suitable airfoil has to be chosen for the rear wing segment which exhibits steady flow at the inflow conditions that provoke buffet on the front wing segment. The geometry of the airfoil needs to be freely accessible so that the results of this analysis can be subsequently released to a broader research community. Moreover, the airfoil should ideally resemble one that can be used for a horizontal tail plane of an actual transport aircraft in terms of its aerodynamic characteristics. This also means that a reasonable loading that would be expected for the tail plane of an aircraft in high-speed cruise conditions should be possible without entering the buffet regime of the airfoil. A purely subsonic flow around the airfoil without a shock is preferred, since this simplifies the validation of wall models for the WM-LES methods developed in the research unit. In addition, requirements regarding the manufacturing of the wind tunnel model and its instrumentation have to be taken into account. As the chord length of 75\,mm of the model is rather small, the installation of pressure sensors in the rear section of the wing is limited due to the available internal space. As a consequence, airfoils with a higher thickness in the rear part are preferred. 
The selection criteria for the airfoil are as follows.
First, the geometry of the airfoil should be openly accessible (criterion 1).
Second, the airfoil should exhibit steady flow conditions without buffet, preferably subsonic flow, for transonic inflow and low to medium aerodynamic loading (criterion 2).
Third, it should be similar to actual tail plane airfoils of transport aircraft with respect to aerodynamic characteristics (criterion 3).
Finally, it should exhibit sufficient thickness in the rear section to provide space for instrumentation (criterion 4).

Simultaneously fulfilling criteria 1 and 3 is challenging, since airfoils used in typical recent actual aircraft are mostly confidential. As the state of the art for jet-powered aircraft wings involves the use of transonic airfoils to minimize wave drag in cruise, the latter represent suitable candidate geometries. However, most publicly available transonic airfoils are designed for wings and therefore exhibit design lift coefficients greater than 0.5~\cite{tm85663,ar138,arccp1386,tp2969}, which is above the loading expected for a tail plane in cruise flight. Such an airfoil would be operated at an off-design condition if used as an HTP. The tail plane lift coefficient in cruise can be roughly estimated to be in the range between $-0.1$ and $-0.4$ for large aircraft based on available information on wing and tail plane geometry, weight and center of gravity, and cruising speed and altitude taken from the manufacturers' websites, (cf. manuals for "airplane/aircraft characteristics for airport (and maintenance) planning" of several Airbus and Boeing aircraft, e.g.~\cite{manual-737,manual-747,manual-787,manual-a320,manual-a330,manual-a380}). In contrast, the NACA series of airfoils for subsonic applications provides a vast selection of airfoils, of which many are suited for low to medium aerodynamic loading. Apart from their well-known application in general aviation aircraft, modified NACA airfoils have been in fact used for tail planes of early turboprop- and jet-powered aircraft, e.g. the Fokker F-27 and F28, the Sud Aviation Caravelle and the Cessna Citation 500~\cite{torenbeek1982}. Yet, their critical Mach numbers tend to be lower and the onset of strong shocks with boundary layer separation typically occurs earlier, so the range of possible lift coefficients where criterion 2 can be satisfied needs to assessed. With regard to that criterion, airfoils from the NACA\,6 digit series, which were originally designed for laminar flow, appear to perform better due to the more aft position of maximum thickness.

\begin{table}[h]
    \begin{center}
    \begin{minipage}{\textwidth}
    \caption{Airfoils considered in the preliminary analysis}\label{tab-prestudy}%
    \begin{tabular}{@{}ll@{}}
    \toprule
    Transonic Airfoils & OAT15A, RAE\,2822, RAE\,5213/5214/5215, NPL\,9510, \\ 
     & RAE\,6-9CK, NASA SC(2)-0410 \\
    NACA 4-digit-series & 1410  \\
    NACA-5-digit-series & 24010, 25010  \\
    NACA 6-digit-series & 63-110/-210/-212/-310/-312, 64-110, 64A109/-110/-111, \\
     & 65-110/-210/-212/-214/-310/-314, 65A210 \\
    Research airfoil & HGR-01 (mod.)~\cite{hgr01}\\
    \botrule
    \end{tabular}
    \end{minipage}
    \end{center}
\end{table}

The airfoils listed in Table \ref{tab-prestudy} were selected for further analysis, including airfoils from the NACA\,4-, 5- and 6-digit series, the research tail plane airfoil HGR-01~\cite{hgr01} and several widely used transonic airfoils~\cite{ar138,arccp1386,tp2969,tm85663}. A numerical study is performed for these airfoils with respect to criterion 2, investigating the range of lift coefficients where they exhibit purely subsonic flow, using the MSES~\cite{mses_drela} solver toolbox in a first step. MSES solves the full potential equations or the Euler equations for two-dimensional flows around airfoils coupled with an integral boundary layer method to account for boundary layers, allowing a fast computation of lift, drag and pressure distributions for several airfoils and inflow conditions.
\begin{figure}[h]%
    \centering
    \includegraphics[clip,trim={60 70 240 190},width=0.5\textwidth]{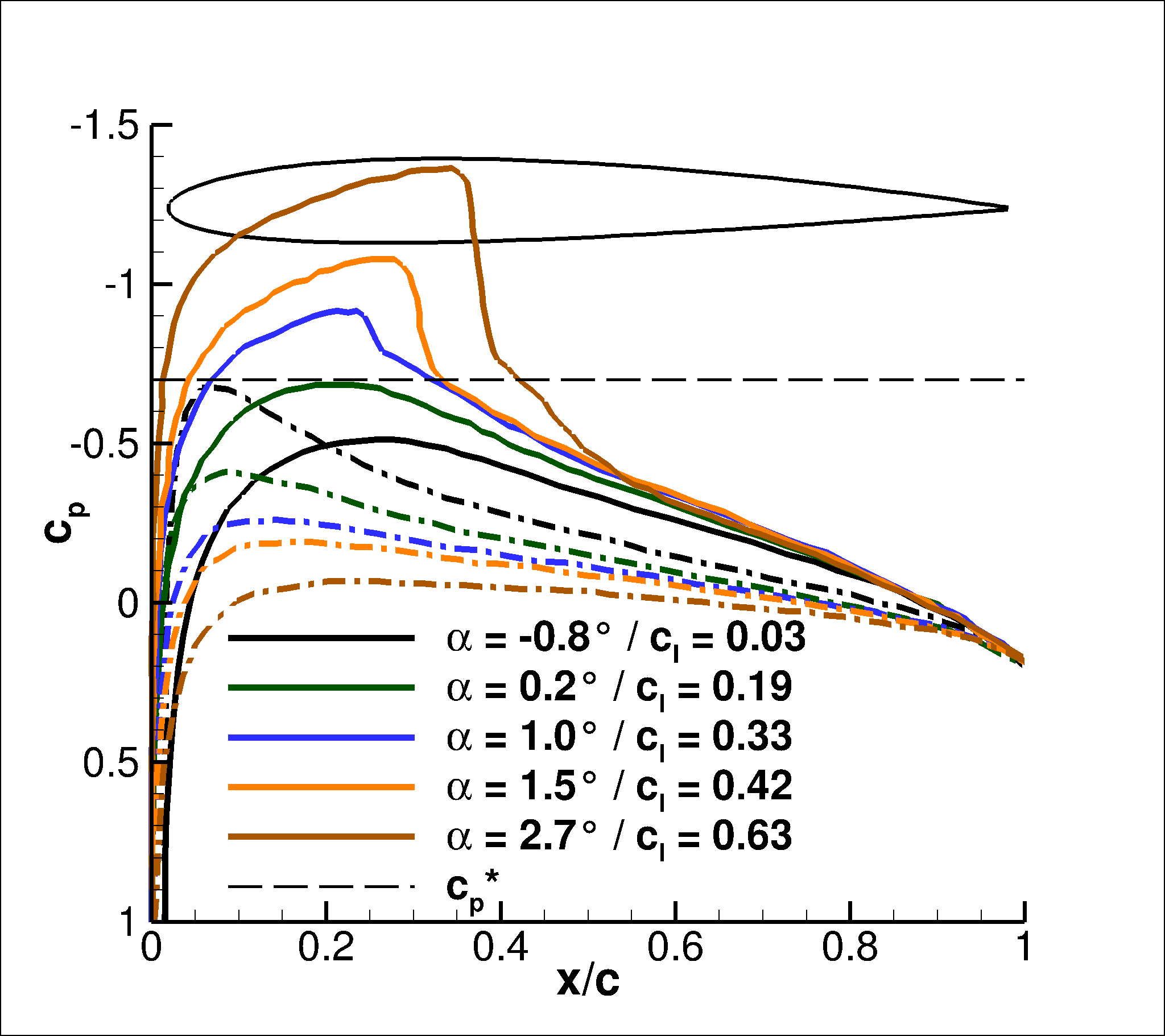}%
    \includegraphics[clip,trim={60 70 240 190},width=0.5\textwidth]{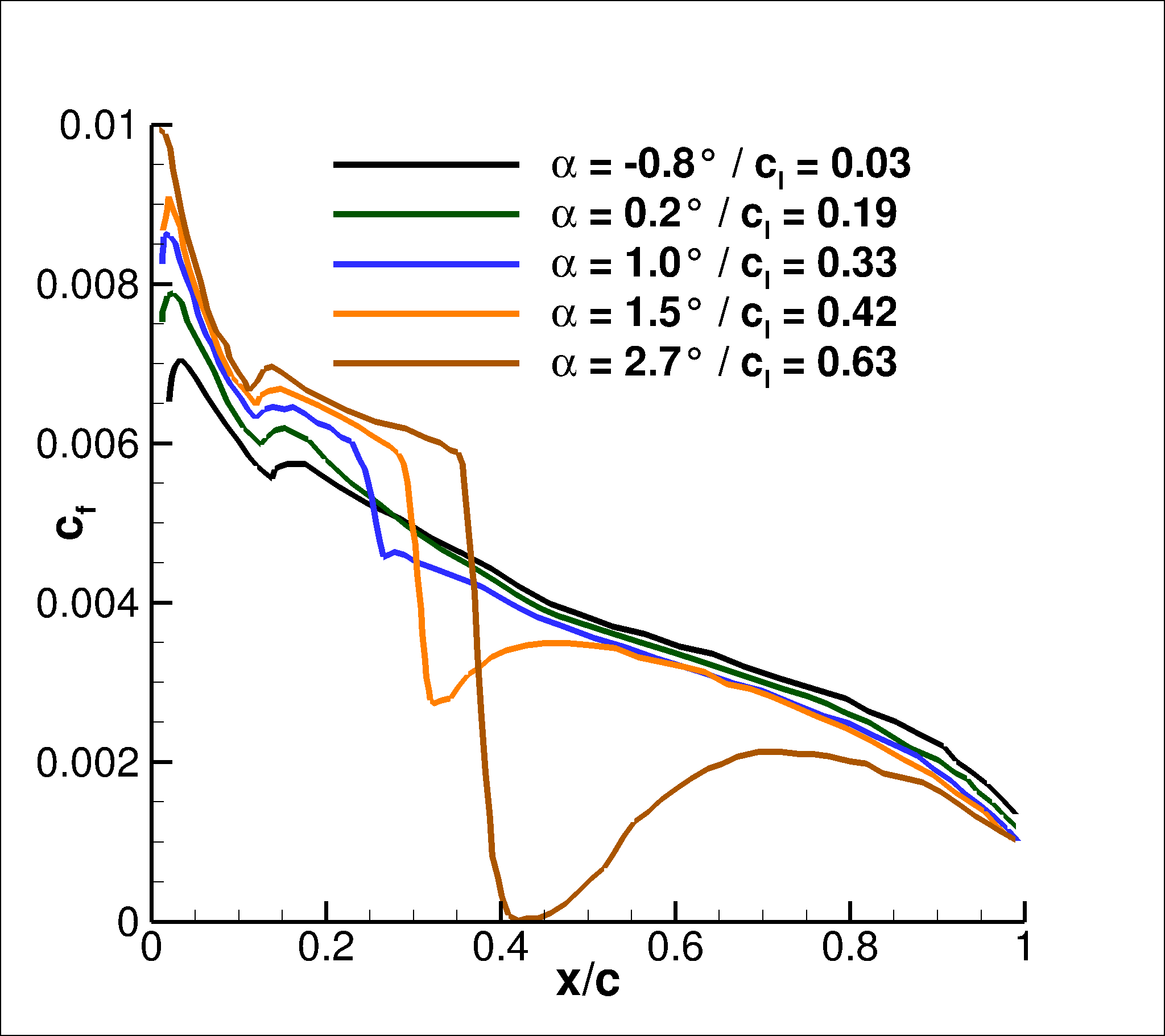}
    \caption{Surface pressure (left) and friction (right) distributions of the NACA\,1410 airfoil, $Re=1.3\cdot10^{6}$, $Ma=0.72$ obtained from MSES. Horizontal dashed line denotes critical pressure coefficient $c_p^{*}$}\label{fig-naca1410}
\end{figure}
Fig.~\ref{fig-naca1410} shows the calculated distributions of surface pressure and friction coefficients using the NACA\,1410 airfoil as an example.
The airfoil shape is also depicted here for reference.
The inflow Mach number is 0.72, with angles of attack varying from -0.8$^{\circ}$ to 2.7$^{\circ}$, and lift coefficients between 0.03 to 0.63. The dashed lines denote the lower side of the airfoil. The Reynolds number is set to 1.3 million corresponding to the value expected for the wind tunnel tests at that time, and the boundary layer is treated as fully turbulent. It can be seen that the critical condition, where a local Mach number of one is attained, is reached at an angle of attack close to 0.2$^{\circ}$ with a lift coefficient of 0.19. Decreasing the angle of attack below -0.8$^{\circ}$ with lift coefficients below 0.03, critical flow is reached on the lower side of the airfoil. At a higher angle of attack to 1$^{\circ}$ (at $c_l=0.33$), on the other hand, a distinct shock forms on the upper side. However, as the shock is relatively weak, the impact on the boundary layer remains small, as there is only a small decrease visible in surface friction $c_f$. The strengthening shock then begins to notably influence the boundary layer profile at around $\alpha=1.5^{\circ}$ and $c_l=0.42$, with a decrease in friction of more than 50\%. Boundary layer separation eventually occurs at around $\alpha=2.7^{\circ}$ and $c_l=0.63$. It should be noted here that the prediction of separation onset using an integral method can only ever be an approximation. Buffer may occur at even higher angles of attack, however this can not be accurately modeled using a steady-state flow solver. Even under this premise, evaluation of transonic buffet occurrence at moderate $\alpha$ can be carried out given the present data. Transonic buffet requires the presence of both a shock and a boundary layer separation. Operating in the subsonic regime should provide enough margin to the buffet boundary, as a small increase in loading may lead to a weak shock, but not necessarily to separation or buffet.
Taking the above into account, the range of lift coefficients usable for the application, i.e. with purely subsonic flow, extends from 0.03 to 0.19. It should be noted that the airfoil would be installed in an inverted manner as an HTP, providing a down force of $-0.19 \leq c_l \leq -0.03$. This available range is rather narrow for the airfoil, as the shock begins to form at a comparably small loading.
\begin{figure}[h]%
    \centering
    \includegraphics[clip,trim={60 70 240 190},width=0.5\textwidth]{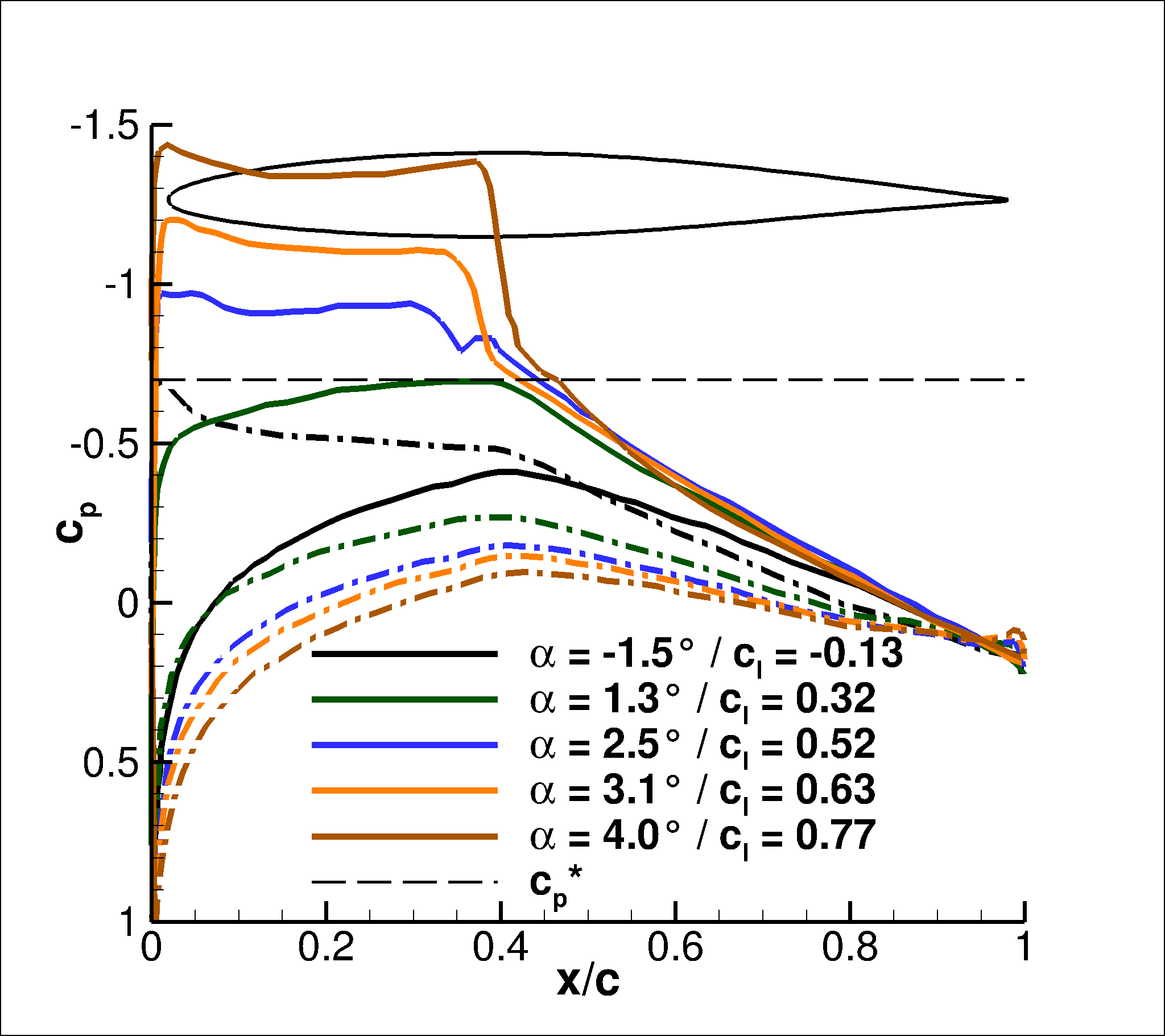}%
    \includegraphics[clip,trim={60 70 240 190},width=0.5\textwidth]{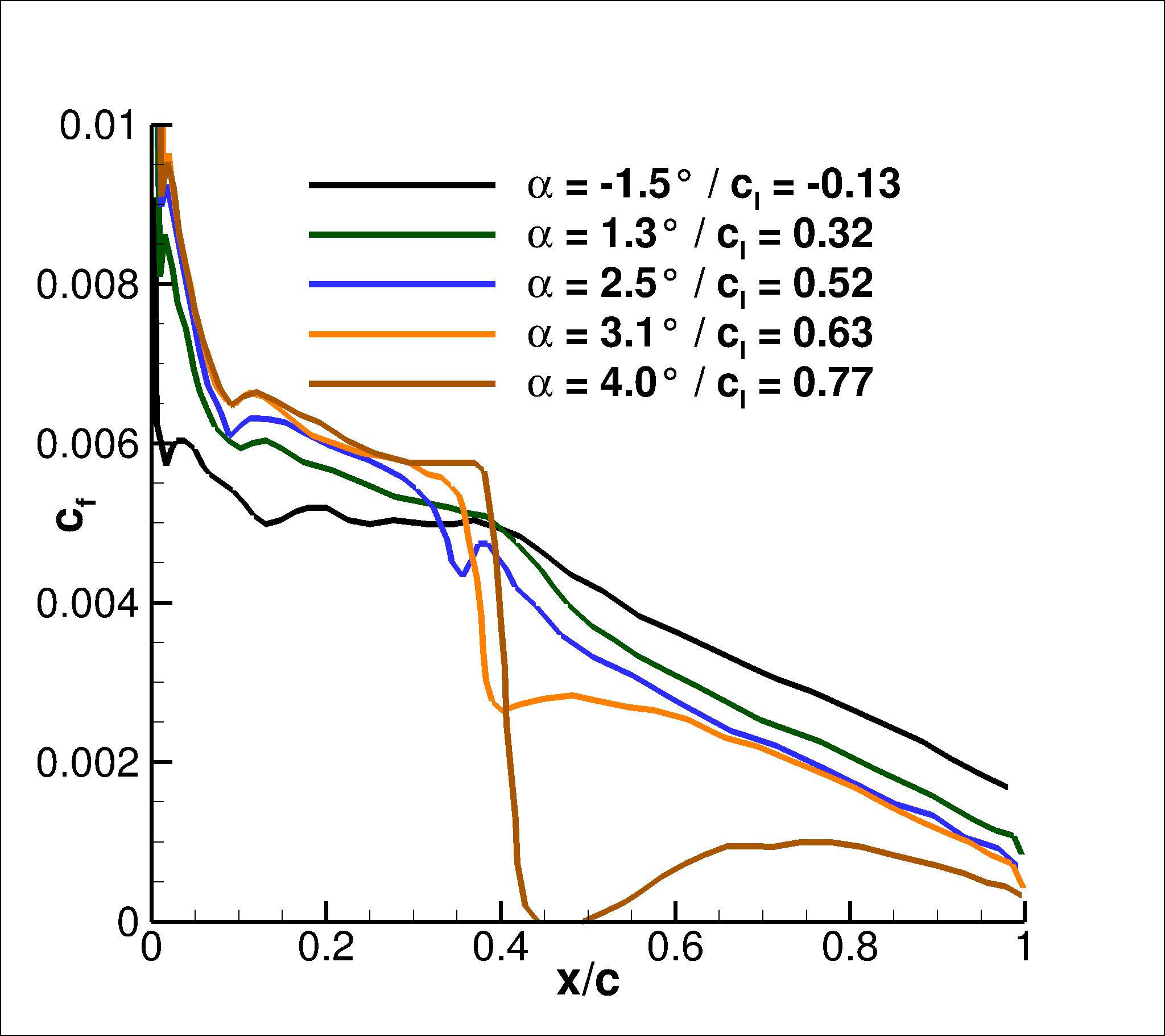}
    \caption{Surface pressure (left) and friction (right) distributions of the NACA\,64A110 airfoil, $Re=1.3\cdot10^{6}$, $Ma=0.72$ (MSES calculation)}\label{fig-naca64a110}
\end{figure}

\begin{figure}[h]%
    \centering
    \includegraphics[clip,trim={60 70 240 190},width=0.5\textwidth]{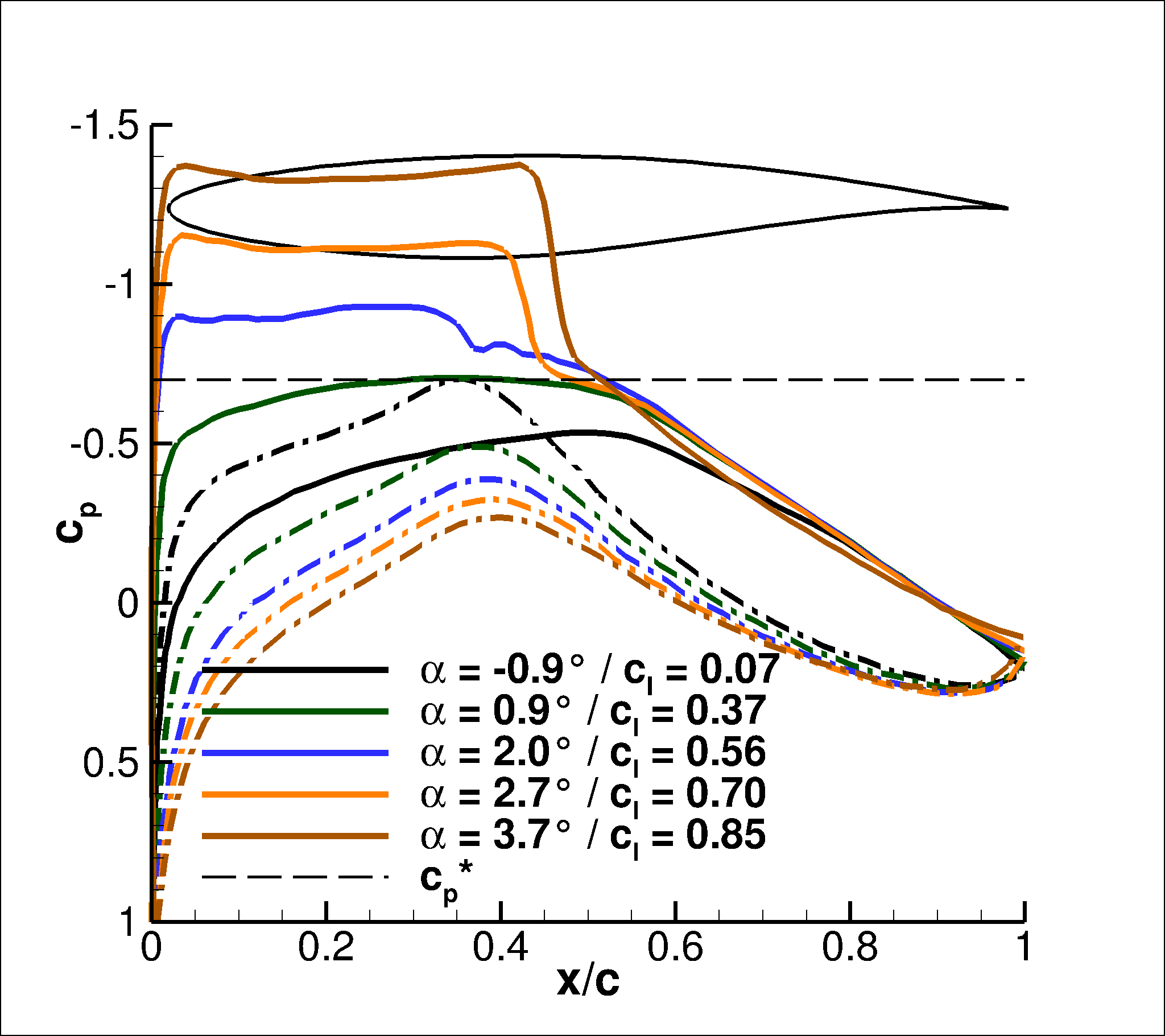}%
    \includegraphics[clip,trim={60 70 240 190},width=0.5\textwidth]{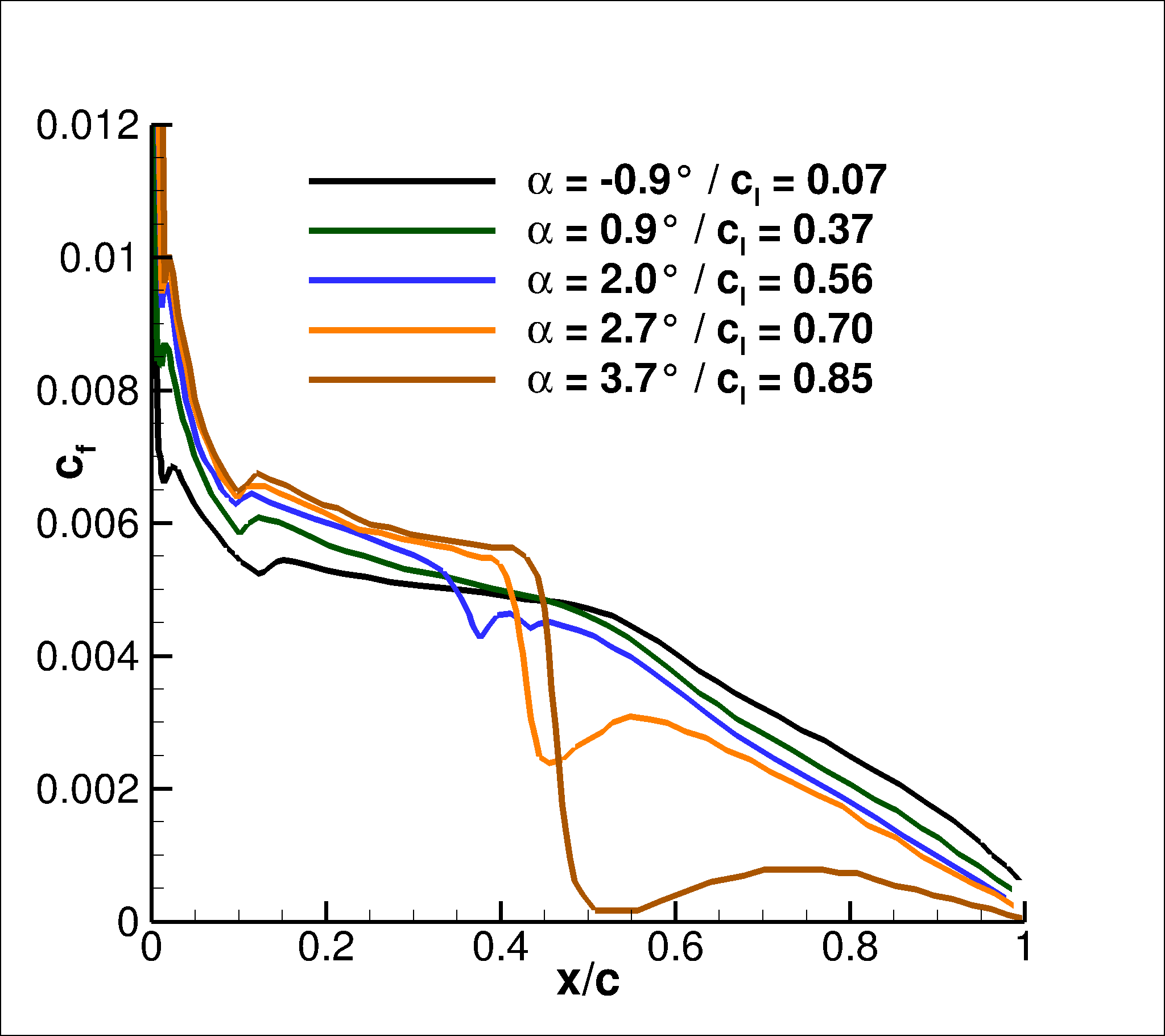}
    \caption{Surface pressure (left) and friction (right) distributions of the RAE\,2822 airfoil, $Re=1.3\cdot10^{6}$, $Ma=0.72$ (MSES calculation)}\label{fig-rae}
\end{figure}

A wider range of $c_l$ can be achieved with the laminar airfoil NACA\,64A110, as seen in the pressure and friction distributions in Fig.~\ref{fig-naca64a110} for the same inflow conditions as above. Purely subsonic flow is present between $\alpha=-1.5^{\circ}$ and $\alpha=1.3^{\circ}$ with lift coefficients between -0.13 and 0.32. Here, a distinct shock appears only above $\alpha=2.5^{\circ}$ or $c_l=0.52$. Notably, a supersonic plateau forms in the pressure distribution on the upper side, similar to that of a transonic airfoil, in contrast to the continuous flow acceleration observed for the NACA\,4-digit airfoil discussed above. This is due to the aft position of maximum thickness and less curvature on the upper surface. Indeed, the pressure distributions resemble those of transonic airfoils quite well, with the exception of the missing aft-loading typical for the latter, cf. Fig.~\ref{fig-rae}. Therefore, criterion 3 can be considered as partially fulfilled, considering that the characteristics of the front part of the airfoil are of greatest importance for wake interaction, as it is being directly impinged by turbulent structures. 
A decrease in surface friction of more than 50\% is found for angles of attack above approximately $3.1^{\circ}$ or $c_l=0.63$ in Fig.~\ref{fig-naca64a110}, and boundary layer separation does not occur until $\alpha=4^{\circ}$ and $c_l=0.77$. Thus, the usable range of lift coefficients according to criterion 2 is $-0.13 \leq c_l \leq 0.32$, which is considerably wider than that of the NACA\,1410.
\begin{figure}[h]%
    \centering
    \includegraphics[clip,trim={60 70 240 190},width=0.5\textwidth]{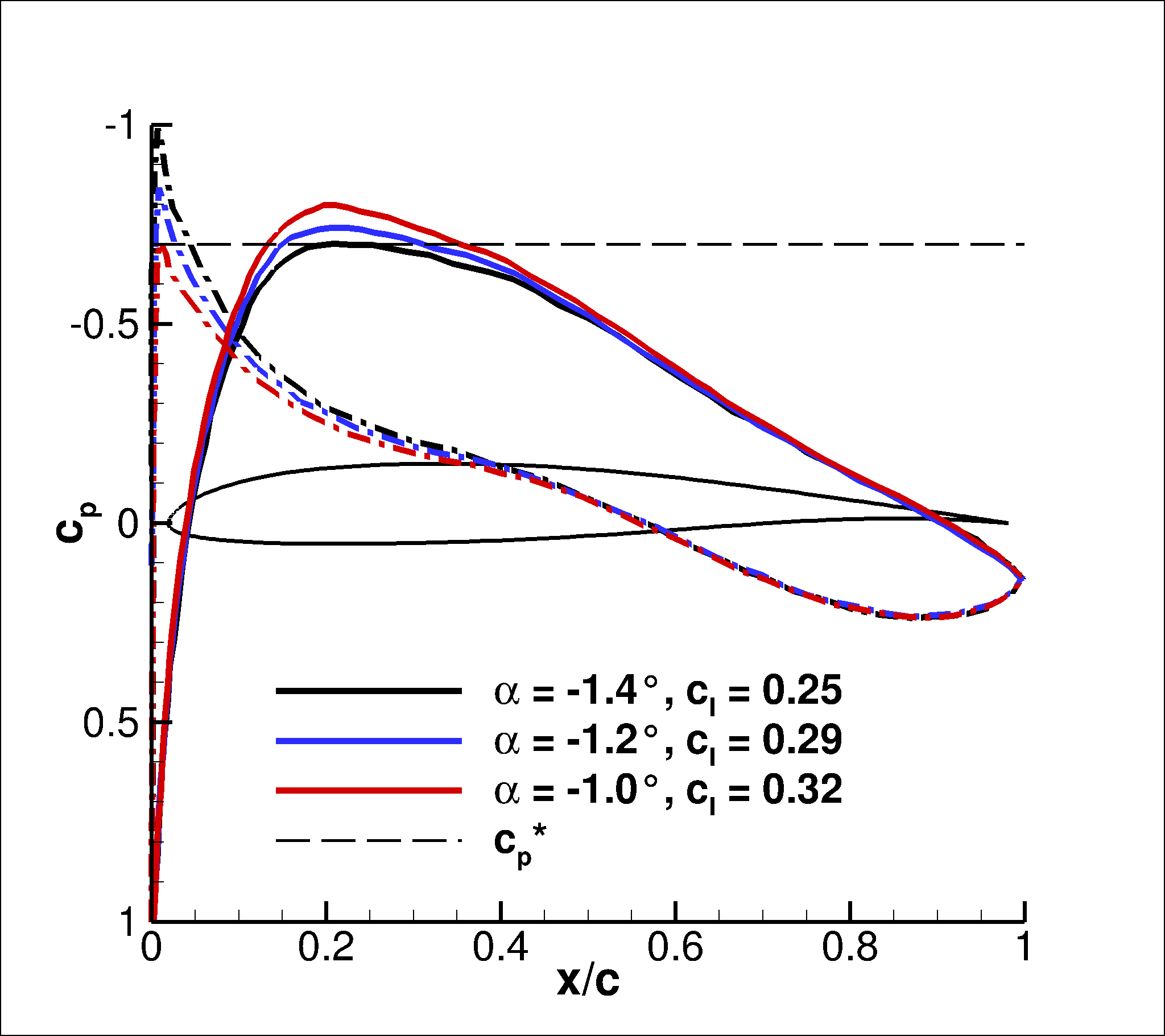}
    \caption{Surface pressure distributions of the HGR-01 research tail plane airfoil, $Re=1.3\cdot10^{6}$, $Ma=0.72$ (MSES calculation)}\label{fig-hgr}
\end{figure}
\begin{figure}[h]%
    \centering
    \includegraphics[clip,trim={60 70 240 190},width=0.5\textwidth]{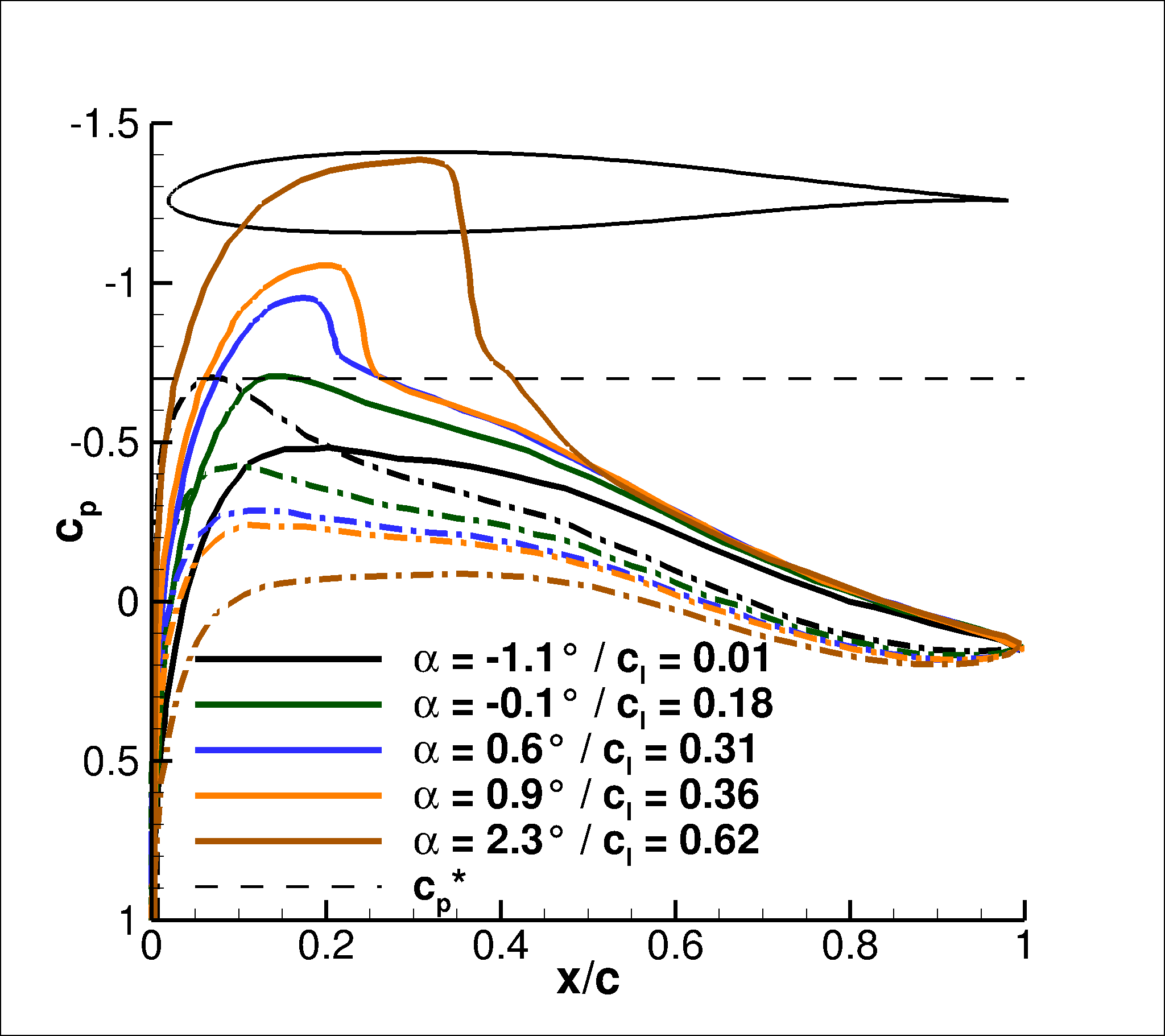}%
    \includegraphics[clip,trim={60 70 240 190},width=0.5\textwidth]{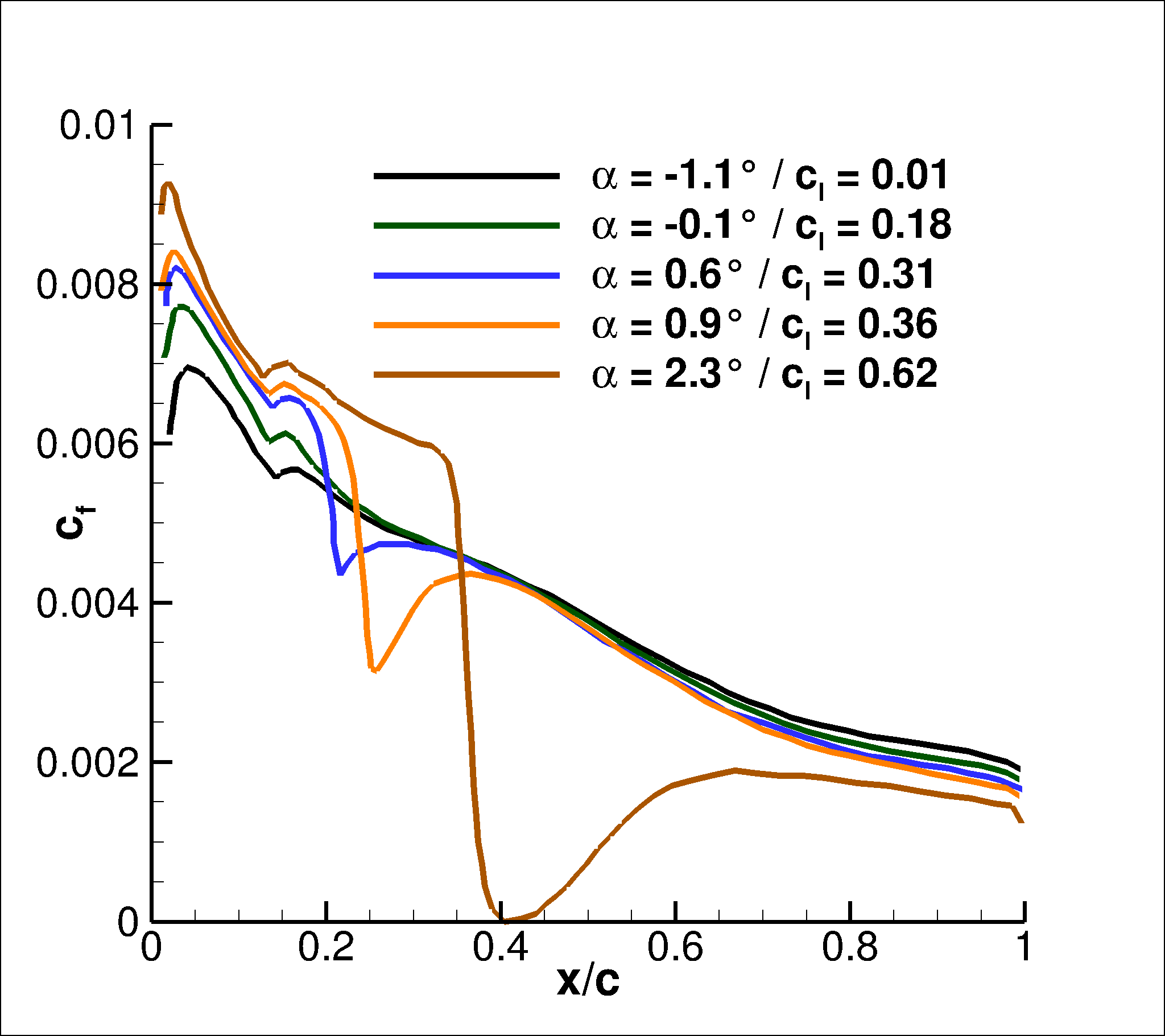}
    \caption{Surface pressure (left) and friction (right) distributions of the HGR-01 airfoil with reduced camber (1\%), $Re=1.3\cdot10^{6}$, $Ma=0.72$ (MSES calculation)}\label{fig-hgr-1perc}
\end{figure}
The research tail plane airfoil HGR-01, designed by the Technische Universit{\"a}t Braunschweig for the analysis of a mixed leading-edge trailing-edge stall behaviour~\cite{hgr01}, in contrast, exhibits no angle of attack with purely subsonic flow. When the angle of attack is decreased to reduce the loading of the upper side, starting from $\alpha=0^{\circ}$, for example, a suction peak forms on the lower side of the airfoil which becomes supersonic before the velocity on the upper side falls below $Ma=1$, as shown in Fig.~\ref{fig-hgr}. Furthermore, flow separation on the lower side is reached at $c_l=0.01$. Yet, this is not the case anymore when the airfoil's camber is scaled down from 2.5\% to 2\% chord or less. Here, cambers of 1\%, 1.5\% and 2\% chord have been investigated. With decreasing camber, the regime of subsonic flow grows. However, the lift coefficient that can be reached without separation becomes smaller. For a camber of 1\% chord, the range of subsonic flow extends from $c_l=0.01$ to $c_l=0.18$, as displayed in Fig.~\ref{fig-hgr-1perc}. A noticeable shock forms for $c_l=0.31$ and separation is finally reached at $c_l=0.62$.

As a representative example for the investigated transonic airfoils, the distributions of pressure and surface friction of the RAE\,2822 airfoil are shown in Fig.~\ref{fig-rae}. Subsonic flow is present between $c_l=0.07$ and $c_l=0.37$. The appearance of a strong shock is evident for $c_l=0.70$ and flow separation begins at about $c_l=0.85$.
\begin{figure}[h]%
    \centering
    \includegraphics[clip,trim={10 10 10 40},width=0.95\textwidth]{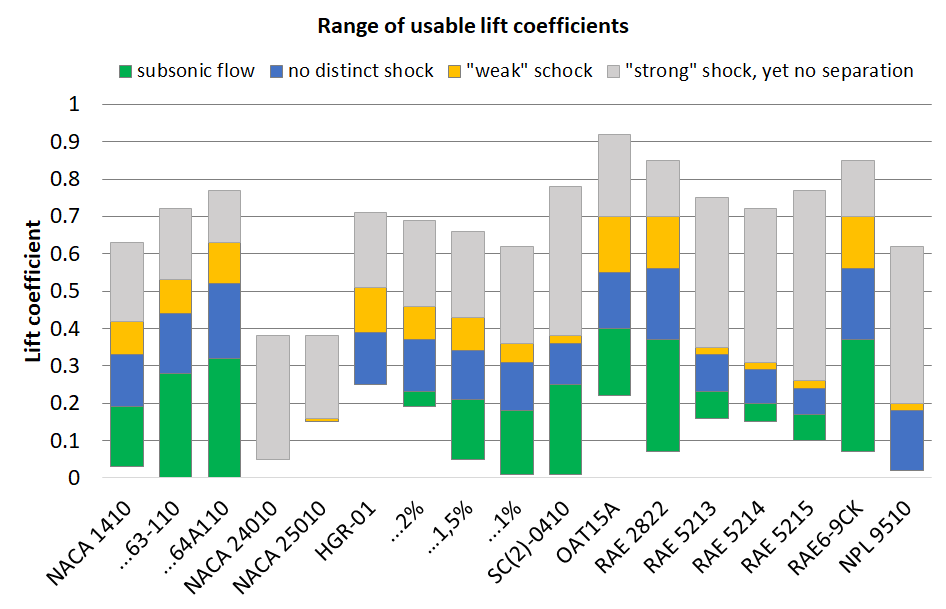}
    \caption{Overview of usable lift coefficients for the analyzed airfoils (MSES calculation)}\label{fig-betriebsbereiche}
\end{figure}

The results of the MSES calculations are summarized in Fig.~\ref{fig-betriebsbereiche} which illustrates the range of usable lift coefficients for the analyzed airfoils. Here, the range of purely subsonic flow is indicated in green and the range where no distinct shock is present is shown in blue. In the yellow and gray areas, a shock is present but no flow separation is observed. Thus, the upper end of the region marked with gray indicates the on-set of flow separation behind the shock. As described above, a further distinction is made between conditions with a relatively “weak” (marked in yellow) or “strong” (marked in gray) shock based on the decline of the surface friction behind the shock. Here, a decrease of more than 50\% is used as a threshold for defining a strong shock. 
Based on these results, the NACA\,64A110 airfoil is selected because of its wide range of lift coefficients with subsonic flow (down to $c_l=0$). Thus, the latter is chosen for all further investigations.

\section{Numerical Setup}\label{sec4}

\subsection{Geometry of the Tandem Wing Configuration}\label{subsec41}

As described above, the investigated tandem configuration consists of two straight, untapered and unswept wing segments. The front wing uses the supercritical OAT15A airfoil, whereas the laminar airfoil NACA\,64A110 is selected for the rear wing segment based on the results of the the work described in section~\ref{chp:airfoil}. The former has been the subject of several experimental and numerical investigations on 2D buffet in the past and is thus chosen for the analysis as well as other studies within the research unit.
\begin{figure}[h]%
    \centering
    \includegraphics[clip,trim={70 40 240 1310},width=0.9\textwidth]{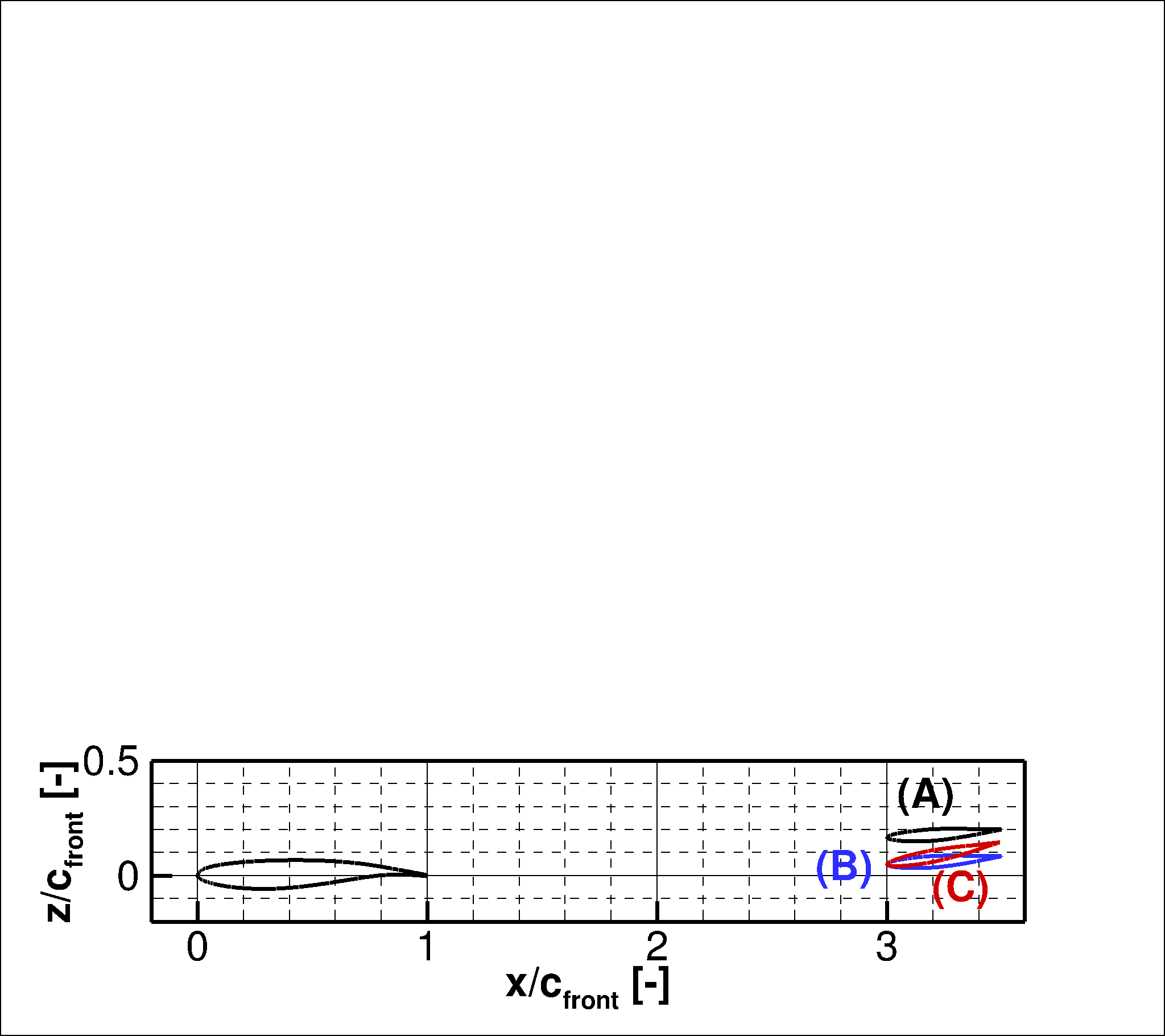}
    \caption{Overview of the analyzed configurations, cf. Table \ref{tab-configurations}}\label{fig-configurations}
\end{figure}
The computational setup includes the front wing segment with a chord length of $c_{\mathrm{front}} = 0.15\,m$ and the rear wing segment with a chord length of 0.075\,m, corresponding to the experimental setup of the investigations performed in the research unit.
The OAT15A airfoil used on the front segment exhibits a blunt trailing edge with a relative thickness of 0.5\%.
For the NACA\,64A110 airfoil applied to the rear segment, the sharp trailing edge has been slightly modified resulting in a blunt shape for meshing reasons. The trailing edge thickness is 0.2\% chord.
The horizontal distance between the trailing edge of the front wing segment and the leading edge of the rear wing segment corresponds to $2\,c_{\mathrm{front}}$, i.e. the latter is positioned at $x/c=3$, as indicated in Fig.~\ref{fig-configurations}.
As sketched in Fig.~\ref{fig-configurations} and listed in Table~\ref{tab-configurations}, two different vertical positions for the rear wing segment are considered, denoted as (A) and (B): In configuration (A), the rear segment is located $1/6\,c_{\mathrm{front}} = 0.025\,m$ above the centerline of the front wing segment, and $1/20\, c_{\mathrm{front}} = 0.0075\,m$ above the centerline in configuration (B), respectively. While the relative size of the rear wing segment and the horizontal separation between the wing segments are roughly based on the geometry of actual transport aircraft (cf. e.g.~\cite{manual-737,manual-a330}),
the vertical position of the rear segment of configuration (A) is chosen so that it is placed directly in the center of the front wing's wake to maximize the interaction effects in order to facilitate the investigation. It should be noted that the trajectory of the wake varies over the buffet cycle so that a fixed position will not be exactly in the center of the wake at every instance in time. In order to analyze the sensitivity with respect to the vertical position, the position of configuration (B) is slightly offset from the wake's center. Since knowledge of the wake's trajectory is required for this placement but unknown a priori, simulations of the isolated front wing segment were performed in advance, denoted as configuration (0). 

In addition, two different angles of incidence ($\alpha_I$) of the rear wing segment are investigated. 
These are realized by rotating the rear segment around its leading edge.
A first setting of $\alpha_I=-4^{\circ}$, corresponding to an angle of $\alpha_{rear}=1^{\circ}$ with regard to the free stream for the selected inflow angle of $\alpha_{\infty}=5^{\circ}$, leads to a lift coefficient of about $c_l =-0.2$ due to the downwash of the front wing segment. This is comparable to a typical tail plane lift coefficient in cruise conditions (cf. section \ref{chp:airfoil}). This setting is used in both configurations (A) and (B). With this $\alpha_I$, buffet occurs only on the front wing segment. It should be noted that the NACA\,64A110 airfoil is mounted in an inverted orientation to represent a typical tail plane section with negative camber optimized for a negative lift coefficient. The second setting of $\alpha_I=-11^{\circ}$, which corresponds to an angle of $\alpha_{rear}=-6^{\circ}$ with regard to the free stream, is chosen to provoke buffet also on the rear wing segment to study the influence of the turbulent wake inflow on the buffet characteristics. This configuration uses the same position of the rear segment as configuration (B), and is denoted as (C). Finally, to allow for a comparison with an undisturbed buffet flow, simulations of the isolated rear wing segment at an angle of attack of $\alpha_{rear}=-6^{\circ}$ are performed as a reference (configuration (D)).

\begin{table}[h]
    \begin{center}
    \begin{minipage}{\textwidth}
    \caption{Overview of the analyzed configurations}\label{tab-configurations}%
    \begin{tabular}{@{}ll@{}}
    \toprule
    Identifier   & Configuration    \\
    \midrule
    (0) & Isolated front wing segment  \\
    (A) &  Tandem configuration with rear segment at $z/c=1/6$ and $\alpha_I=-4^{\circ}$ \\
    (B) & Tandem configuration with rear segment at $z/c=1/20$ and $\alpha_I=-4^{\circ}$ \\
    (C) & Tandem configuration with rear segment at $z/c=1/20$ and $\alpha_I=-11^{\circ}$\\
    (D) & Isolated rear wing segment\\
    \botrule
    \end{tabular}
    \end{minipage}
    \end{center}
\end{table}

\subsection{Computational Grids and Boundary Conditions}\label{subsec42}

\begin{figure}[h]%
    \centering
    \includegraphics[clip,trim={0 400 0 400},width=0.95\textwidth]{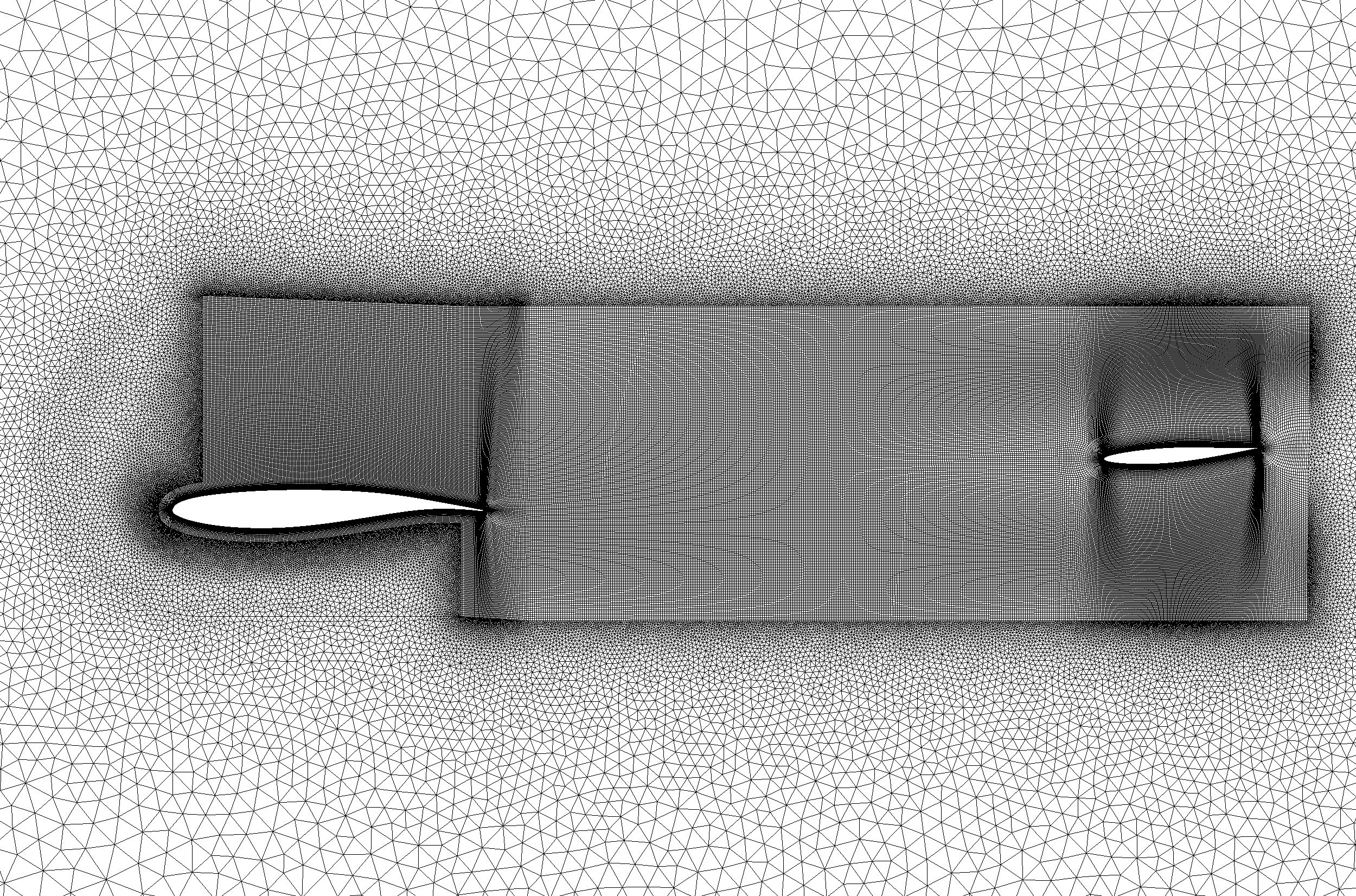}
    \caption{Main section of the computational grid used for the hybrid RANS/LES simulations of the tandem wing configuration (configuration (A))}\label{fig-mesh}
\end{figure}

The computational grids used for the simulations are hybrid meshes consisting of hexahedrons in the structured and triangular prisms in the unstructured regions, respectively, and are created using the meshing software Pointwise. The grids of all configurations share the same topology, meshing parameters and cell sizes. Fig.~\ref{fig-mesh} visualizes the mesh for configuration (A). The grids employ farfield boundary conditions and do not include the wind tunnel geometry. All boundary layers, as well as the region of the shock above the front wing segment and the area between the wing segments are discretized with hexahedrons to assure a high cell quality. This minimizes numerical dissipation for the resolution of the shock movement and the turbulent structures in the wake. The height of the first cell layer above all surfaces is chosen so that a $y^{+}$ of less than 0.4 is achieved as recommended for the application of Reynolds stress models. A total of 607 points is used the surface of the front wing segment, with a cell size of 0.4\% chord over most of the airfoil. The surface of the rear wing segment is discretized with 388 points and an average cell size of 0.8\% of its chord, respectively. For the wake region between the two segments, a resolution of 0.7\% of the front wing's chord is chosen.
The spanwise extent of the computational domain is set to 0.0735\,m for all cases, which corresponds to 49\% of the chord length of the front wing segment. To allow for the resolution of three-dimensional turbulent structures present in the wake, the spanwise dimension is discretized with 70 cell layers, created by an extrusion of the sectional 2D meshes in spanwise direction. Therefore, the cells in the area between the wing segments are nearly cubic with a resolution of 0.7\% of the front wing's chord. Periodic boundary conditions are applied at the spanwise boundaries of the domain. The cylindrical far field boundary has a radius of 50\,m, which is equal to about 95 times the overall length of the tandem wing configuration. In total, the meshes for the different configurations consist of around 16 million points. 

\subsection{Flow Conditions}
All simulations of the tandem wing configuration are performed for an inflow Mach number of 0.72 and a Reynolds number of 2 million with respect to $c_{\mathrm{front}}$ (i.e. 1 million based on $c_{\mathrm{rear}}$), with an inflow angle of $\alpha_{\infty}=5^{\circ}$ relative to the front segment's centerline. These conditions are well inside the buffet regime of the OAT15A airfoil. The boundary layers of both segments are tripped at 7\% of their respective chord, corresponding to the experimental setup.
Tripping is realized in TAU by switching the production term in the transport equations of the Reynolds stresses off in the user-defined laminar regions, and on in the turbulent regions, respectively.

\subsection{Numerical Setup}

The CFD simulations used for further analysis in the present work are performed with the finite-volume code TAU (version 2018.1.0)~\cite{schwamborn2008} provided by the German Aerospace Center (Deutsches Zentrum f{\"u}r Luft- und Raumfahrt, DLR), applying the zonal hybrid RANS/LES method AZDES (Automated Zonal Detached Eddy Simulation) developed in the authors' working group at the Institute of Aerodynamics and Gas Dynamics (IAG) and described by Ehrle et al.~\cite{ehrle2020, ehrle20202}. This method introduces a fixed, zonal separation of RANS and LES regions based on geometric parameters and flow properties obtained during a precursor unsteady RANS (URANS) simulation. This enables a robust treatment of the attached boundary layer and the region of shock-boundary layer interaction in RANS mode, as well as an improved resolution of turbulence in regions of separated flow, when compared to purely geometry-dependent zonal approaches (e.g. Deck's Zonal Detached Eddy Simulation (ZDES)~\cite{deck2005}). The scale-resolving behavior is realized by activating the Detached Eddy Simulation (DES) model (in this case a Delayed Detached Eddy Simulation (DDES)~\cite{spalartDDES}) outside of user-defined, zonal RANS areas.

The definition of these RANS areas is controlled by the user based on wall distance dependent parameters and involves a threshold value for the integral turbulent length scale $L_{t}$ which is a priori accumulated in the precursor URANS simulation for the same flow conditions. $L_{t}$ is calculated using the quantities of the turbulence model and is supposed to give an estimate of the overall extent of flow separation over time.
It is computed as
\begin{equation*}
	L_t = \max_{\{\text{time}\}}(\sqrt{k_t}/(c_\mu\,\omega))\text{,}
\end{equation*}
where $k_t$ is the turbulent kinetic energy, $\omega$ denotes the specific turbulence dissipation rate estimated by the turbulence model and $c_\mu$ is a model constant.
In areas where $L_{t}$ reaches high values, large turbulent structures are to be expected that can be resolved by the mesh. The threshold value for $L_t$, above which the DES model is activated, denoted as $L_c$, can be adjusted by the user.
The latter implies that regions with $L_t < L_c$ are marked as RANS zones, whereas areas with $L_t > L_c$ are covered in DDES mode in the following actual hybrid RANS/LES simulation.
This is realized~\cite{ehrle20202} by modifying the hybrid length scale for the hybrid simulation from the (D)DES length scale 
\begin{align*}
	L_{DES} &= \min(L_{RANS},\,C_{DES}\,\Delta)\text{~or} \\
	L_{DDES} &= L_{RANS} - f_d \cdot (L_{RANS}-C_{DES}\,\Delta)\text{~\cite{spalartDDES}},
\end{align*}
with the RANS length scale $L_{RANS}$ (which is equal to the wall distance in SA based models or calculated from $k_t$ and $\omega$ for others), the hybrid model constant $C_{DES}$, the filter width $\Delta$ (originally the largest cell edge's length) and the DDES shielding function $f_d$,
to 
\begin{equation*}
	L_{AZDES} = L_{RANS} \cdot (1 - f_a) + L_{(D)DES} \cdot f_a\text{.}
\end{equation*}
Here, $f_a$ is a blending function based on the ratio $L_t/L_c$,
\begin{equation*}
	f_a = 0.5\cdot(1+\tanh(8\cdot[L_t/L_c-1]))\text{,}
\end{equation*}
i.e.~$f_a = 1$ for $L_t/L_c \gg 1$ and $f_a = 0$ for $L_t/L_c \ll 1$.
Additionally, all regions further away from surfaces than a user-defined wall distance are forced into DES mode regardless of $L_t$.

\begin{figure}[h]%
	\centering
	\includegraphics[clip,trim={70 50 550 940},width=0.5\textwidth]{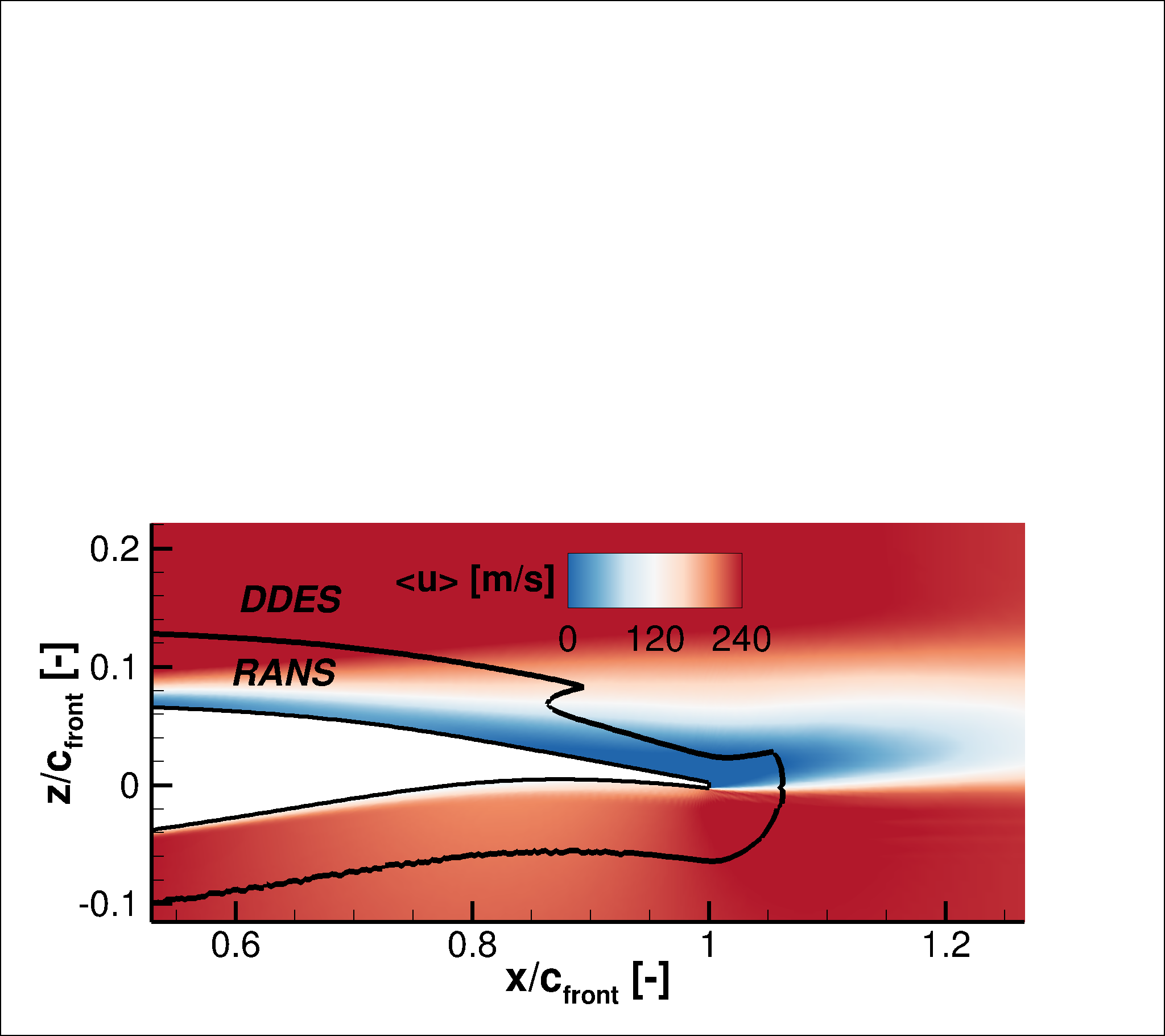}%
	\includegraphics[clip,trim={70 50 550 940},width=0.5\textwidth]{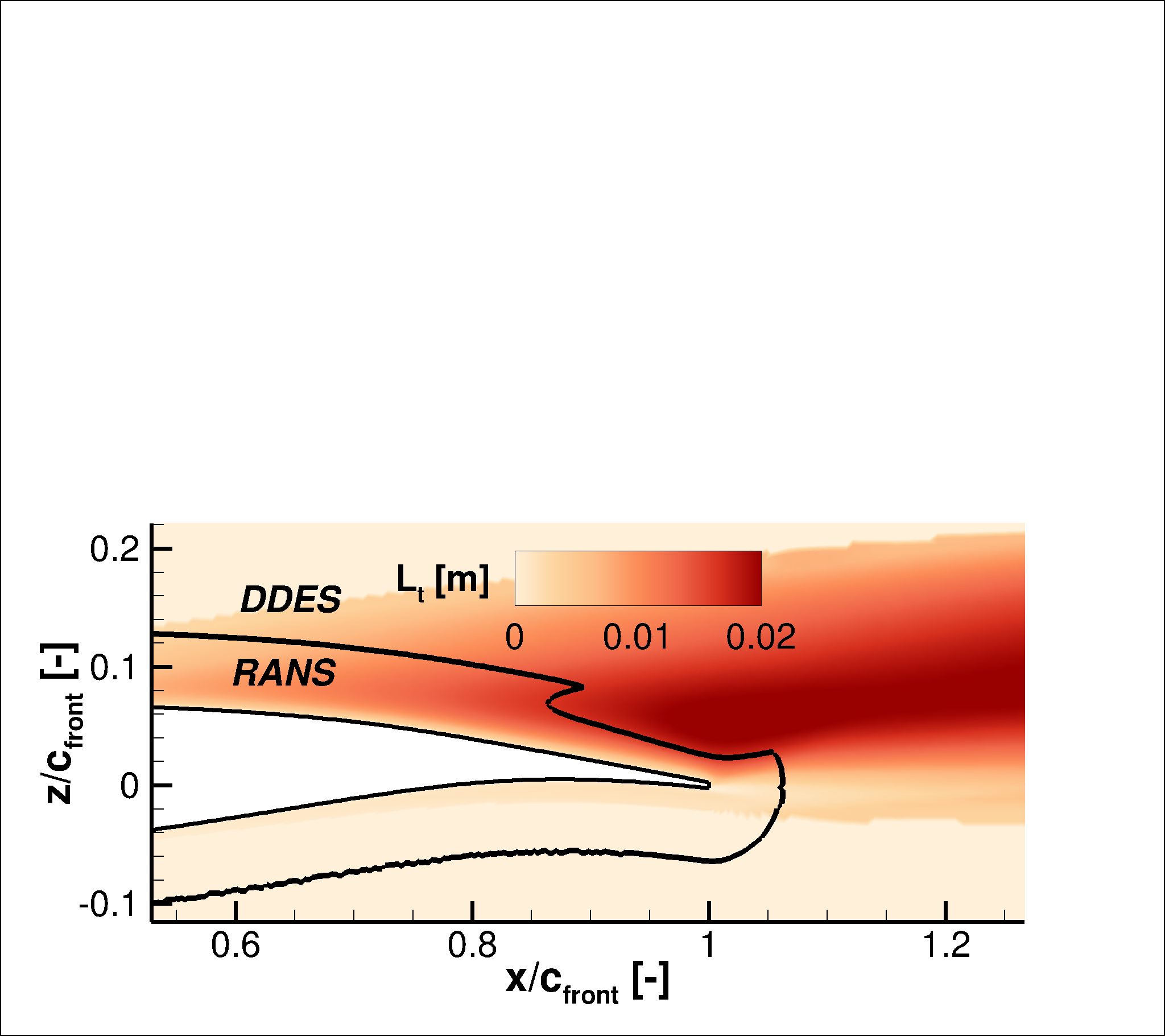}
	\caption{RANS/LES zone division in the flow field around the front wing section, together with the averaged streamwise velocity $u$ (left) and the integral turbulent length scale $L_t$ (right) from the precursor URANS simulation}\label{fig:zone-division}
\end{figure}

For the present simulations, a threshold value of $L_c/c=11\%$ is selected in order to enable the switching from RANS to LES mode at the most upstream position possible, without influencing the periodic shock motion of transonic buffet at the same time. Regions further away from the airfoil's surface than $6\%$ of the chord are forced into DES mode by applying the aforementioned wall distance dependent parameters.
These settings are based on recommendations of previous buffet studies~\cite{ehrle2020,ehrle20202,illi2016} that suggest a threshold length scale $L_t$ of at least 6\%\,chord and a DES switching wall distance of at least 4\%\,chord, respectively, to shield the region of the shock and attached boundary layers sufficiently. Thus, grid-induced separation is avoided, which would otherwise lead to a non-physical dampening of the shock motion.
During precursor simulations with different values for $L_c$, the settings above were verified for the present configuration based on the respective buffet amplitudes observed.
The choice of the threshold length scale is also dependent on the grid resolution. It should be set according to the size of smallest turbulent eddies that can be resolved. As the central scheme used in the present work is of second order, the number of grid points required to resolve a flow structure can roughly be estimated to be in the range of ten to twenty~\cite{spalart2001,Swanson1992,loewe2016}. Based on the grid resolution of 0.7\% chord, this corresponds to 7\% to 14\% chord, a range that fits to the value chosen.
The resulting zone division is shown in Fig.~\ref{fig:zone-division}. The shock (foot) itself and the begin of the shock-induced separation are shielded and treated in RANS mode, wheres a DDES is performed beginning in the region above the trailing edge and in the wake area between the wing segments, as intended within the AZDES approach.

All simulations are second order accurate in space and time by applying a central flux approximation for the convective terms and an implicit dual time stepping scheme based on the second Backward Differentiation Formula (BDF\,2) for time integration, respectively. For the discretization of the turbulence model equations, a second order Roe scheme is used, and gradients are evaluated using the Green Gauss Theorem.
The quality of the numerical results heavily depends on the error induced by excessive artificial dissipation. Therefore, the numerical parameters are chosen to enable a stable, converging simulation with a minimum of artificial dissipation, as recommended by DLR~\cite{TAU2018,probst2015}. This is realized by using TAU's matrix type artificial dissipation in combination with a hybrid flux blending. 95\% matrix dissipation and a fourth order dissipation coefficient k(4) of 1/1024 are applied the scale resolving areas. 80\% matrix dissipation and k(4)= 1/64 in RANS areas ensure numerical stability. Furthermore, a low dispersion scheme, as described by Loewe et al.~\cite{loewe2016}, is applied in scale resolving areas, which minimizes numerical dispersion errors in these regions. The locally kinetic energy conserving skew symmetric scheme~\cite{kok2009}, which is the preferred scheme for hybrid RANS/LES computations with low-dissipation requirements, is chosen for the mean flow fluxes.  

A physical time step of $\Delta t = 4.4\cdot10^{-6}\,s$ representing 150 time steps per convective time unit $\Tilde{t} = c_{\mathrm{front}}/u_{\infty}$ is used, leading to a local Courant-Friedrichs-Lewy (CFL) number of approximately one in the wake region and a sufficient temporal resolution of the shock motion.
One buffet cycle is resolved with about 1900 time steps.
It should be noted that the time step size is not primarily determined by the shock motion, which can be captured properly with a coarser temporal resolution, but by the requirements of the LES in the wake, for which a local CFL number of around one is generally recommended for accuracy~\cite{spalart2001,kornhaas2008,mocket2014}.
The SSG/LRR-$\omega$ Reynolds stress model~\cite{eisfeld2005} serves as turbulence model for the RANS regions and as subgrid scale model in the LES regions. In combination with 150 inner iterations per time step, a sufficient convergence of the force and moment coefficients within one time step is achieved. 
For the inner iterations, the implicit backward Euler scheme is employed together with a 2v geometrical multigrid method for convergence acceleration, based on the full approximation storage scheme (FAS)~\cite{brandt77}.
The corresponding (pseudo) CFL number for the inner iterations is set to 4. On the coarser grid levels of the multigrid scheme, it is reduced to 2, and further down to 1 in areas of high pressure gradient to guarantee numerical stability.

The setup of the time accurate hybrid RANS/LES simulations is started with a steady state computation with gradually increased angle of attack until reaching $\alpha = 5^\circ$. Subsequently, an unsteady RANS simulation is performed over 50 convective times $\Tilde{t}$. This serves as precursor simulation for the AZDES method to accumulate the turbulent length scale for the subsequent RANS/LES zone distribution. The ${\Delta}_{max}$-filter width definition is used.
The time series used for statistics and analyses in the following section comprises around 30 to 40 convective time scales, depending on the case, which represents two to three buffet cycles on the main wing segment.
The sampling rate corresponds to the applied time step.

\section{Results}\label{sec5}

The results presented in the following sections \ref{subsec51} and \ref{subsec52} correspond to the simulations with the first setting of the angle of incidence of the rear wing segment, i.e. the case where buffet occurs only on the front wing segment (configurations (A) and (B)). First, the transonic buffet and the corresponding unsteady flow separation from the front wing segment are characterized in section~\ref{subsec51}. The separated turbulent wake is then investigated including a spectral analysis of the pressure and velocity fluctuations. The interaction of the separated wake with the rear wing segment is analyzed in detail in section~\ref{subsec52}. The corresponding findings for the second incidence setting (configuration (C)), for which buffet is also present on the rear segment, are then discussed in section~\ref{subsec53}. 

\subsection{Buffet Flow and Wake Development of the Front Wing Segment}\label{subsec51}

\begin{figure}[h]%
    \centering
    \includegraphics[clip,trim={20 20 200 180},width=0.5\textwidth]{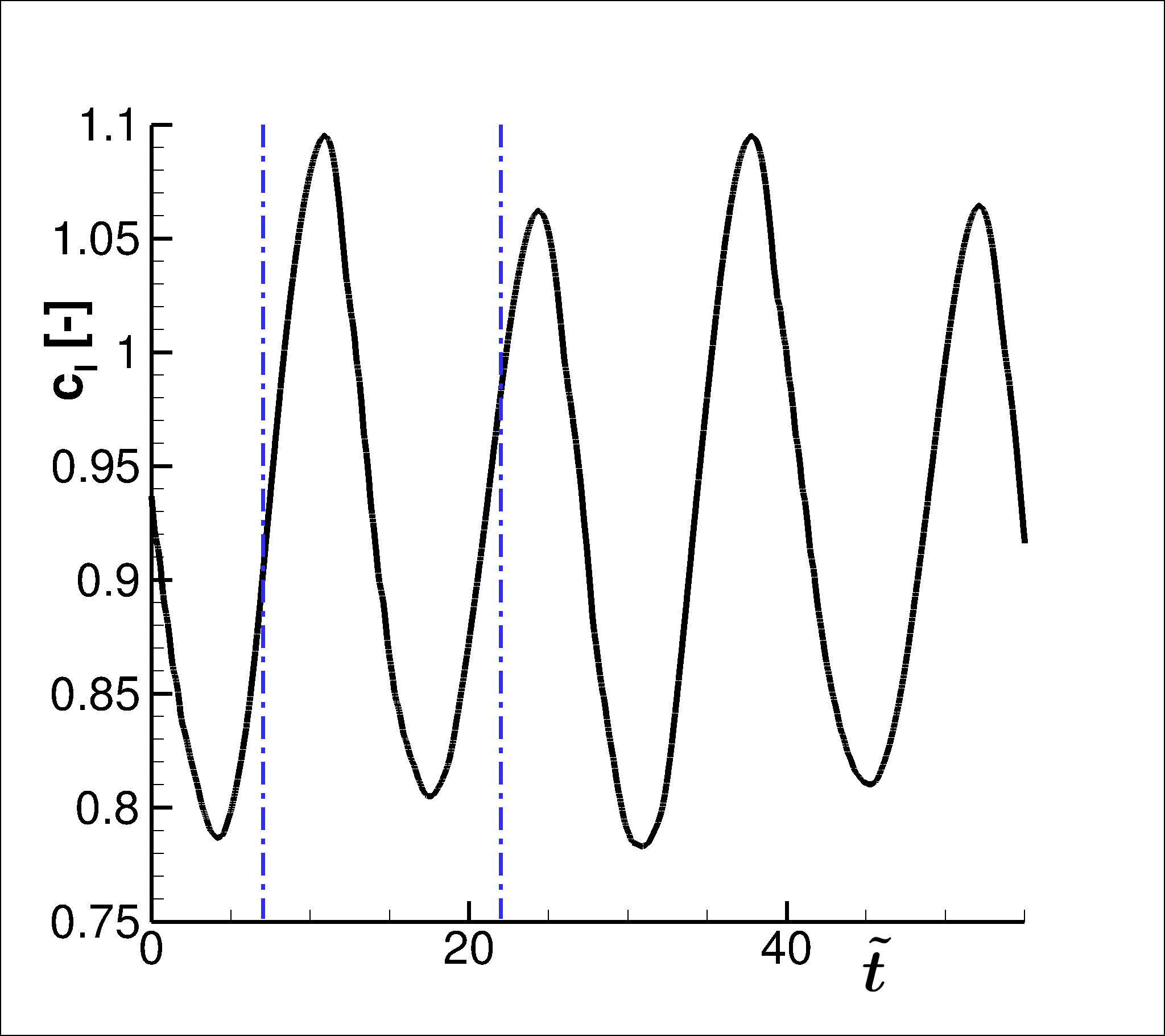}%
    \includegraphics[clip,trim={20 20 200 180},width=0.5\textwidth]{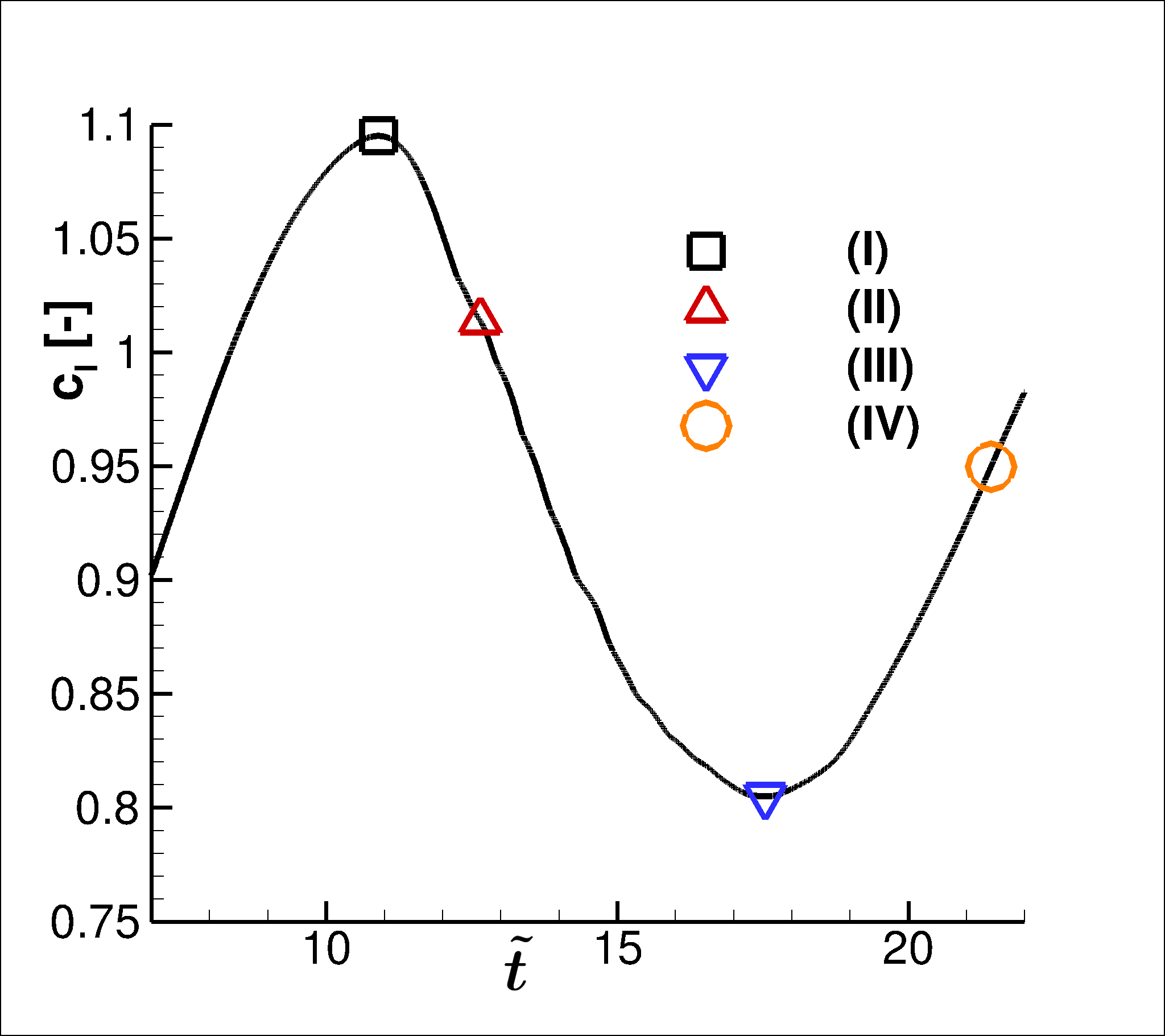}
    \caption{Lift coefficient $c_l$ of the front wing segment over several buffet cycles}\label{fig:cl-signal}
\end{figure}

\begin{figure}[h]%
    \centering
    \includegraphics[clip,trim={40 30 240 880},width=0.5\textwidth]{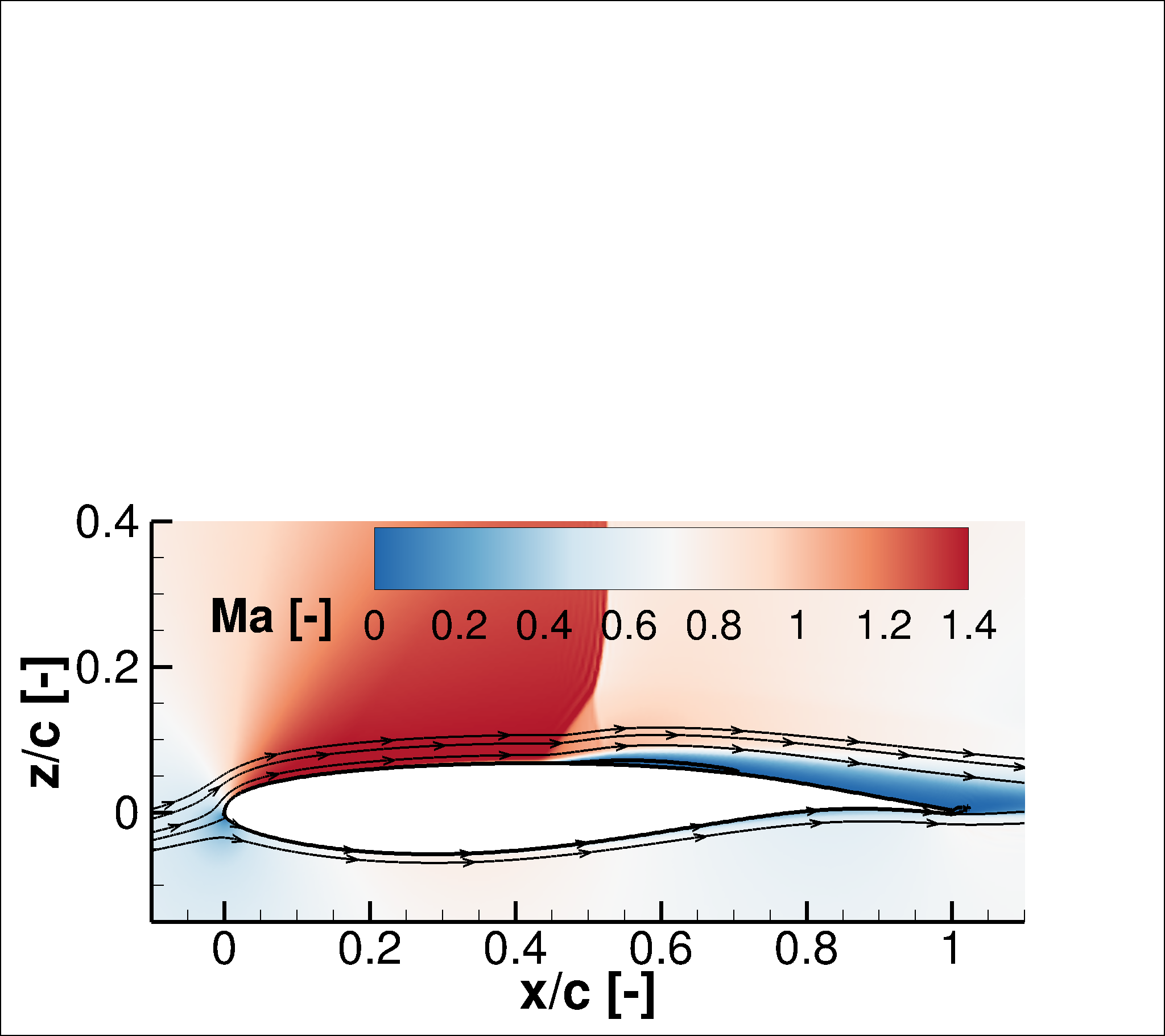}%
    \includegraphics[clip,trim={40 30 240 880},width=0.5\textwidth]{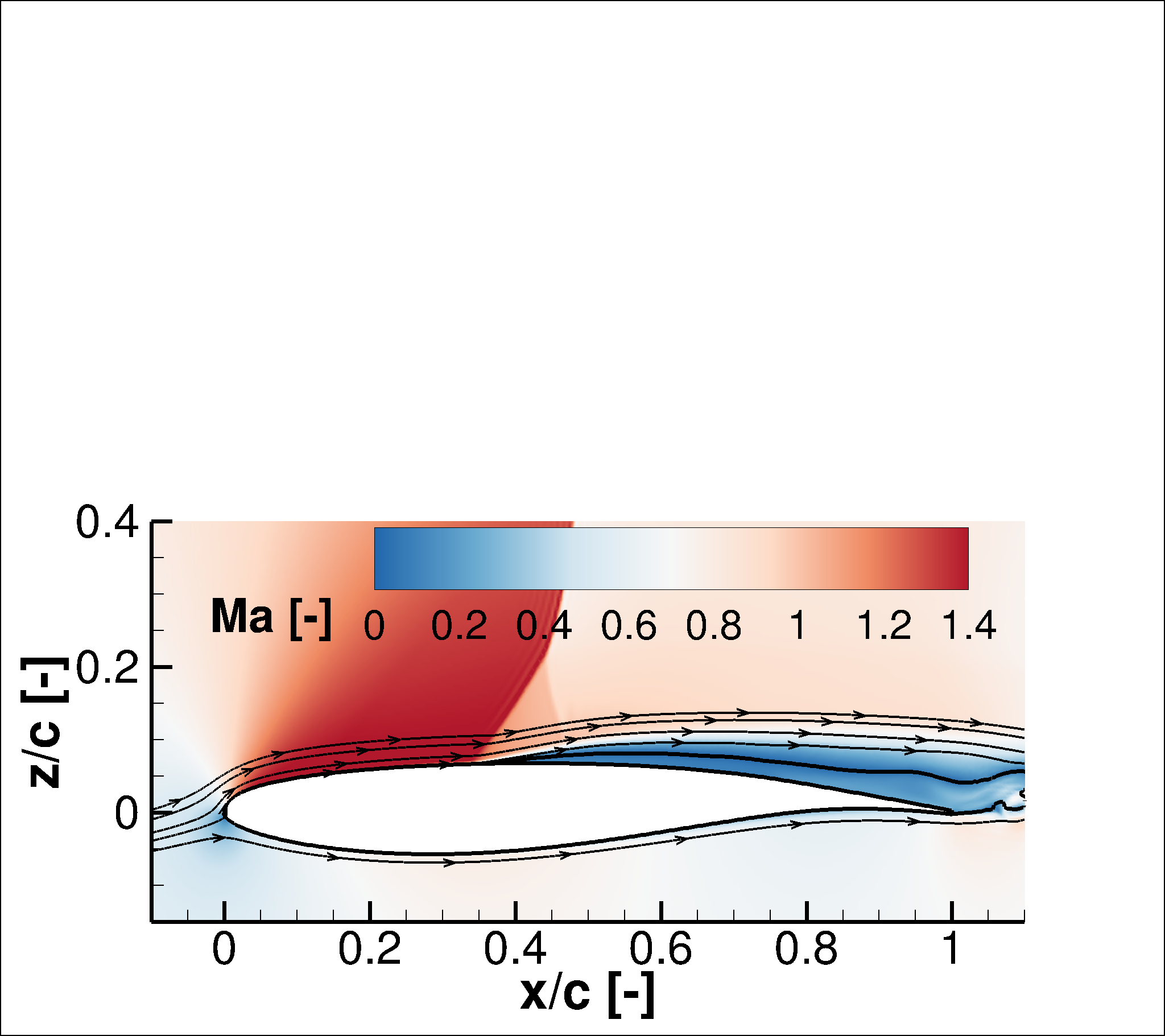}
    \includegraphics[clip,trim={40 30 240 880},width=0.5\textwidth]{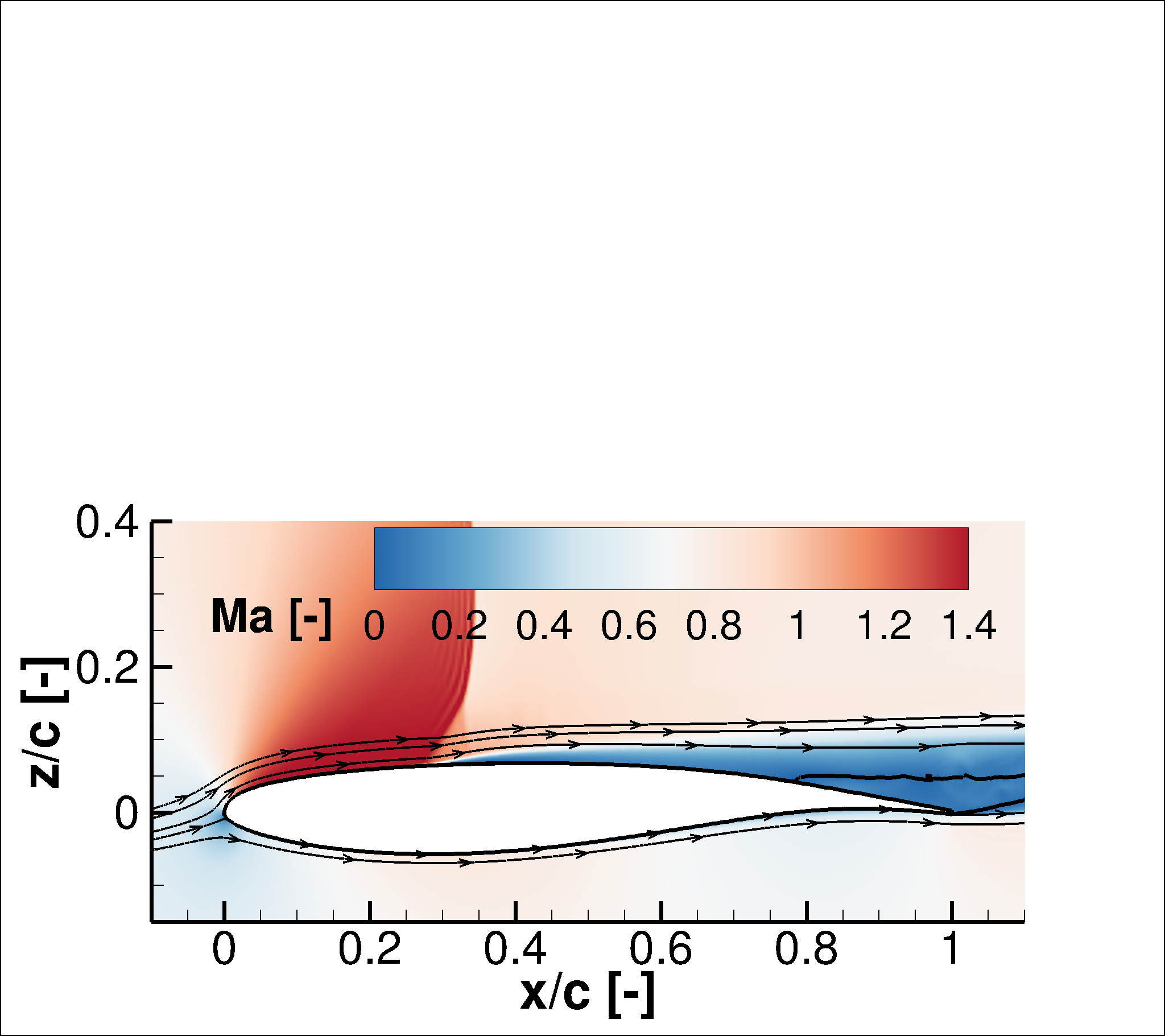}%
    \includegraphics[clip,trim={40 30 240 880},width=0.5\textwidth]{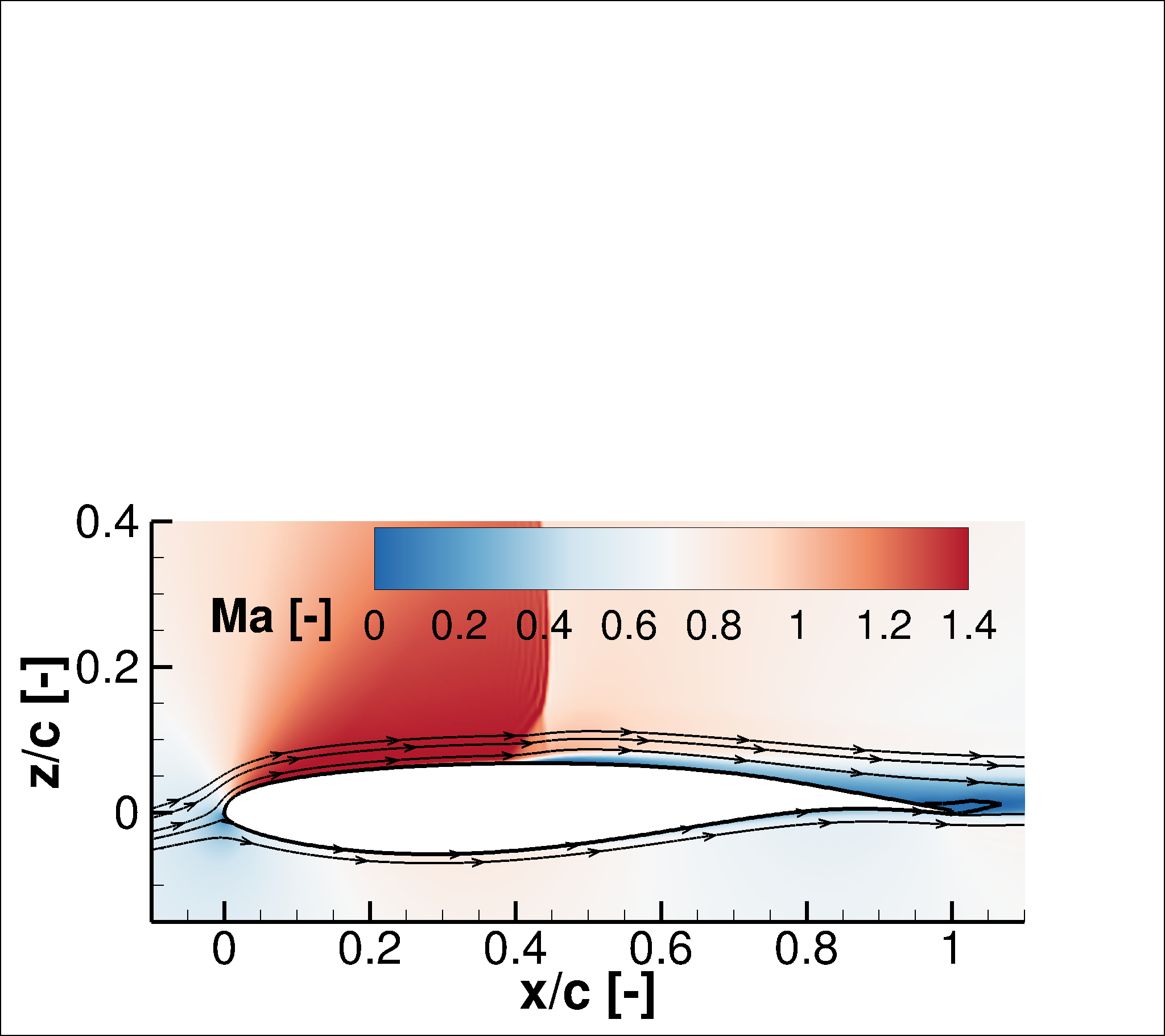}
    \caption{Mach number in the flow field surrounding the front wing segment; at the time of the most downstream shock position (I) (top left), during the upstream shock motion (II) (top right), at the time of the most upstream shock position (III) (bottom left), and during the downstream shock motion (IV) (bottom right).}\label{fig-separation-front}
\end{figure}

For the considered inflow conditions, a pronounced shock oscillation is present on top of the upper side of the front wing segment. Accompanying the movement of the shock, the lift of the front wing segment varies strongly during the buffet cycle, as displayed in Fig.~\ref{fig:cl-signal}, which shows the development of the lift coefficient over several buffet periods. A zoomed-in view of one exemplary cycle is shown on the right. The lift coefficient oscillates between a minimum of $c_l=0.81$ and a maximum of $c_l=1.10$ during this cycle, corresponding to an amplitude of $\hat{c_l}=0.145$, and with a frequency of 118.5\,Hz or a Strouhal number of $Sr=f\cdot c/U_{\infty}=0.0745$, based on the chord of the front segment, which is in the range generally found for 2D buffet.
For example, Jacquin et al. found $Sr=0.068..0.075$~\cite{jacquin2009}, Schrijer et al. reported $Sr=0.071$~\cite{schrijer2018,aguanno21} and Accorinti et al. listed $Sr=0.056..0.076$~\cite{accorinti2022} for the OAT15A airfoil.
Thus, the buffet period is equal to 13.4 convective time units. The corresponding experimental investigations performed in the research group (cf. section~\ref{chp:tandem_intro}) of the isolated OAT15A airfoil yield a comparable Strouhal number of $Sr=0.0714$ for the same inflow conditions, as reported by Schauerte et al.~\cite{schauerte2022}. The mean lift coefficient averaged over one buffet period equals to $c_l=0.94$. 
\begin{figure}[h]%
    \centering
    \includegraphics[clip,trim={20 20 200 180},width=0.5\textwidth]{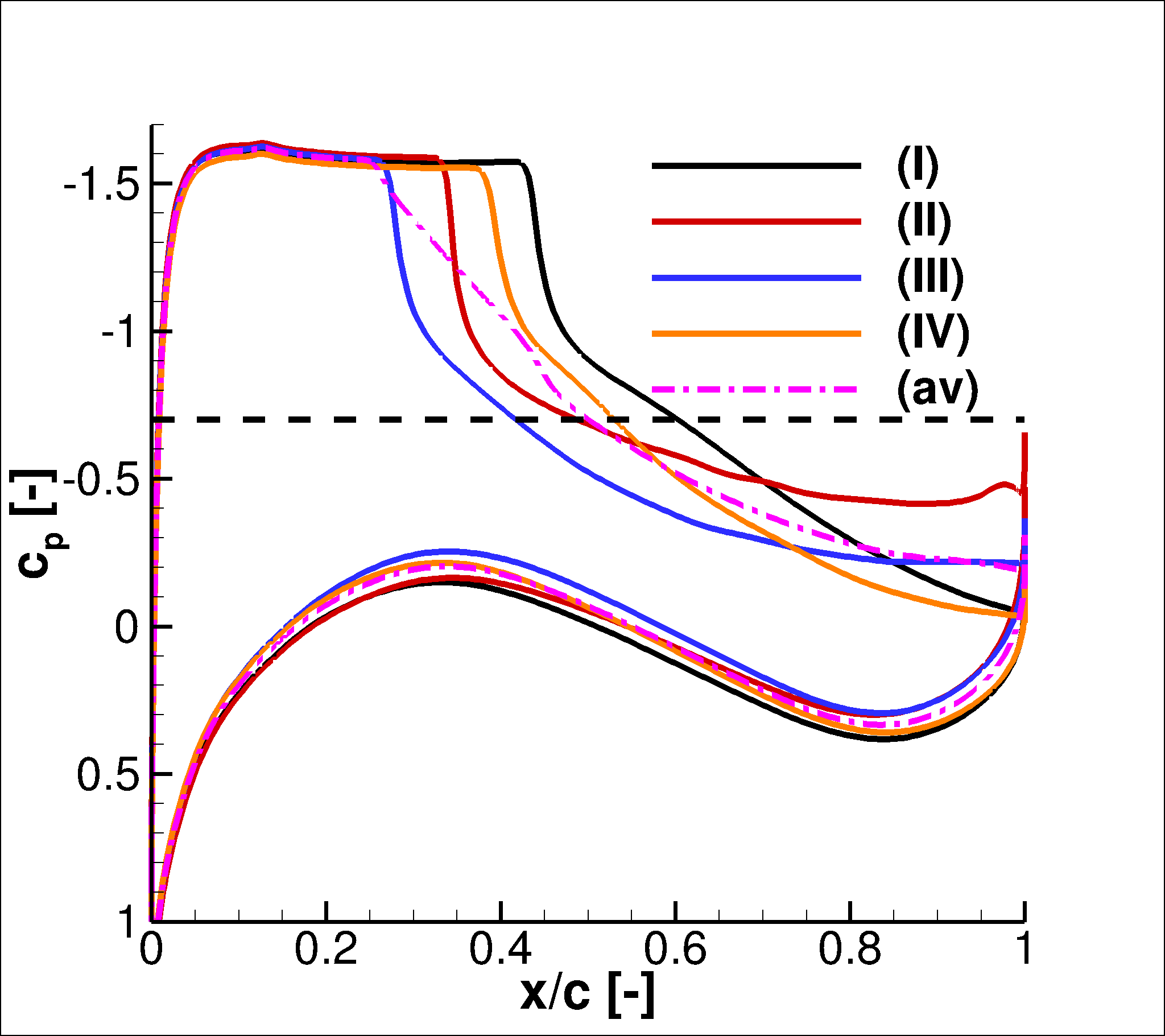}%
    \includegraphics[clip,trim={20 20 240 180},width=0.5\textwidth]{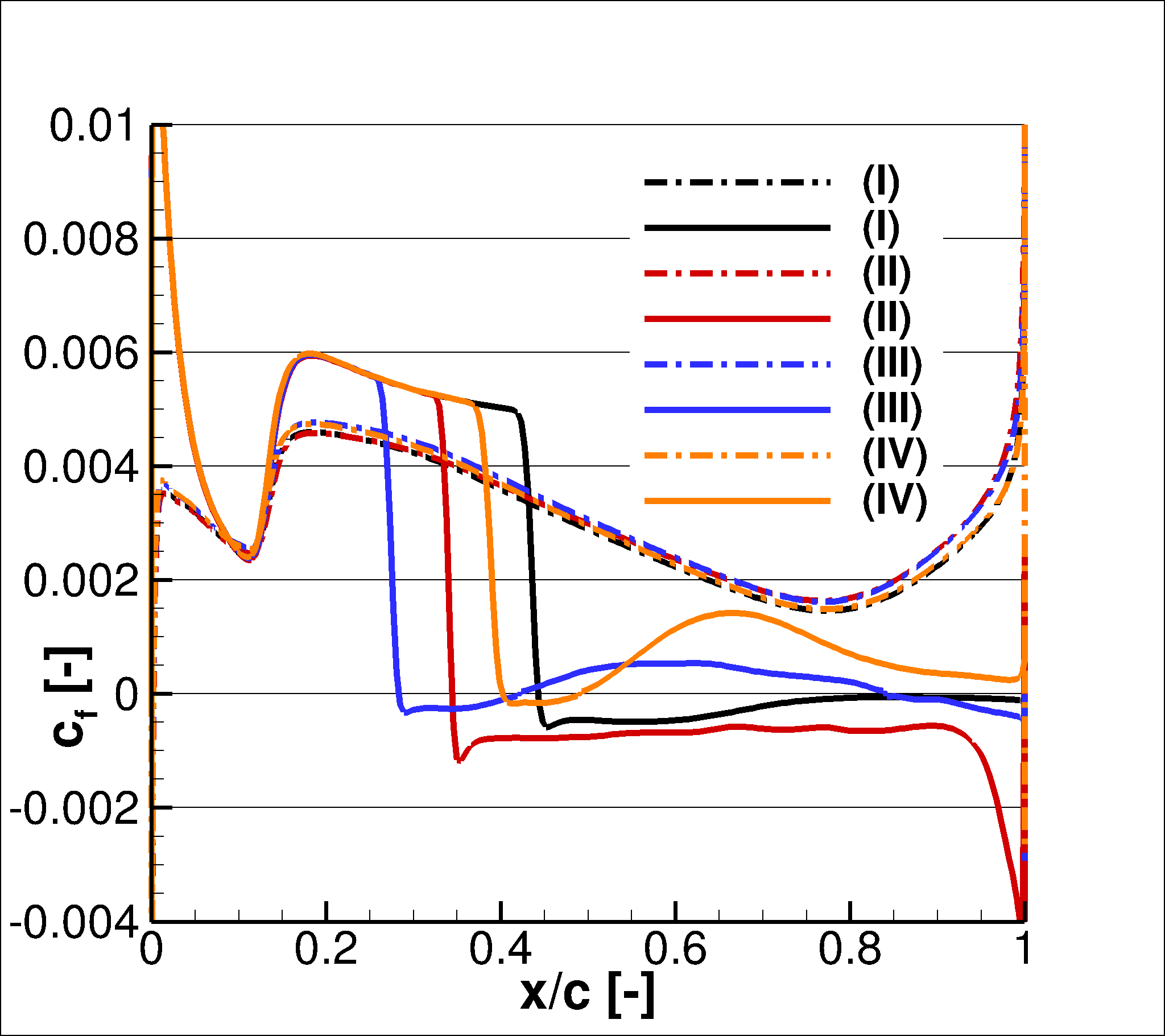}
    \caption{Distribution of the pressure (left) and friction (right) coefficient of the front wing segment at different moments in time during one buffet cycle}\label{fig:shock-motion}
\end{figure}
The range of the shock motion is displayed in Fig.~\ref{fig:shock-motion} (left), which shows the distribution of the pressure coefficient of the front wing segment at four distinct moments in time during one buffet period. Here, the most downstream position of the shock is marked with (I) and its most upstream position with (III). Snapshots during the upstream and downstream movement of the shock are denoted as (II) and (IV), respectively. These four moments in time are also marked in Fig.~\ref{fig:cl-signal}, for reference. Additionally, the mean pressure distribution averaged over one buffet cycle is included, marked with (av). The shock position (as defined by the steepest pressure gradient) is found to move between $x/c=0.28$ and $x/c=0.44$, i.e. a range of 16\% chord which is close to the experimental value measured in the Trisonic Wind Tunnel of 16.5\% \cite{schauerte2022}. As the amount of lift generated is largely dominated by the extent of the low pressure region on the upper side, the most downstream position of the shock coincides approximately with the maximum, and the most upstream shock position with the minimum of the lift coefficient, respectively.
The different levels of pressure recovery towards the trailing edge already indicate different amounts of flow separation for the different shock positions, which becomes evident in Fig.~\ref{fig:shock-motion} (right) that shows the corresponding distributions of the surface friction coefficient. Here, the upper surface is indicated with solid lines, and the lower surface with dashed lines. 
The varying amount of separation can also be seen in Fig.~\ref{fig-separation-front}, which shows the Mach number in the flow field around the front wing segment together with streamlines close to its surface for the four moments in time introduced above.
The flow exhibits a (relatively) small amount of separation when the shock is located at its most downstream location (I) due to a comparatively small shock strength.
However, during the upstream movement of the shock (II), the flow behind the shock is completely separated until the trailing edge. As the shock moves upstream, the velocity of the fluid relative to the shock is increased, which leads to a greater shock strength, forcing the point of separation to move forward on the airfoil. 
When the shock temporarily comes to a stop at its most upstream position (III), this effect fades, so both the shock strength and the amount of separation are reduced. The flow begins to reattach behind a shock-induced separation bubble at around $x/c=0.4$. At this moment in time, a trailing edge separation is still present in the last ten percent of the chord.
Finally, the velocity of the fluid relative to the shock is decreased during the downstream movement of the shock (IV). Therefore, the resulting strength of the shock is reduced allowing for a further reattachment of the flow during this phase. Whereas the trailing edge separation vanishes almost completely, the separation bubble behind the shock becomes smaller but does not disappear fully, however, and quickly grows back towards the trailing edge when the shock reaches it most downstream location again. 
Notably, the boundary layer on the lower surface remains attached at all times, and only small changes in pressure and surface friction occur during the buffet cycle.
The pronounced variation of the amount of separation within the buffet cycle strongly influences the characteristics of the wake and its interaction with the rear wing segment, as discussed later.
\begin{figure}[h]%
    \centering
    \includegraphics[clip,trim={40 30 240 200},width=0.5\textwidth]{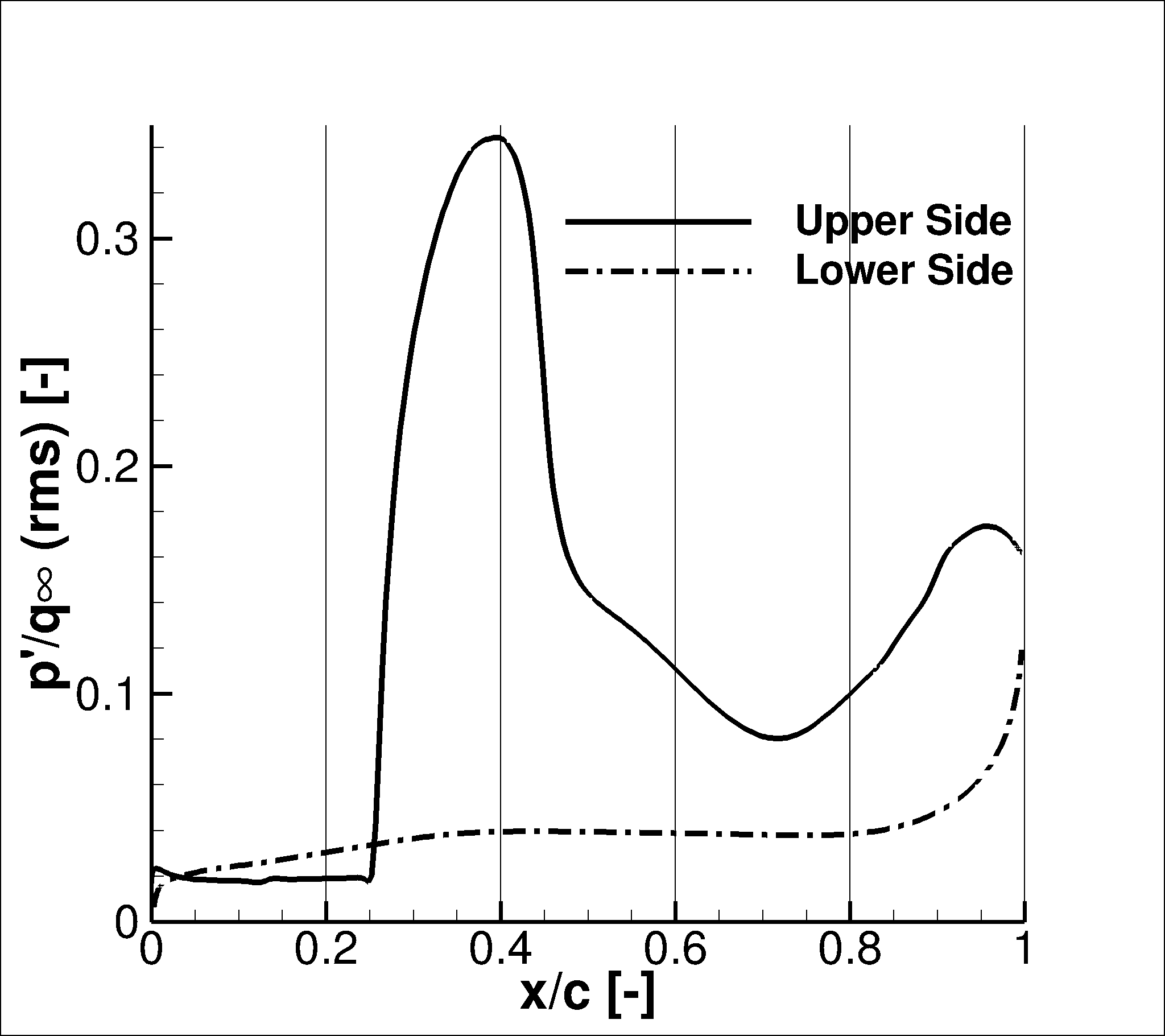}
    \caption{Root mean square (rms) of the pressure fluctuations on the front wing segment}\label{fig:cp-rms-vorne}
\end{figure}
\begin{figure}[h]%
    \centering
    \includegraphics[clip,trim={10 20 240 170},width=0.5\textwidth]{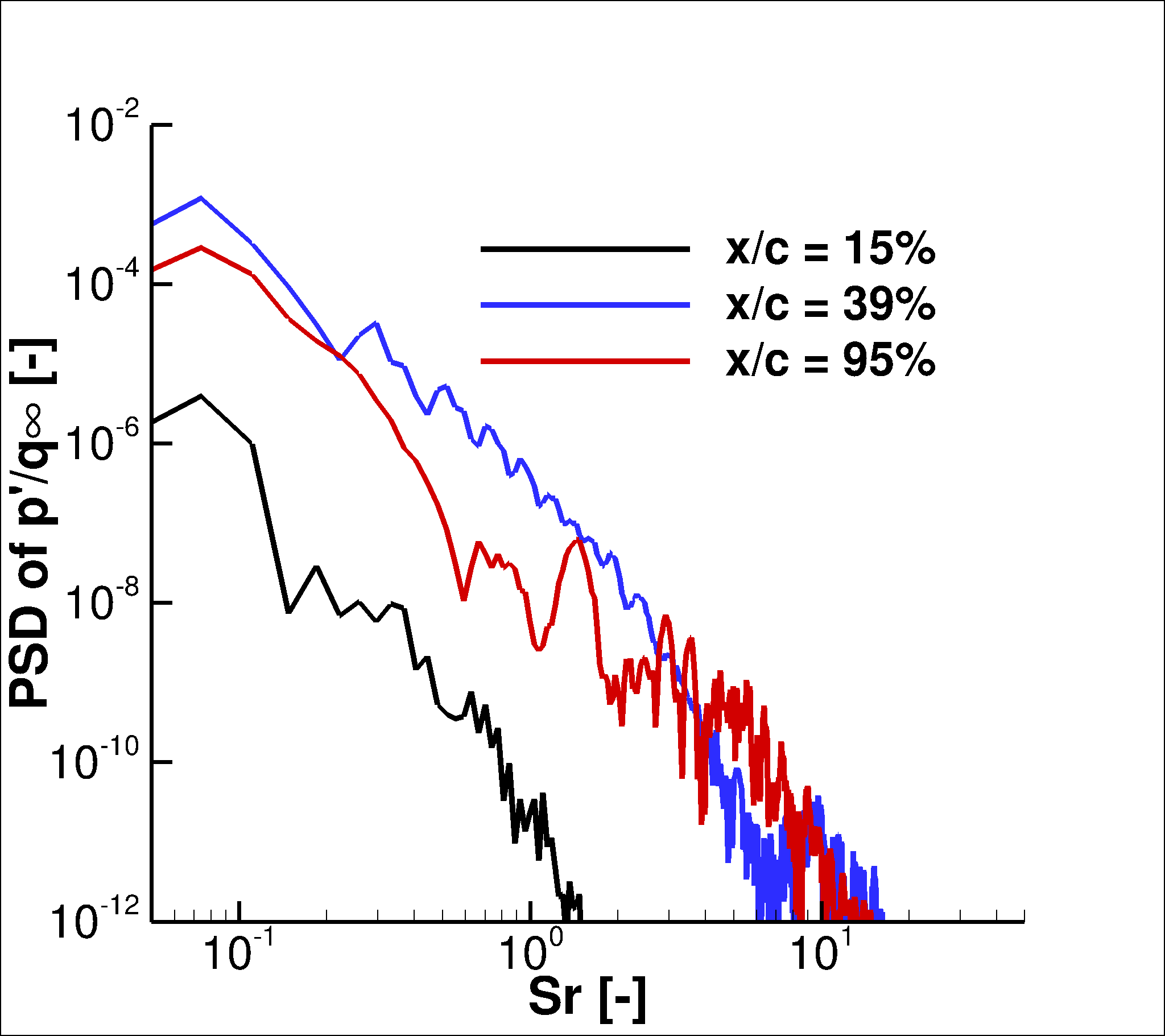}
    \caption{Power spectral density (PSD) of the surface pressure fluctuations on the front wing segment's upper side at $x/c=0.15$, $x/c=0.39$ and $x/c=0.95$}\label{fig:PSD-front}
\end{figure}
Both shock movement and flow separation induce pressure fluctuations on the surface of the front wing segment, which are shown in Fig.~\ref{fig:cp-rms-vorne} in terms of root mean square (rms) values over the chord position, normalized with the dynamic pressure of the inflow, i.e.~$p'/q_{\infty}$. The highest fluctuation levels can be seen between $x/c=0.25$ and $x/c=0.50$, corresponding to the range of shock motion. Due to the strong pressure gradient at the shock, a point on the surface experiences a sharp pressure increase or decrease every time the shock passes, leading to a high temporal variation of the pressure. In contrast, the pressure fluctuations in the supersonic region upstream of the shock are comparatively small. The area downstream of the shock, however, exhibits elevated values that suggest strong pressure disturbances and turbulent fluctuations in the separated flow. On the lower side that shows no shock or flow separation, the fluctuations are comparatively small, however, an increase of unsteadiness is evident close to the trailing edge, which can be attributed to pressure disturbances propagating from the upper side. 
A corresponding power spectral density (PSD), computed with Welch's method~\cite{welch1967}, of the normalized surface pressure fluctuation is displayed in Fig.~\ref{fig:PSD-front} for three points on the upper surface, at $x/c=0.15$, $x/c=0.39$ and $x/c=0.95$, respectively. The first point is located in the region upstream of the shock at all times, the second one at the position of the rms maximum and the third one downstream of the shock. It is evident that all spectra are dominated by the buffet frequency. Corresponding to the small rms of the pressure fluctuations noted above, the amplitudes are significantly smaller for the position in front of the shock. 

\begin{figure}[h]%
	\centering
	\includegraphics[clip,trim={10 10 240 250},width=0.5\textwidth]{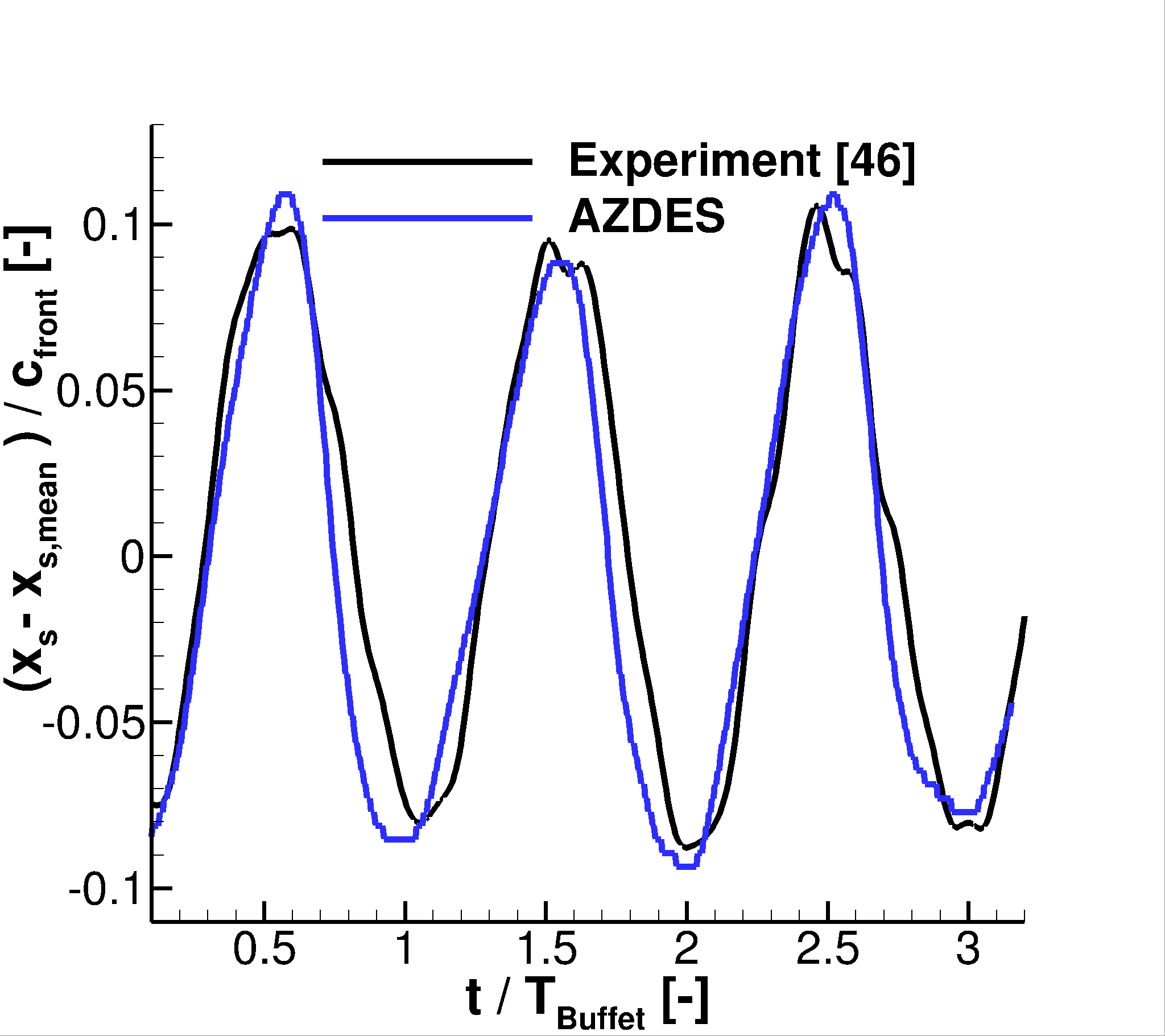}
	\caption{Time history of the shock position $x_s$ with respect to its mean position $x_{s,mean}$, simulation and experiment~\cite{schauerte2022,schauerte2023}}\label{fig:shock-comparison}
\end{figure}
\begin{figure}[h]%
	\centering
	\includegraphics[clip,trim={70 40 240 1180},width=0.5\textwidth]{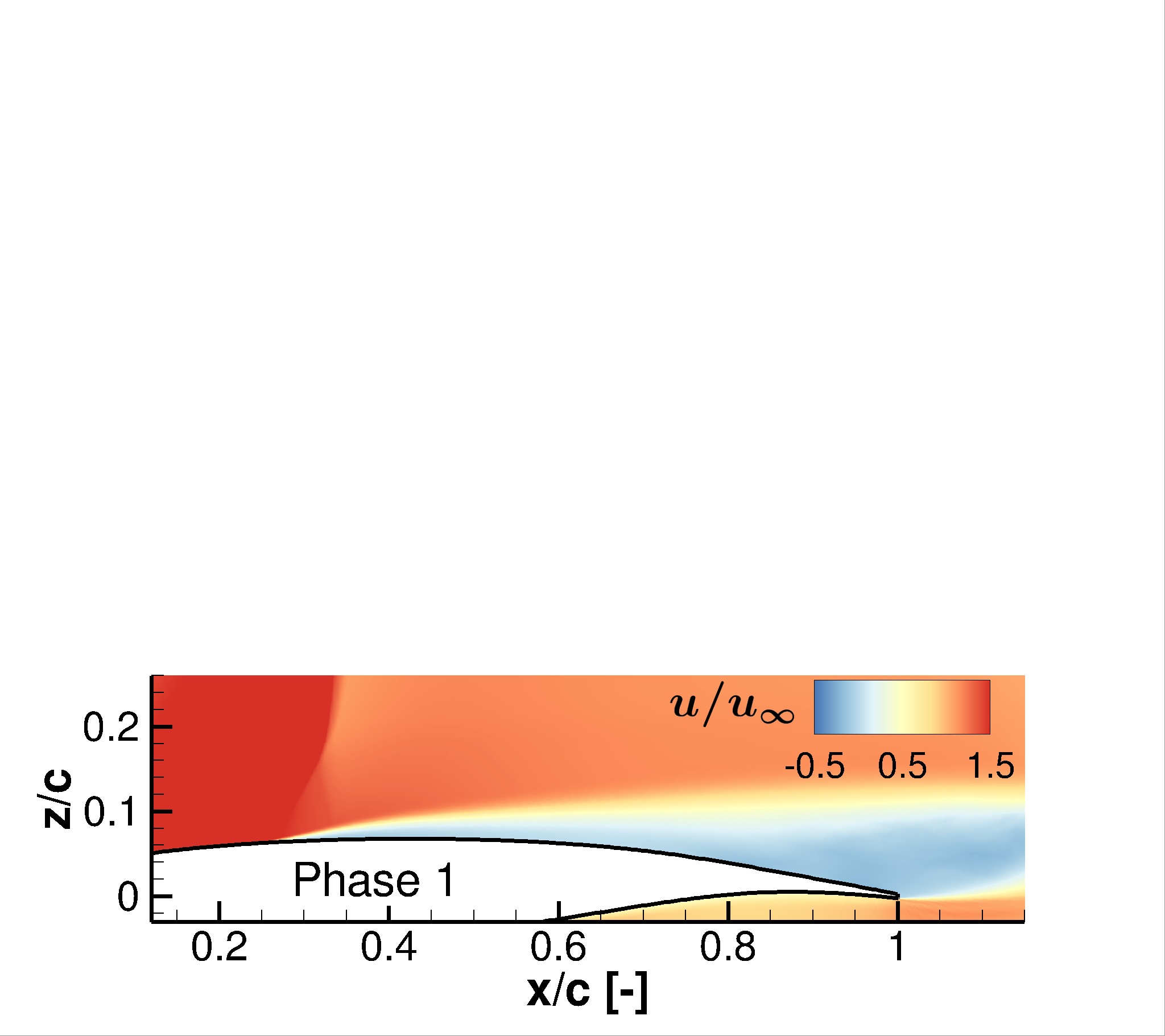}%
	\includegraphics[clip,trim={70 40 240 1180},width=0.5\textwidth]{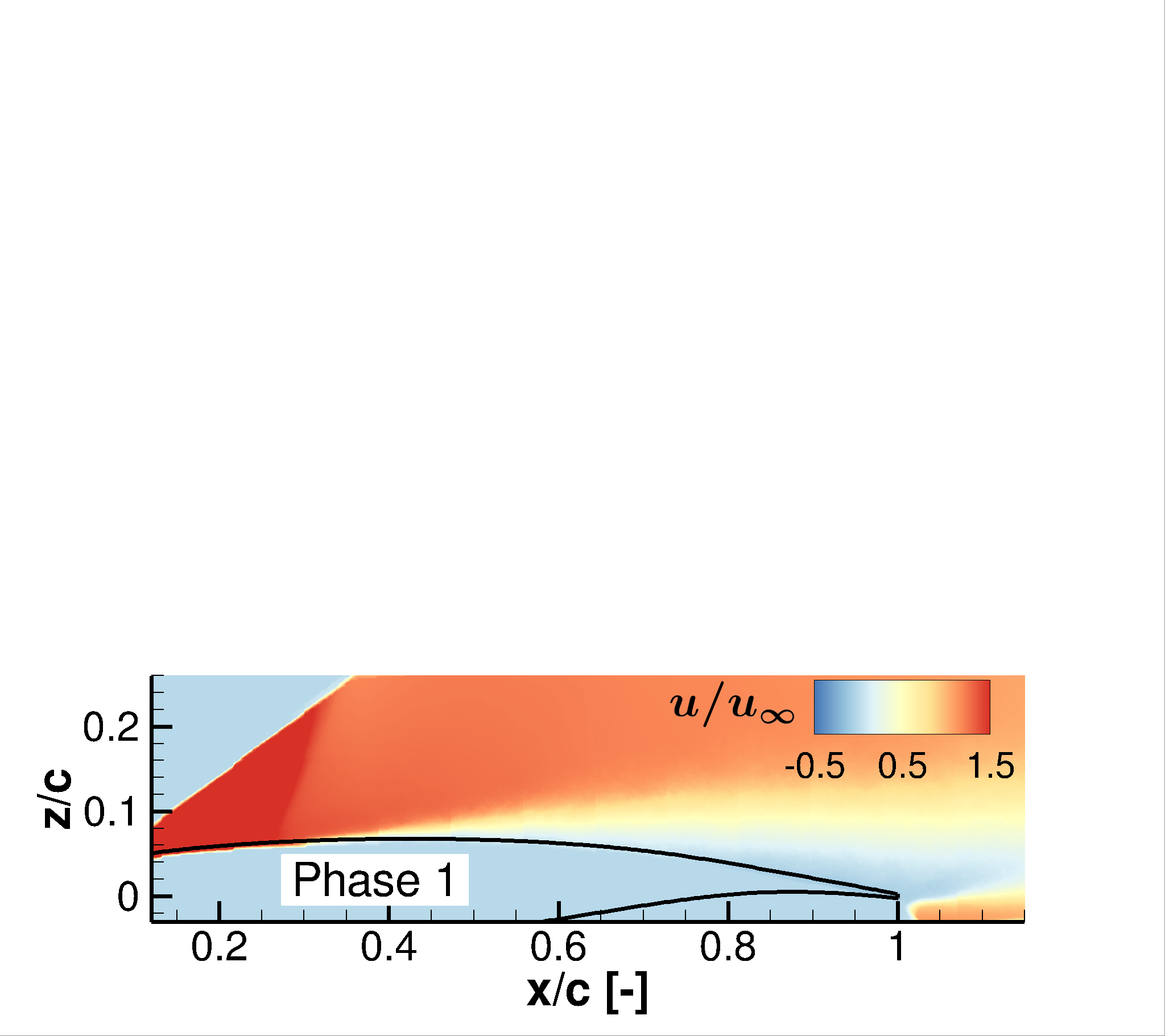}
	\includegraphics[clip,trim={70 40 240 1180},width=0.5\textwidth]{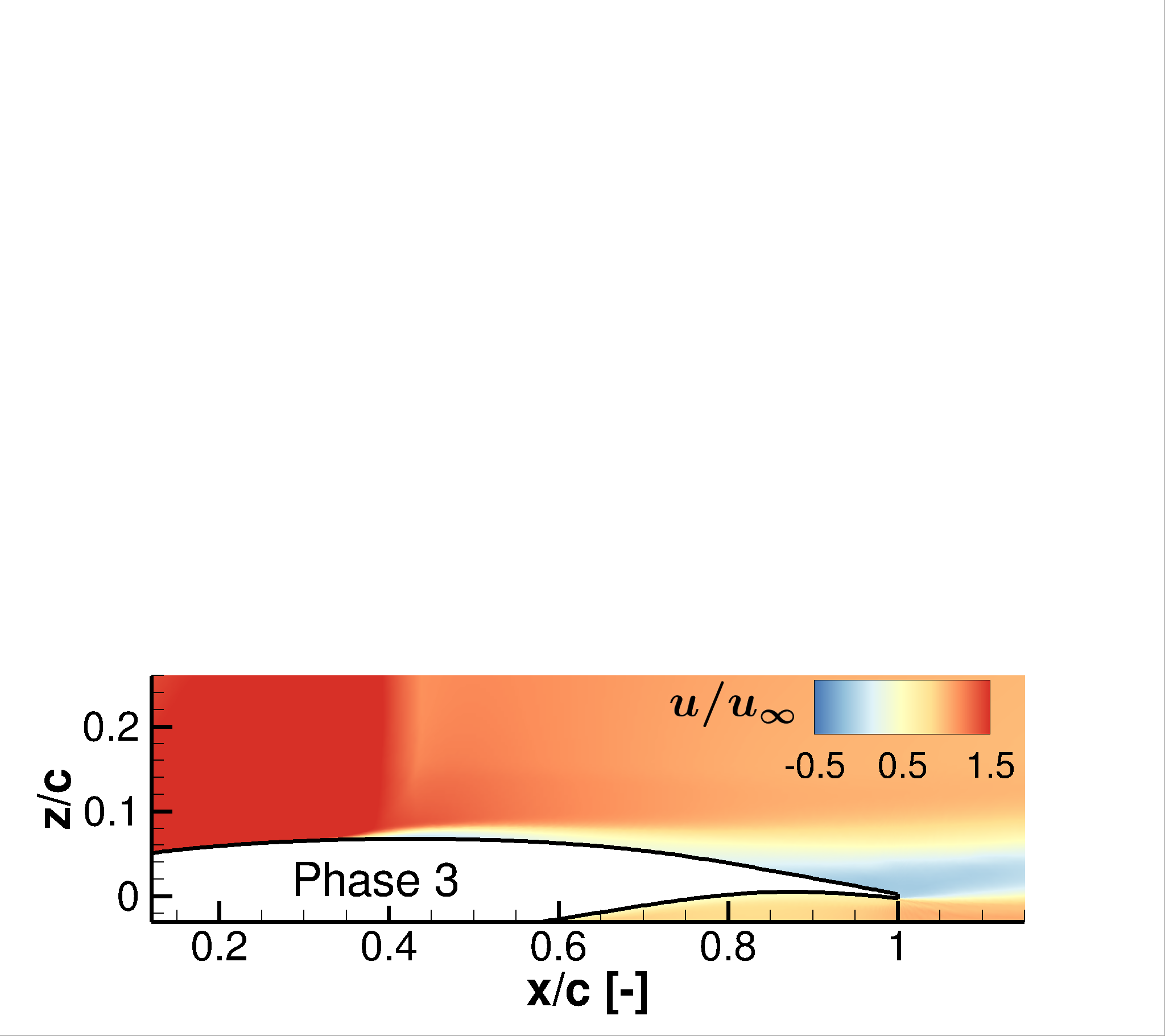}%
	\includegraphics[clip,trim={70 40 240 1180},width=0.5\textwidth]{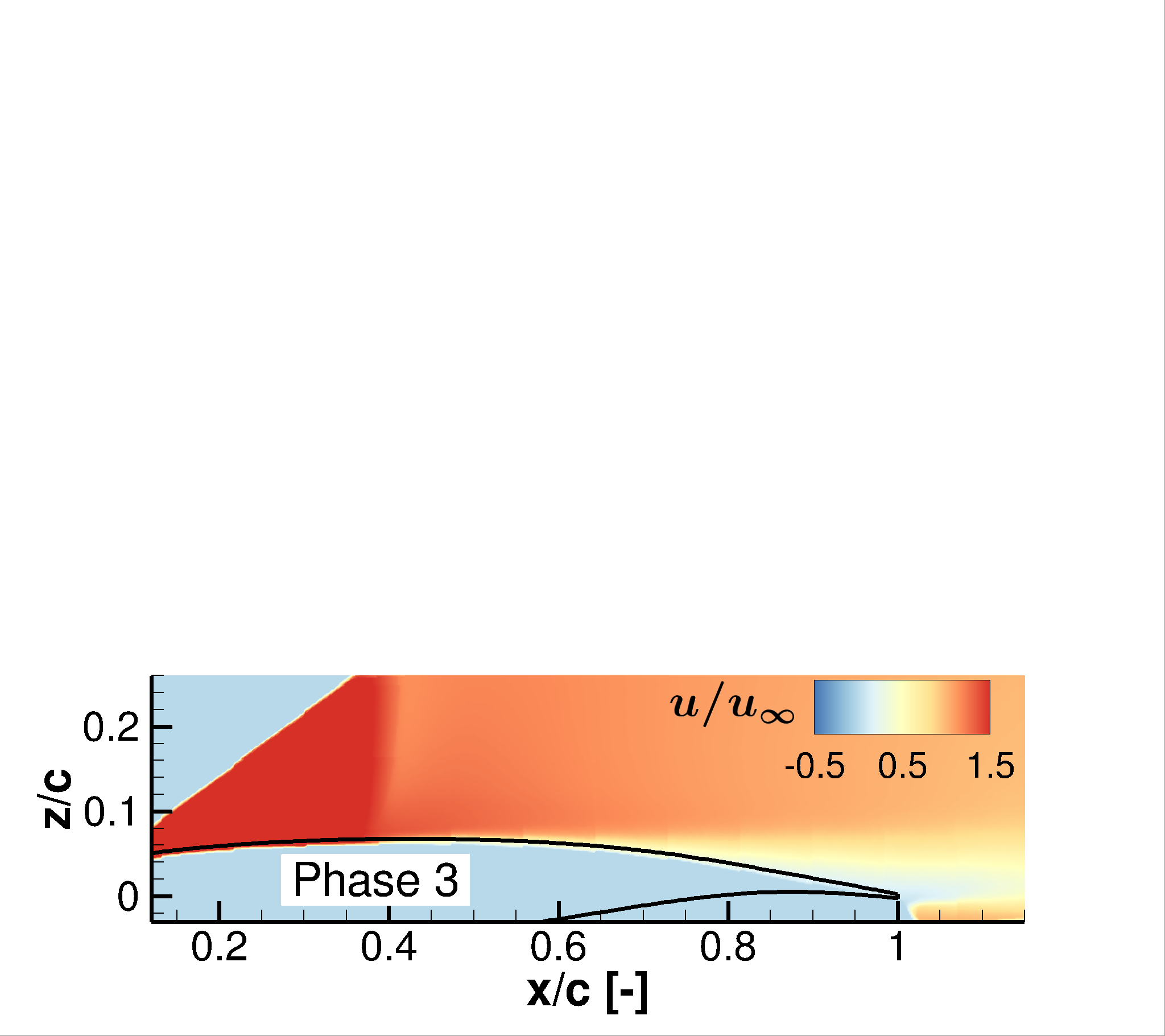}
	\includegraphics[clip,trim={70 40 240 1180},width=0.5\textwidth]{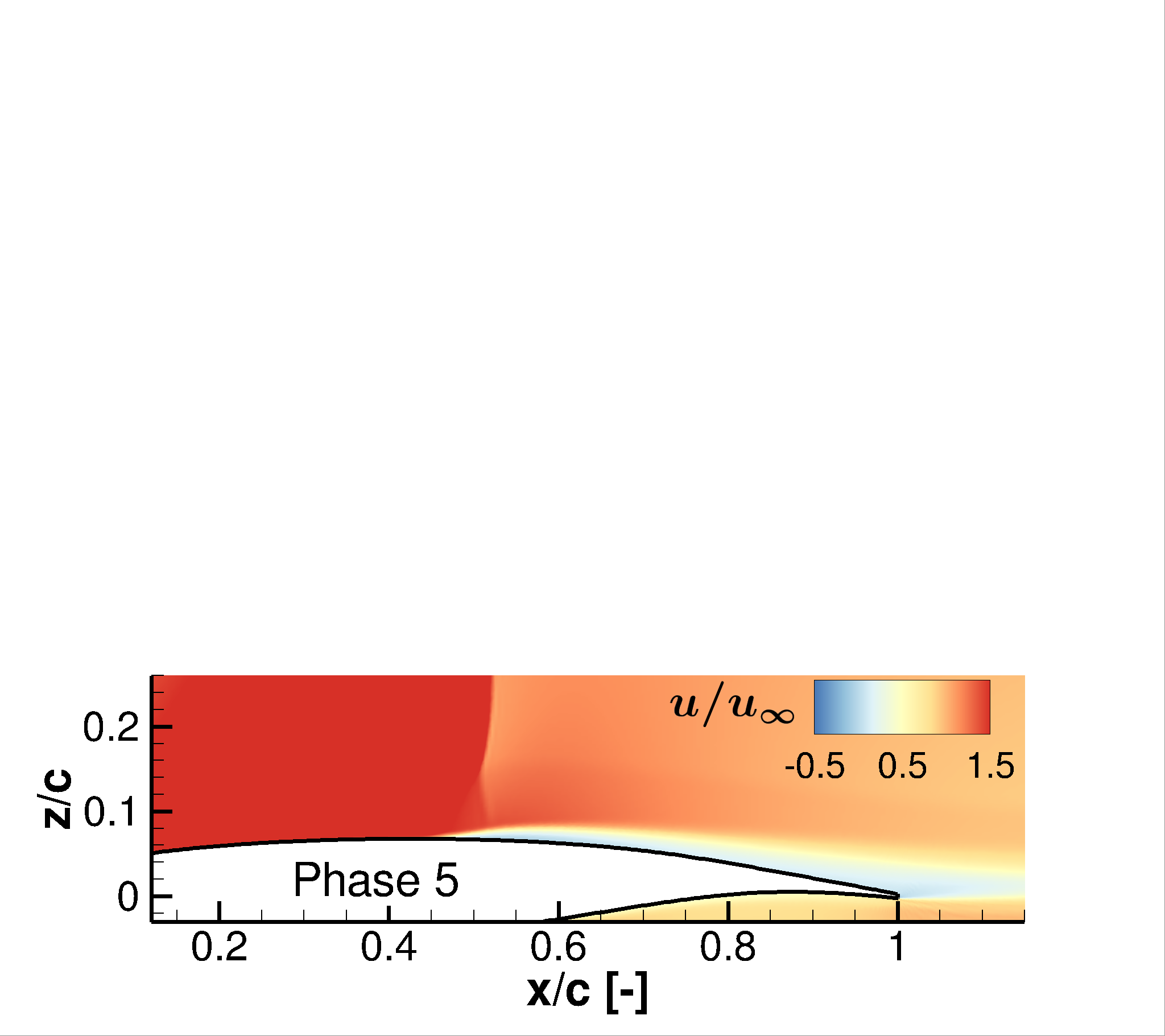}%
	\includegraphics[clip,trim={70 40 240 1180},width=0.5\textwidth]{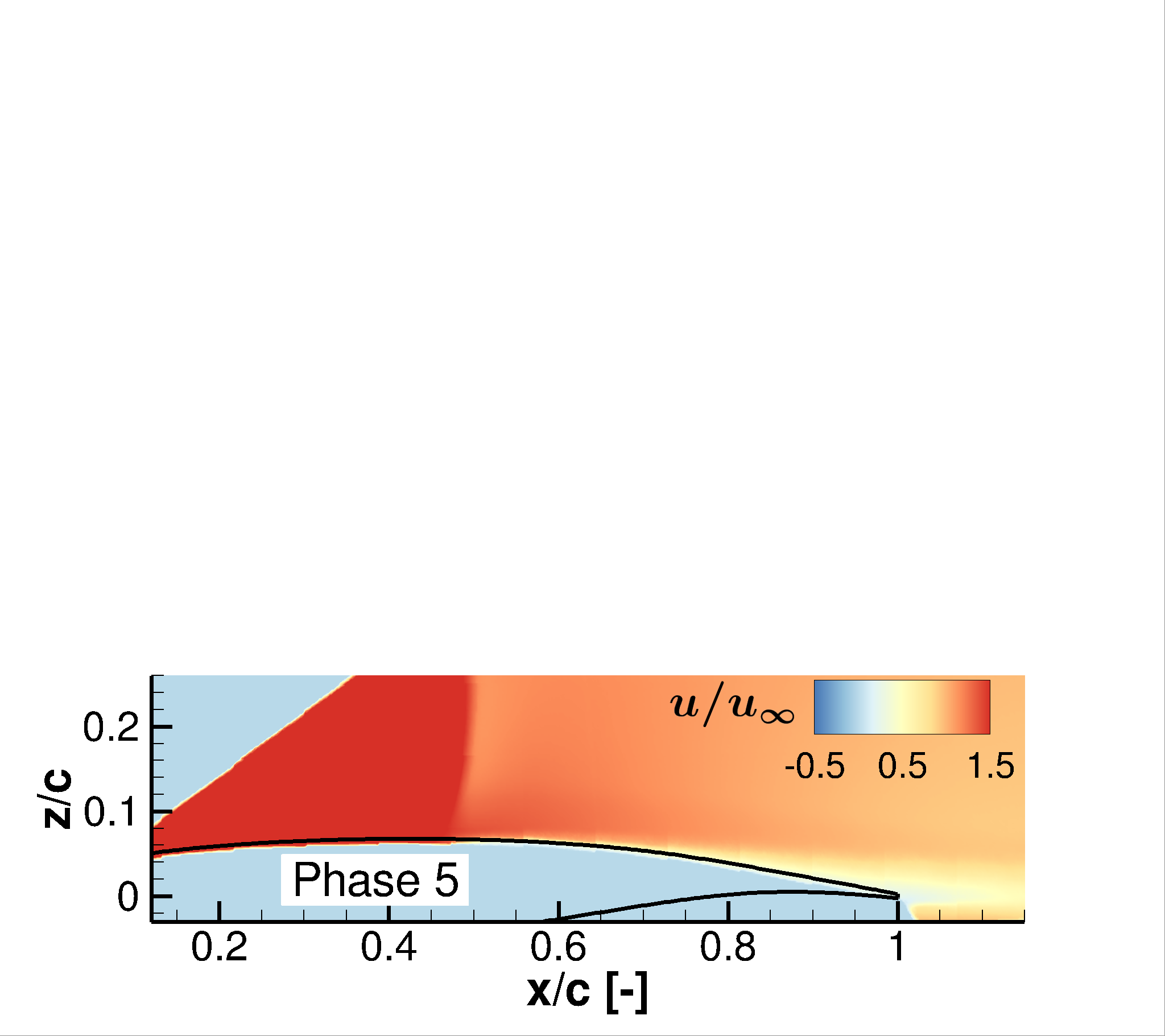}
	\includegraphics[clip,trim={70 40 240 1180},width=0.5\textwidth]{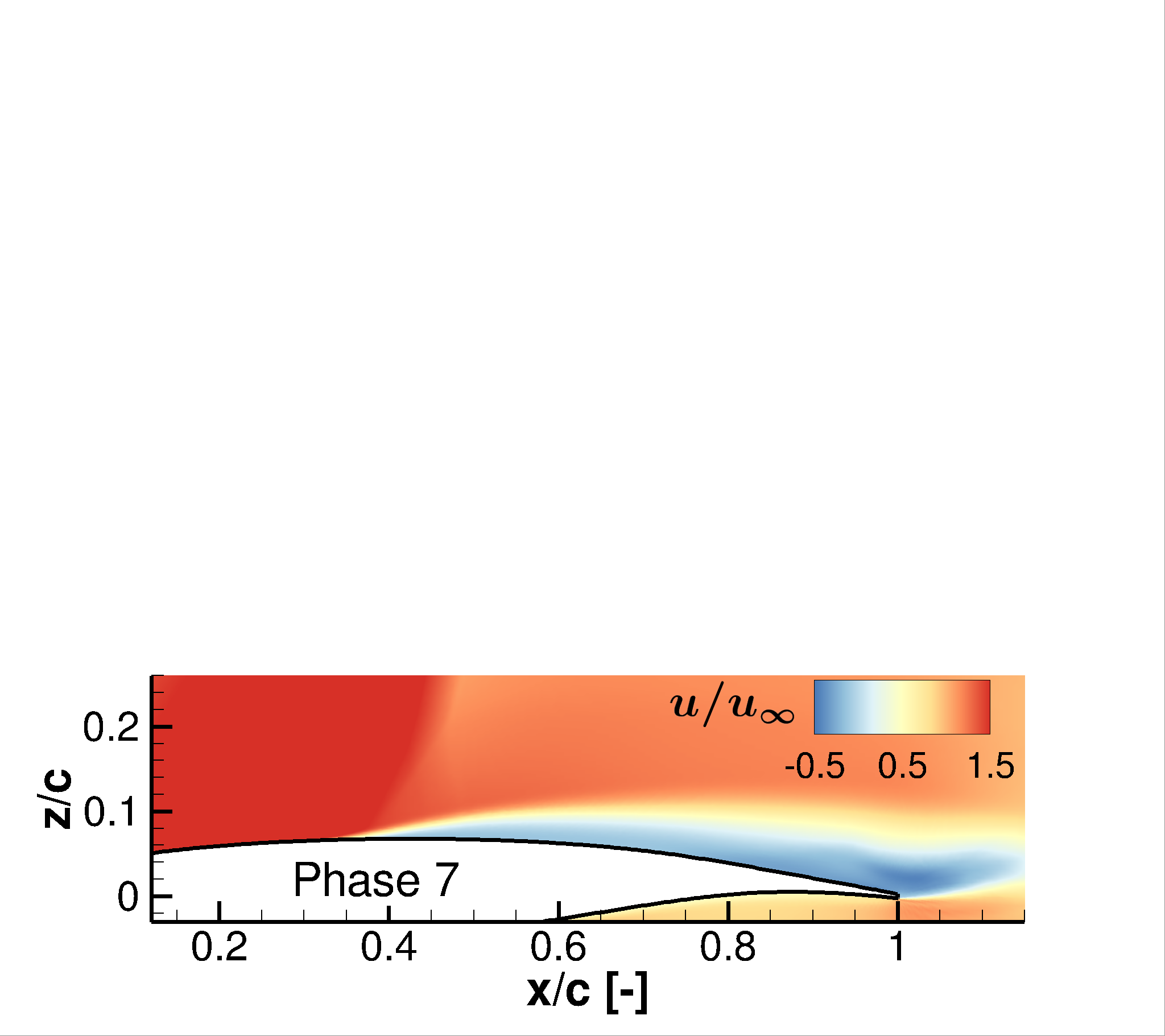}%
	\includegraphics[clip,trim={70 40 240 1180},width=0.5\textwidth]{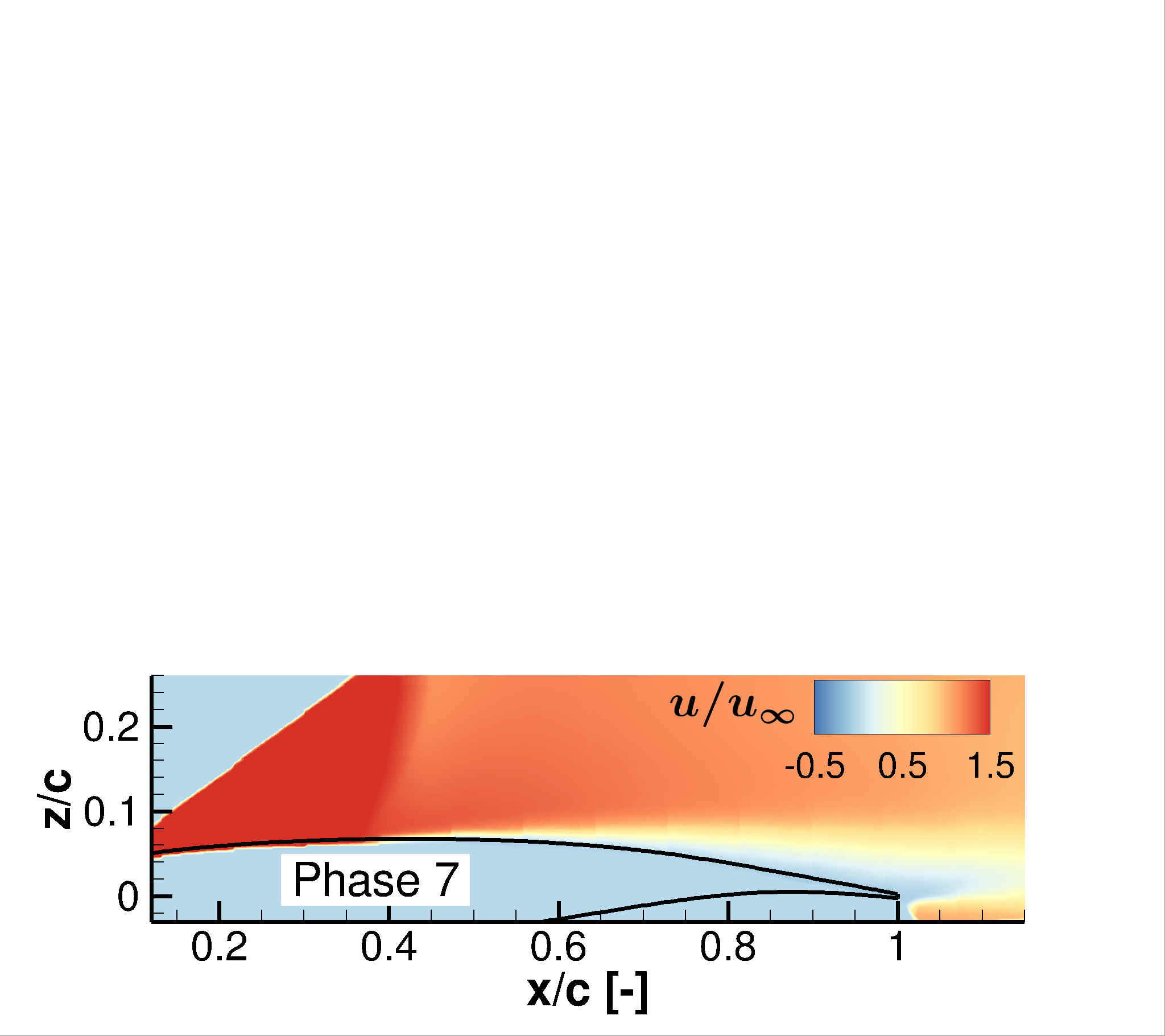}
	\caption{Phase-averaged flow field surrounding the front wing segment, simulation (left) and PIV measurement (right)~\cite{schauerte2022,schauerte2023}; phase 1: most upstream shock position, phase 3: downstream shock motion, phase 5: most downstream shock position, phase 7: upstream shock motion}\label{fig-piv-comparison}
\end{figure}
A brief comparison of the buffet flow around the front wing segment, as simulated with the AZDES method, with experimental data from corresponding measurements of the isolated OAT15A airfoil performed in the Trisonic Wind Tunnel~\cite{schauerte2022,schauerte2023} is shown in the following.
Fig.~\ref{fig:shock-comparison} depicts the time history of the chordwise shock position $x_s$ with respect to its mean position $x_{s,mean}$ as a fraction of the chord length $c_{front}$ for both the simulation and the experiment, determined by the maximum of the density gradient at a line $0.1\,c_{front}$ above the airfoil surface (at $z/c=0.1678$).
It is evident that simulation and experiment agree quite well considering the overall shock motion, its absolute range and extreme positions. 
It should also be noted that there is a notable amount of cycle-to-cycle variation for both cases (please see~\cite{schauerte2023} for a longer time series).
Whereas the downstream motion is comparable, the simulated upstream motion consistently appears to be slightly shifted to an earlier moment in time of the cycle in relation to the experimental data. 
In~\cite{schauerte2022,schauerte2023}, PIV measurements of the flow field around the OAT15A airfoil were performed and analyzed. In order to take the significant variation of the flow field over the buffet cycle into account, the latter was divided in eight phases with regard to the shock location relative to its most upstream and downstream position, and an averaging was performed for these phases over several buffet periods (for details we refer to~\cite{schauerte2023}).
The same phase-averaging is performed with the simulation data in this work, and compared to the PIV results in Fig.~\ref{fig-piv-comparison} for four selected phases - at the time of the most upstream and downstream shock position (phase 1 and 5, respectively) and during the downstream and upstream shock motion (phase 3 and 7, respectively). 
Again, a good agreement between simulation and experiment is evident considering the respective shock locations. Additionally, the curvature of the shock line is matched quite well in the simulation. Furthermore, the varying extent of the separated flow region behind the shock is in agreement as well as the flow velocity and direction at the trailing edge. 
However, differences inside the separated flow region are visible, where the flow velocities are smaller or more negative in the simulation than in the experiment. Also, the thickening of the boundary layer is slightly overpredicted, especially for the phases with strong separation.
Nevertheless, the overall agreement can be considered satisfactory.

\begin{figure}[h]%
    \centering
    \includegraphics[clip,trim={80 20 240 980},width=0.7\textwidth]{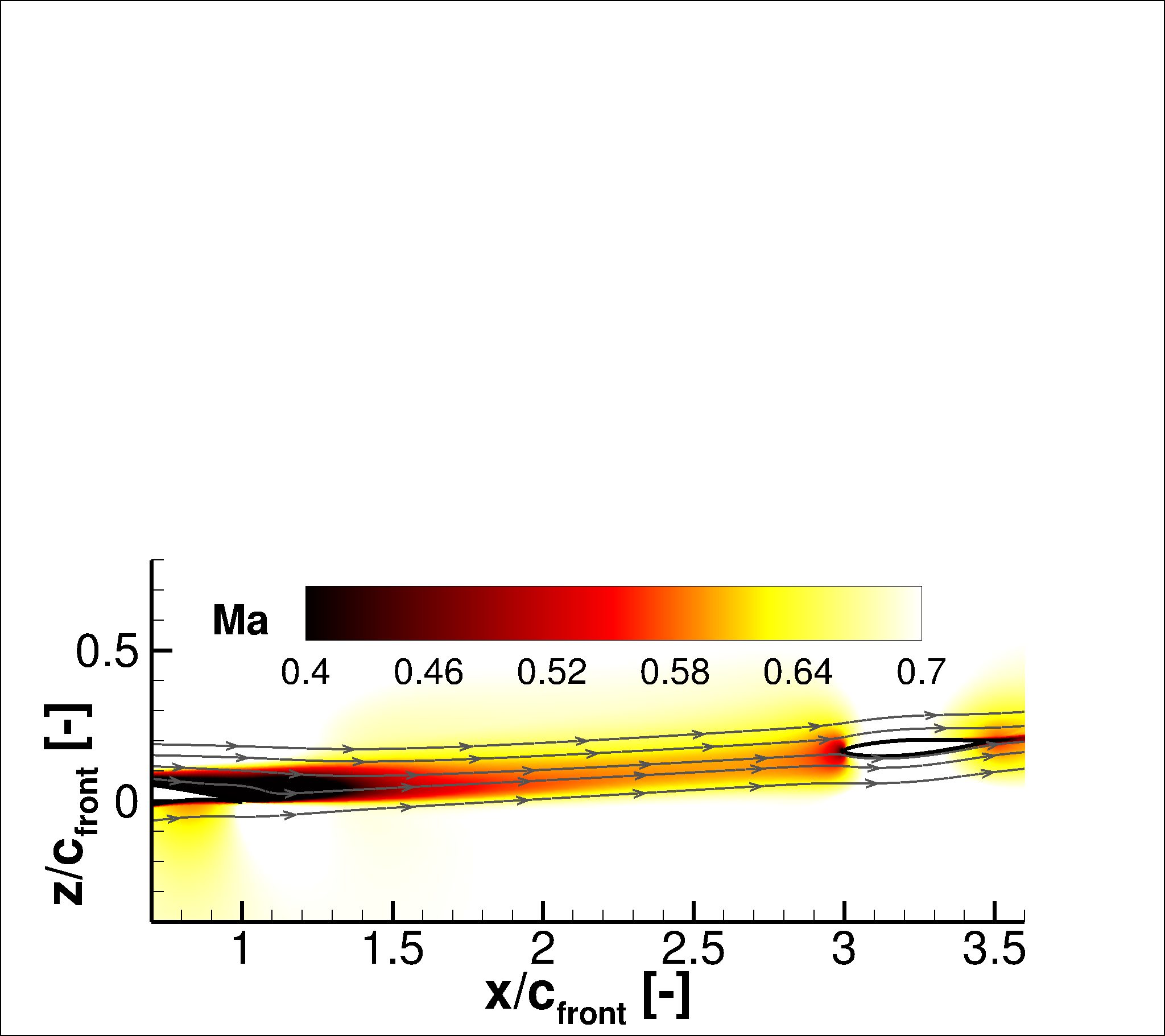}
    \caption{Time-averaged wake behind the front wing segment in terms of Mach number}\label{fig:mean-wake}
\end{figure}
\begin{figure}[h]%
    \centering
    \includegraphics[clip,trim={80 20 240 980},width=0.5\textwidth]{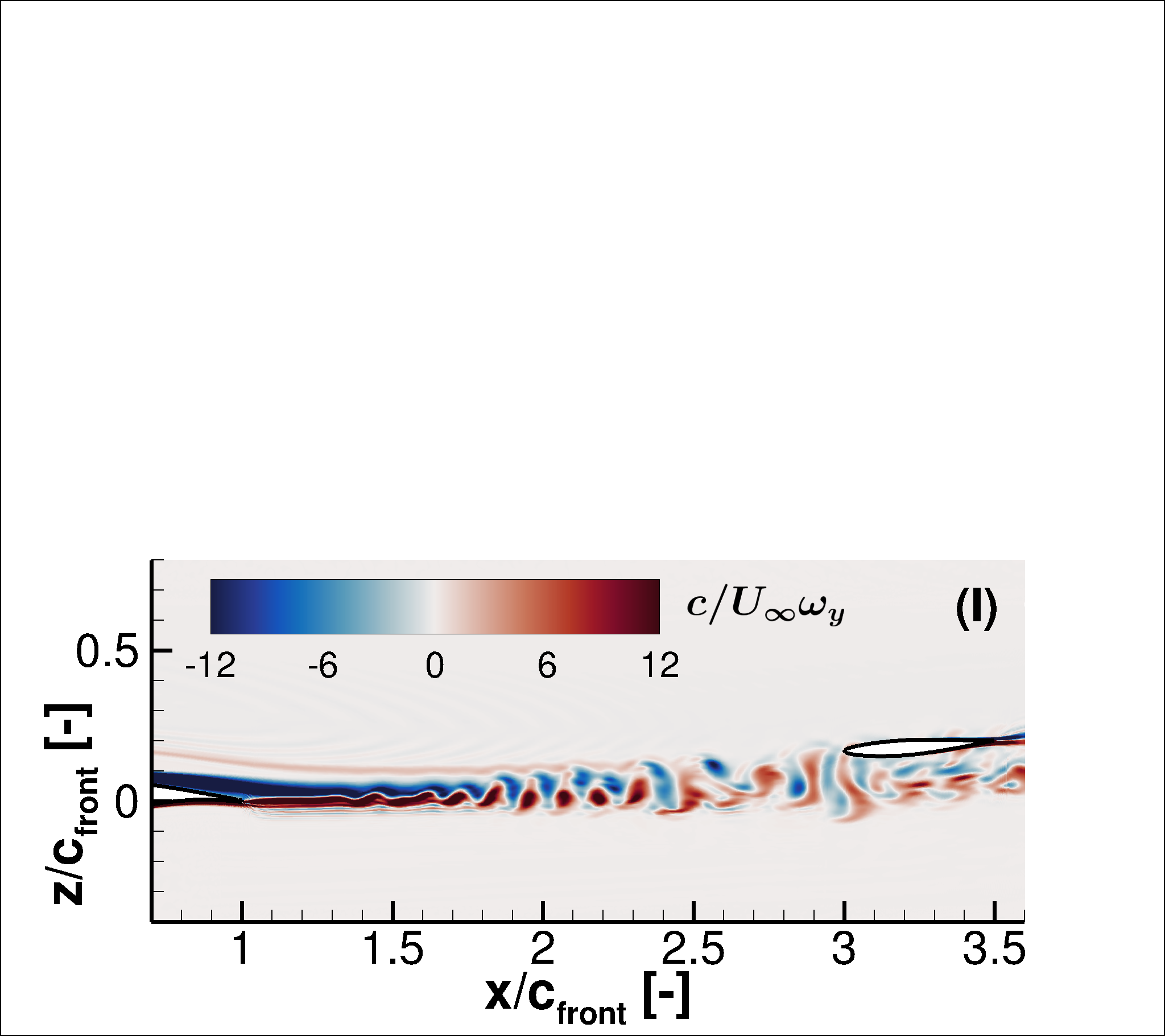}%
    \includegraphics[clip,trim={80 20 240 980},width=0.5\textwidth]{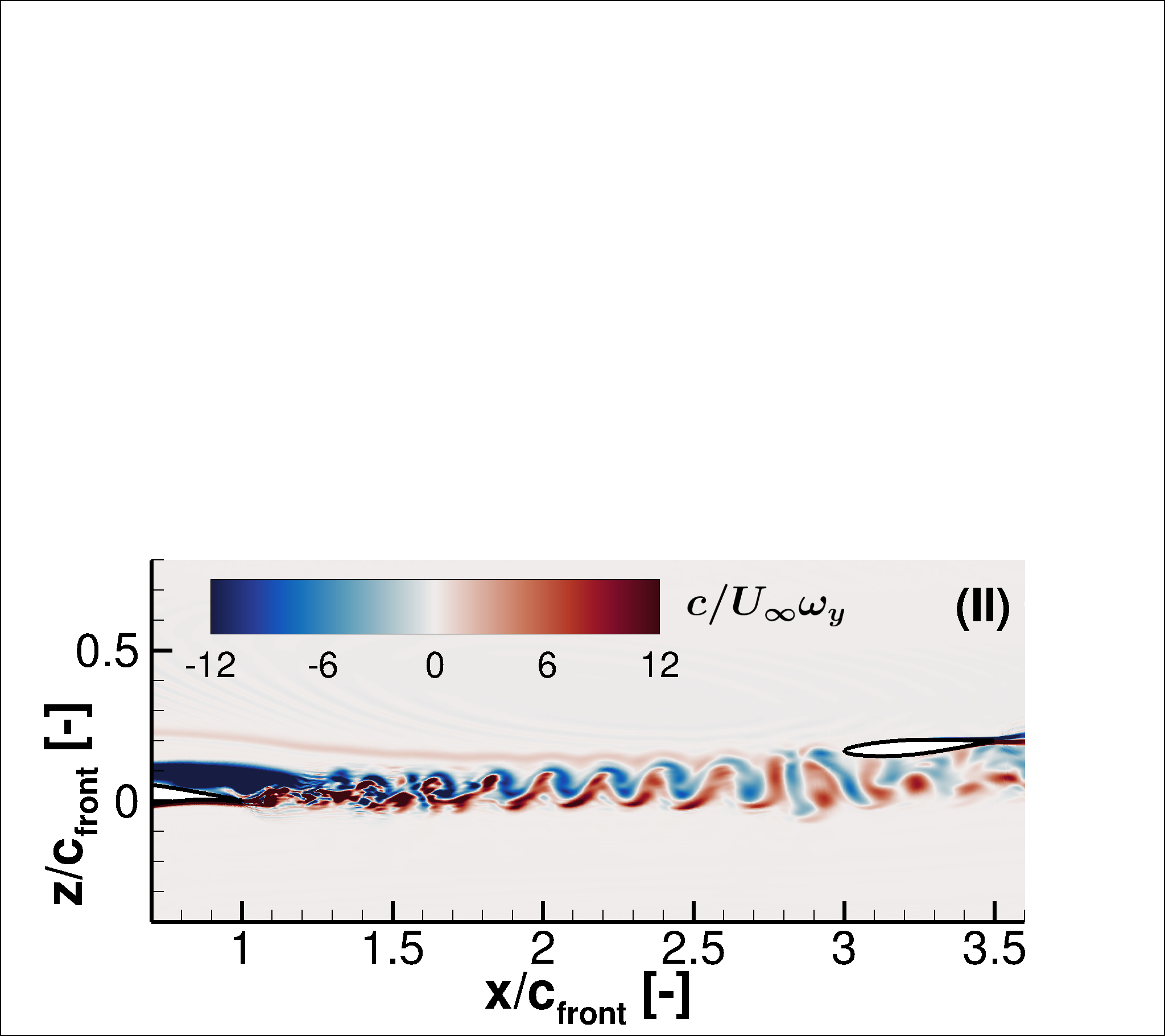}
    \includegraphics[clip,trim={80 20 240 980},width=0.5\textwidth]{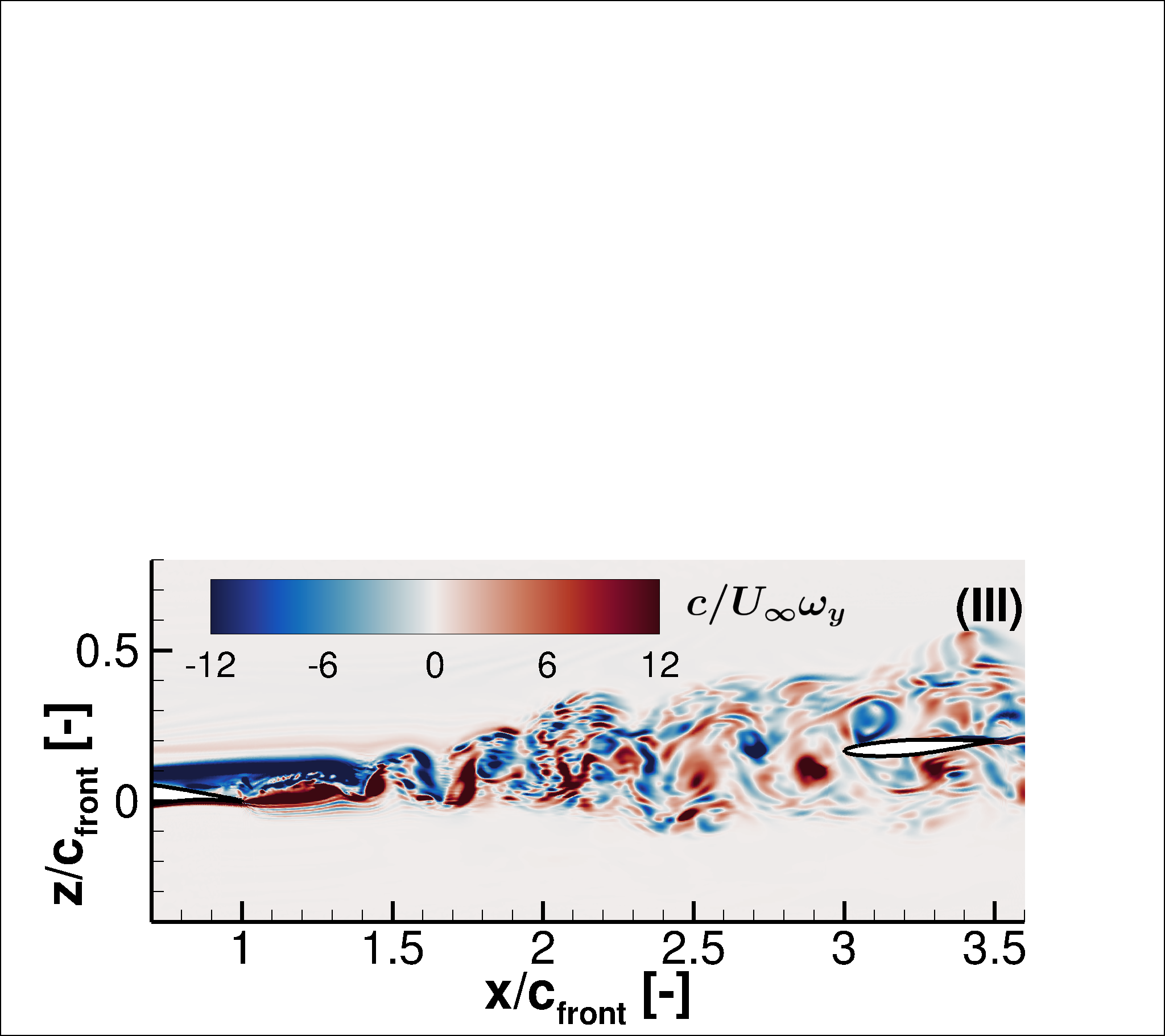}%
    \includegraphics[clip,trim={80 20 240 980},width=0.5\textwidth]{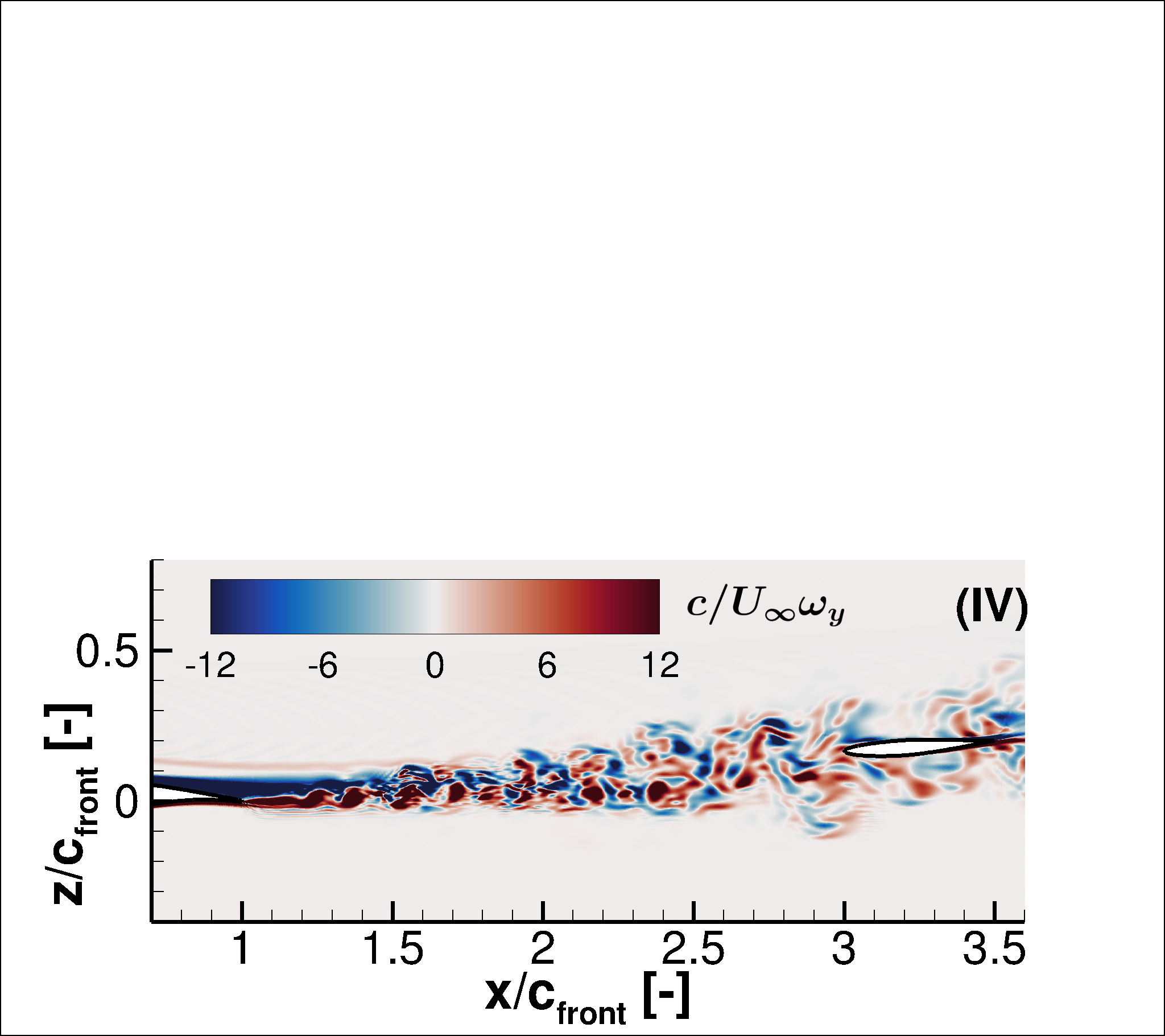}
    \caption{Spanwise vorticity in the wake behind the front wing segment at different moments in time during one buffet cycle}\label{fig:wake-t}
\end{figure}
Fig.~\ref{fig:mean-wake} shows the time-averaged wake downstream of the front wing segment. Here, the mean Mach number in the flow field is shown, together with time-averaged streamlines. The separated wake appears as a region of reduced Mach number downstream of the front wing segment. 
Remarkably, the wake flow is orientated almost parallel to the chord close to the trailing edge, despite the incidence of the inflow of 5$^{\circ}$, which is due to the downwash created by the front wing segment. Further downstream, at distances greater than roughly one half of a chord away from the trailing edge ($x/c>1.5$), the trajectory of the wake is slightly curved upwards in $z$-direction as the strength of the downwash declines, and the wake flow becomes more and more aligned with the inflow. 
Additionally, an increase of the mean velocity in the wake with increasing distance from the trailing edge is evident, which indicates a progressive dissipation of the wake.
The development of the wake over one buffet period is depicted in Fig.~\ref{fig:wake-t}, 
which shows the instantaneous dimensionless spanwise vorticity $(c/U_\infty)\cdot\omega_y$ in the flow field behind the front wing segment for the four moments in time discussed above. 
Large turbulent vortices are formed in the separated flow behind the trailing edge, evolving from the shear layers at the wake boundaries.
Notably, the characteristics of the wake change significantly within the buffet cycle. During the time when there is only a small amount of separation or attached flow, the wake appears thin and small vortices are generated, which is the case for the most downstream shock position (I) and during the downstream movement of the shock (IV). When the amount of flow separation behind the shock is large, however, larger vortices are generated from the thick wake, which is the case for the most upstream position (III) and during the upstream movement of the shock (II). 
Strikingly, the amount of separation is largest during the upstream movement of the shock; yet, the biggest vortices in the wake are found at a moment later in time when the shock has already reached its most upstream position (III). This is because of the time that the separated flow, starting from the shock location, needs to reach the considered downstream position. The time shift between the flow situation at the front wing segment and the appearance of the corresponding vortices increases with increasing distance. 
It is also evident that the pattern of the vortices alternates during the buffet period. When the amount of separation is comparatively small (I)/(II), pairs of alternating vortices can be seen in the wake, similar to those in the wake of a bluff body. However, when the amount of separation is large, the pattern seems to be more irregular (III), which is also the case during the transitional phase (IV). The vortices begin to break up almost immediately and the pattern becomes increasingly chaotic further downstream.

\begin{figure}[h]%
    \centering
    \includegraphics[width=1\textwidth]{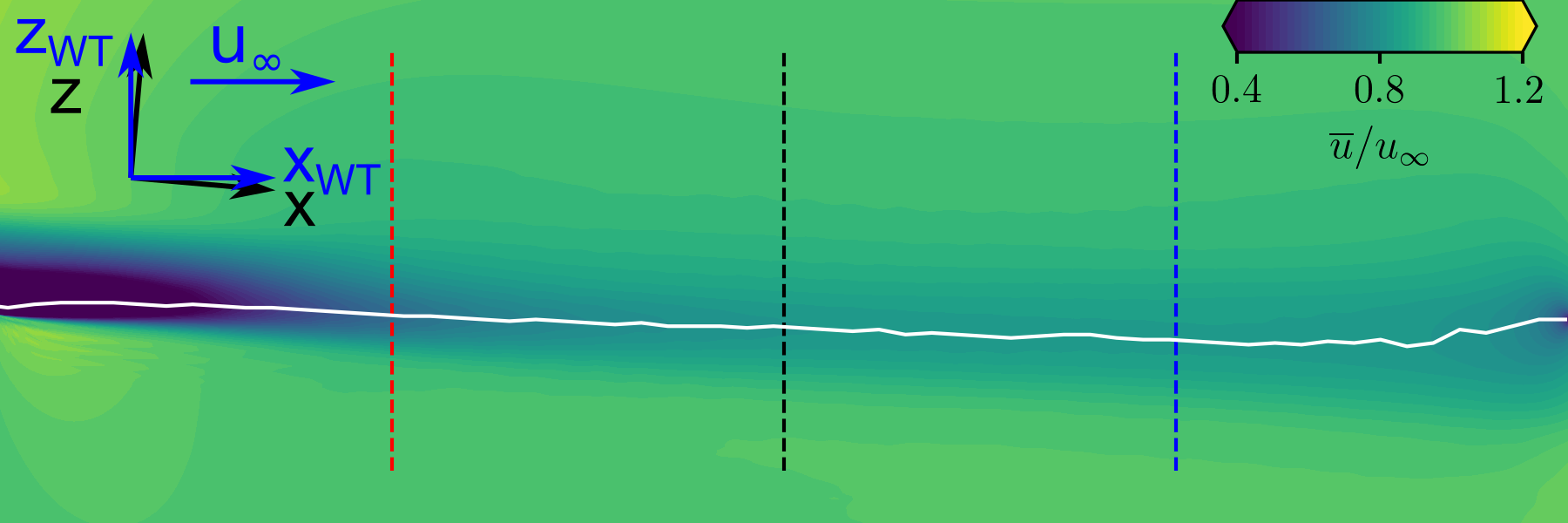}
    \caption{Time-averaged streamwise velocity overlaid with the mean wake centerline position. The data extraction locations at 0.5$c$, 1$c$ and 1.5$c$ behind the trailing edge are indicated with dashed lines. Shown wake data extends from the front wing's trailing edge to the rear wing's leading edge.}\label{fig:centerline_pos_umean}
\end{figure}

\begin{figure}[h]
\centering
\begin{subfigure}{.49\textwidth}
  \centering
  \includegraphics[width=1\linewidth]{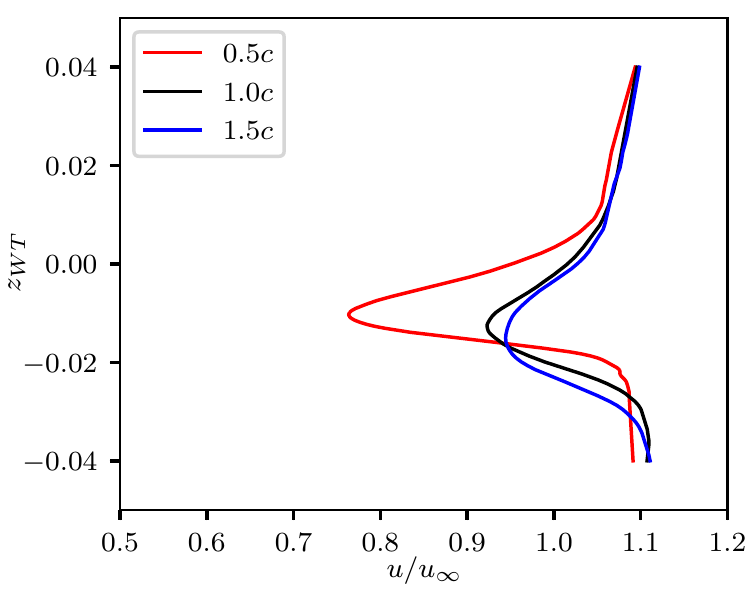}
\end{subfigure}
\begin{subfigure}{.49\textwidth}
  \centering
  \includegraphics[width=1\linewidth]{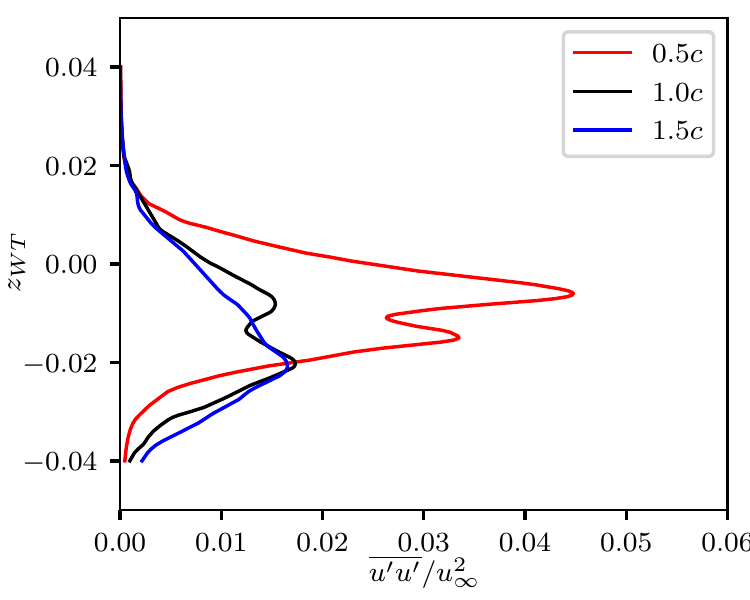}
\end{subfigure}%

\begin{subfigure}{.49\textwidth}
  \centering
  \includegraphics[width=1\linewidth]{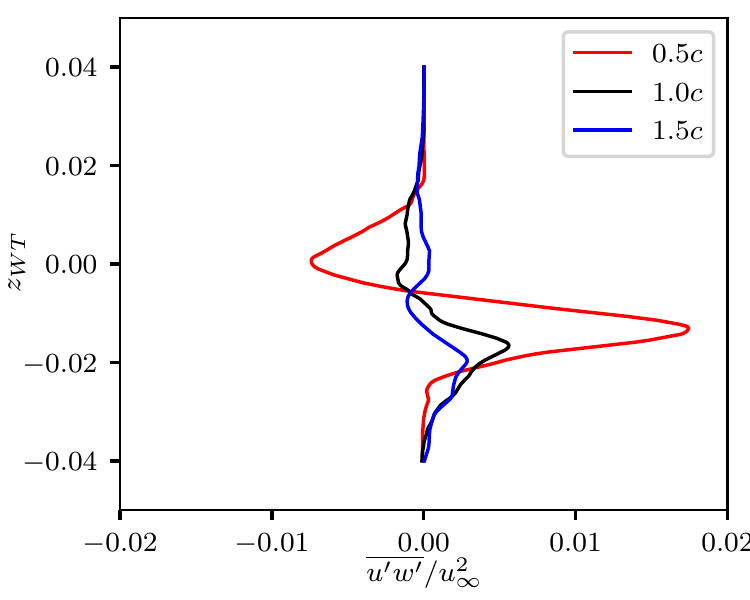}
\end{subfigure}
\begin{subfigure}{.49\textwidth}
  \centering
  \includegraphics[width=1\linewidth]{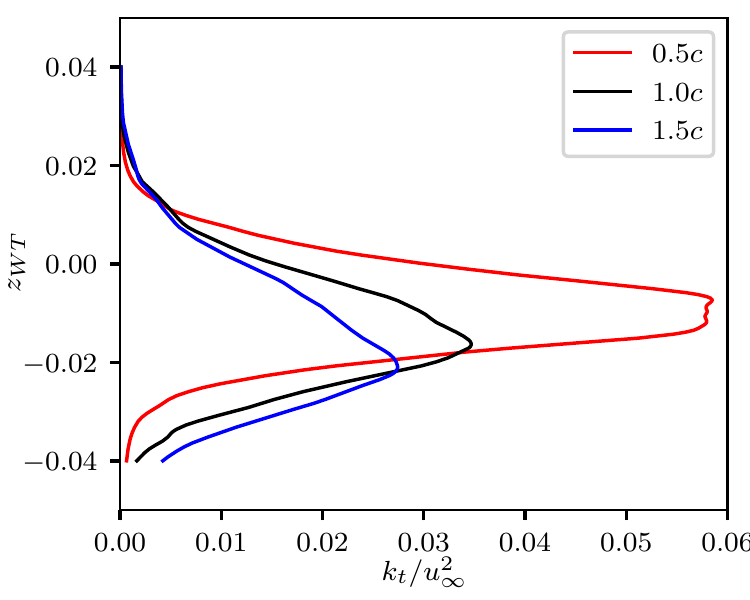}
\end{subfigure}%
  \caption{Profiles of wake properties extracted at the three positions in the wake shown in Fig.~\ref{fig:centerline_pos_umean}, at 0.5$c$, 1$c$ and 1.5$c$ behind the front wing's trailing edge.}
  \label{fig:profiles}
\end{figure}

The development of the wake velocity characteristics is shown in Fig.~\ref{fig:profiles}. The data is extracted at three locations shown in Fig.~\ref{fig:centerline_pos_umean}. The latter also visualizes the location of the wake centerline, which is computed at every streamwise plane in the wake (normal to the inflow direction) by locating the vertical position ($z_{WT}$) of the minimum of the (averaged) axial velocity ($\overline{u})$.
For the following discussion, a local wake coordinate system is introduced, denoted by the subscript "$WT$" (for "wind tunnel"), which is rotated around the spanwise axis (y) compared to the body-fixed coordinate system shown above in Fig.~\ref{fig-separation-front} such that the new x axis $x_{WT}$ is aligned with the inflow direction, as visible in Fig.~\ref{fig:centerline_pos_umean}.
The wake velocity deficit is evident in the shape of $\overline{u} / u_{\infty}$ in Fig.~\ref{fig:profiles}, with a sharp minimum at the upstream location which becomes wider further downstream. The velocity deficit dissipates and the vertical velocity gradients decrease. The downward displacement of the velocity minimum between successive positions is consistent with the centerline shape in Fig.~\ref{fig:centerline_pos_umean}, which trends downward due to the front wing producing lift.

The streamwise normal component of the Reynolds stress $\overline{u^{\prime}u^{\prime}}/u_{\infty}^2$ in Fig.~\ref{fig:profiles} shows two distinct maxima at the upstream position at 0.5$c$, which is consistent with a strong velocity deficit bounded by shear layers associated with high speed flow regions from above and below the wing segment and the wake. The fluctuations are stronger in the upper portion of the wake. The opposite can be observed for $\overline{u^{\prime}w^{\prime}}/u_{\infty}^2$ in Fig.~\ref{fig:profiles}, where the positive values in the lower wake portion dominate at the upstream position. These values indicate a strong exchange between the high speed outer flow below the wake and the low speed wake core. The overall turbulent kinetic energy $k_t$ in Fig.~\ref{fig:profiles} exhibits a maximum near the wake center, which shows a consistent downward displacement with streamwise distance. All fluctuation quantities in Fig.~\ref{fig:profiles} as well as the velocity deficit decrease with streamwise distance in the wake, reflecting momentum exchange and mixing of outer and inner flow, which eventually leads to the dissipation of the wake.

\begin{figure}[h]%
    \centering
    \includegraphics[width=1\textwidth]{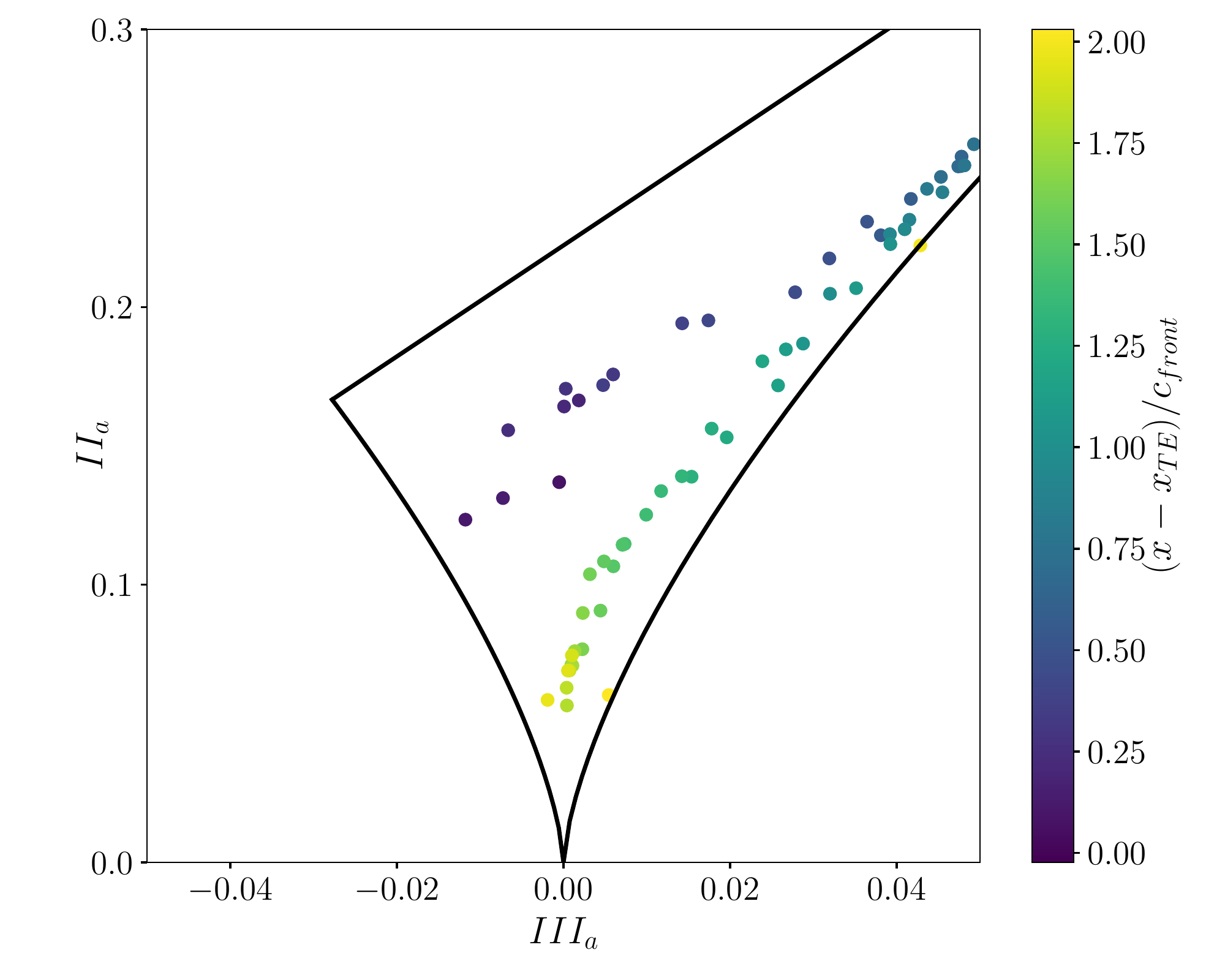}
    \caption{Lumley's anisotropy invariant map showing different locations along the wake centerline. Points are extracted at equidistant streamwise positions, colors represent streamwise distance from the front wing's trailing edge.}\label{fig:anisotropy_triangle}
\end{figure}

The anisotropy of the Reynolds stress tensor can give an indication of the dominant processes occurring in the wake. The anisotropy invariants computed from eigenvalues of the anisotropy tensor~\cite{lumley:1977} along the wake centerline are shown in Lumley's turbulence triangle in Fig.~\ref{fig:anisotropy_triangle}. 
The latter visualizes the anisotropy of turbulence by mapping each turbulent state to a location in an invariant map, depending on the local relation between the magnitudes of the turbulent fluctuations in the three dimensions in space.
Every (physically) possible state falls inside of a triangle (giving the map its name), whose borders are indicated here by solid lines.
The three corners of the triangle mark three particular cases: A "one-component" state, where fluctuations exist only in one direction (top corner), an "axisymmetric two-component" state with fluctuations in two directions of the same magnitude (left corner), and the isotropic state, which exhibits fluctuations in all three directions of the same strength (bottom corner).
Furthermore, the borders connecting those corners indicate a "two-component" state with fluctuations in only two directions (top border), an "axisymmetric expansion" (or "rod-like") state, where the fluctuations in a single direction notably exceed the other two (right border), and an "axisymmetric contraction" (or "disc-like") state, which is characterized by fluctuations in two directions of the same magnitude that are considerably stronger than those in the third direction (left border).
More details on the theory behind and the involved algorithms can be found e.g.~in~\cite{lumley:1977,choi2001}.
Fig.~\ref{fig:anisotropy_triangle} shows a zoomed-in view of the lower part of the triangle near the state of isotropy, which is represented by $III_a = II_a = 0$. The dark blue points in the wake immediately downstream of the trailing edge are located near the center of the shown view, with the anisotropic state trending toward the right-hand boundary beginning at about $(x-x_{TE})/c = 0.5$. This boundary is representative of axisymmetric expansion of the tensor, i.e. with one normal component being significantly larger than the other two, as noted above. In the present case, the vertical normal stress component $\overline{w^{\prime}w^{\prime}}/u_{\infty}^2$ dominates and is significantly larger than $\overline{u^{\prime}u^{\prime}}/u_{\infty}^2$ or $\overline{v^{\prime}v^{\prime}}/u_{\infty}^2$.
This axisymmetric expansion state remains over a large part of the wake propagation distance. The dissipation of $\overline{w^{\prime}w^{\prime}}/u_{\infty}^2$ causes movement toward isotropic equilibrium, where all three normal components approach equality.

\begin{figure}[h]%
    \centering
    \includegraphics[width=0.5\textwidth]{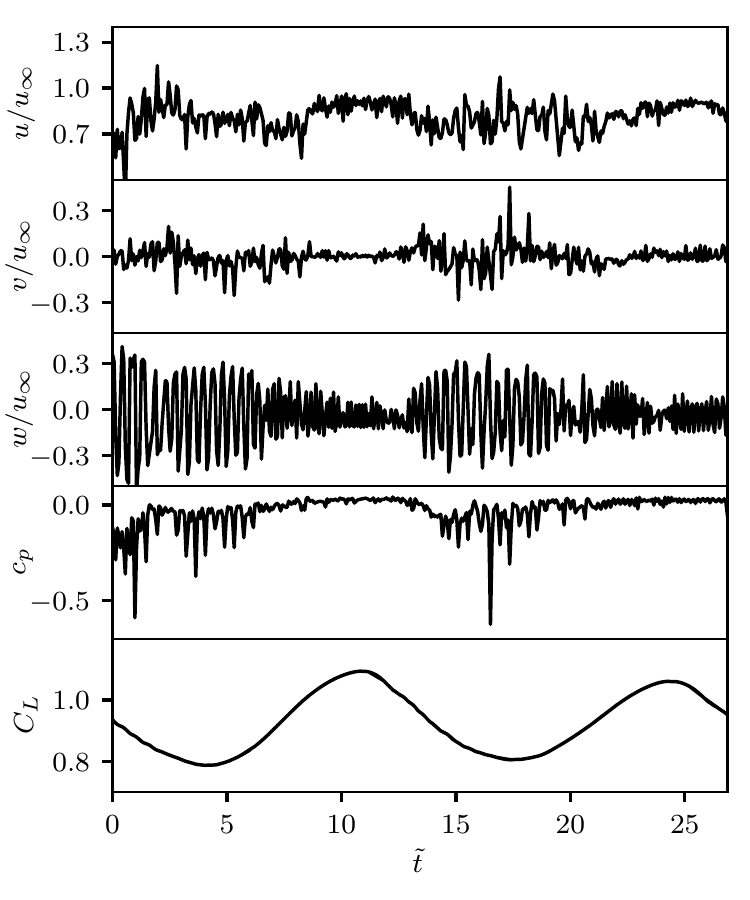}
    \caption{Time series of the velocity components and $c_p$ at the wake centerline one chord downstream of the trailing edge, together with the lift coefficient of the front wing segment.}\label{fig:wake_point_cl_signals}
\end{figure}

The wake is characterized by high velocity fluctuations, as shown for a representative location, one chord behind the trailing edge, in the center of the wake in Fig.~\ref{fig:wake_point_cl_signals} over two buffet periods. In all three components, phases with comparatively low amplitudes of velocity fluctuations alternate with phases of high fluctuations during a single buffet cycle. Periods with high lift correspond to the least amount of separation on the front wing segment, which correlates with low levels of wake turbulence; and vice versa. 

\begin{figure}[h]%
    \centering
    \includegraphics[width=0.5\textwidth]{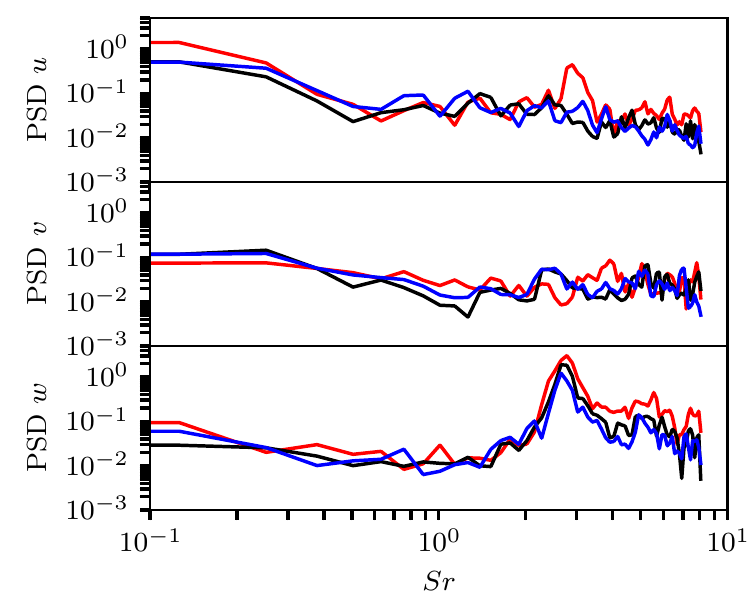}
    \caption{Power spectral densities of the velocity components $u,v,w$. Colors correspond to streamwise position, as in Fig.~\ref{fig:profiles}}\label{fig:wake_point_PSD}
\end{figure}

Spectral densities of the time signals of the velocity components at three positions in the wake center at 0.5$c$, 1$c$ and 1.5$c$ are shown in Fig.~\ref{fig:wake_point_PSD}. The high frequency region is focused upon, as the short time series does not permit a high resolution of the region of the buffet oscillation around $Sr \approx 0.075$. Therefore, a variance-reducing averaging using Welch's method~\cite{welch1967} and 5 overlapping segments is used for power spectral density estimation. The high frequency content decreases with streamwise distance, especially in the streamwise and vertical components. Apart from high amplitudes at low frequencies near the frequency resolution limit at $Sr = 0.1$ caused by the shock oscillation, there is a distinct peak at about $Sr \approx 2.4$ in the spectrum of the axial velocity $u$ at the upstream position at $0.5c$. This spectral feature dissipates downstream and is not discernible in the spectra of the spanwise velocity $v$. The spectra of the vertical velocity $w$, however, consistently show this peak at all three positions. This high-frequent wake fluctuation propagates consistently over significant distances in the wake. The spanwise velocity oscillation in the center panel of Fig.~\ref{fig:wake_point_PSD} does not show characteristic peaks to a degree similar to the other two components. The spectra generally reflect the insights gained via the anisotropy invariants, which showed that the streamwise and vertical fluctuations predominate in the flow behind the front wing trailing edge, with $\overline{w^{\prime}w^{\prime}}/u_{\infty}^2$ remaining dominant thereafter. While the Reynolds and anisotropy tensors contain information which is integrated over the entire frequency range, Fig.~\ref{fig:wake_point_PSD} shows that the redistribution and shifts in anisotropy invariant space occur predominantly between $Sr=2$ and $Sr=4$.

Fig.~\ref{fig:wake_point_cl_signals} shows how the spectral characteristics change over time during the buffet phases. Fourier transformation based methods cannot resolve this when applied to the entire time series. Future work on the interpretation of spectral characteristics of the wake may involve wavelet transformation in order to isolate temporal variation in the spectra, as the dynamics in the wake are strongly dependent on the buffet phase.

Especially the spectrum of the vertical velocity $w$ shows strong fluctuations in the wake at frequencies comparatively high in relation to the buffet frequency in the region of $Sr \approx 1.5$ to $Sr \approx 8$, that can be attributed to the vortices or turbulent structures present in the wake. The broadband characteristic in the spectrum mirrors the variation of the size and frequency of those structures. 
As the pairs of counter-rotating vortices forming from the shear layers are accompanied by the alternation of positive and negative vertical velocities, their footprint is most notable in the $w$-spectrum.

\subsubsection*{Modal Analysis of the Wake Flow}
\begin{figure}[h]%
 \begin{subfigure}{.49\textwidth}
      \centering
      \includegraphics[width=1\linewidth]{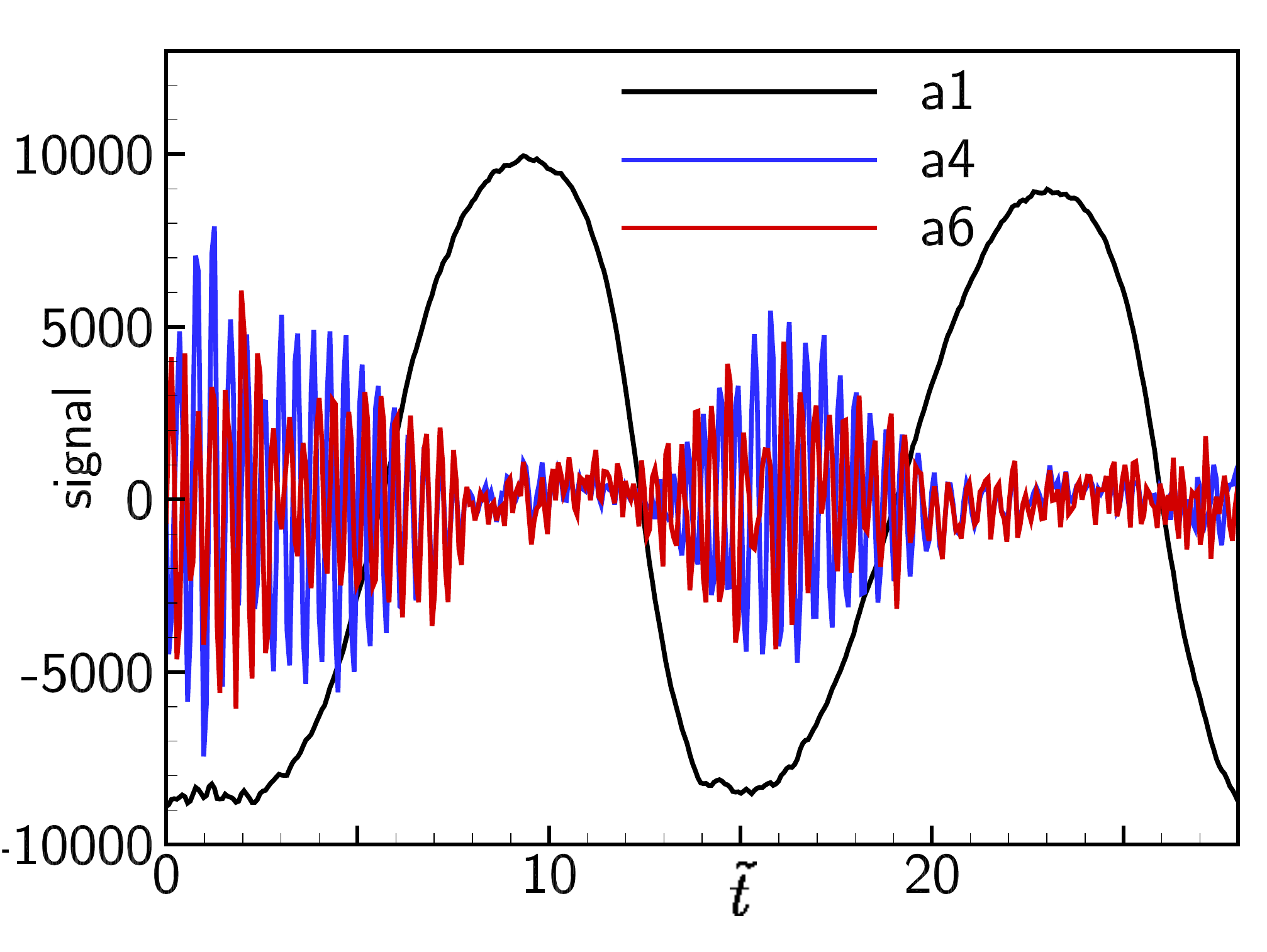}
      \caption{Time series of selected POD mode coefficients}
      \label{fig:PODts}
\end{subfigure}
     \begin{subfigure}{.49\textwidth}
      \centering
      \includegraphics[width=1\linewidth]{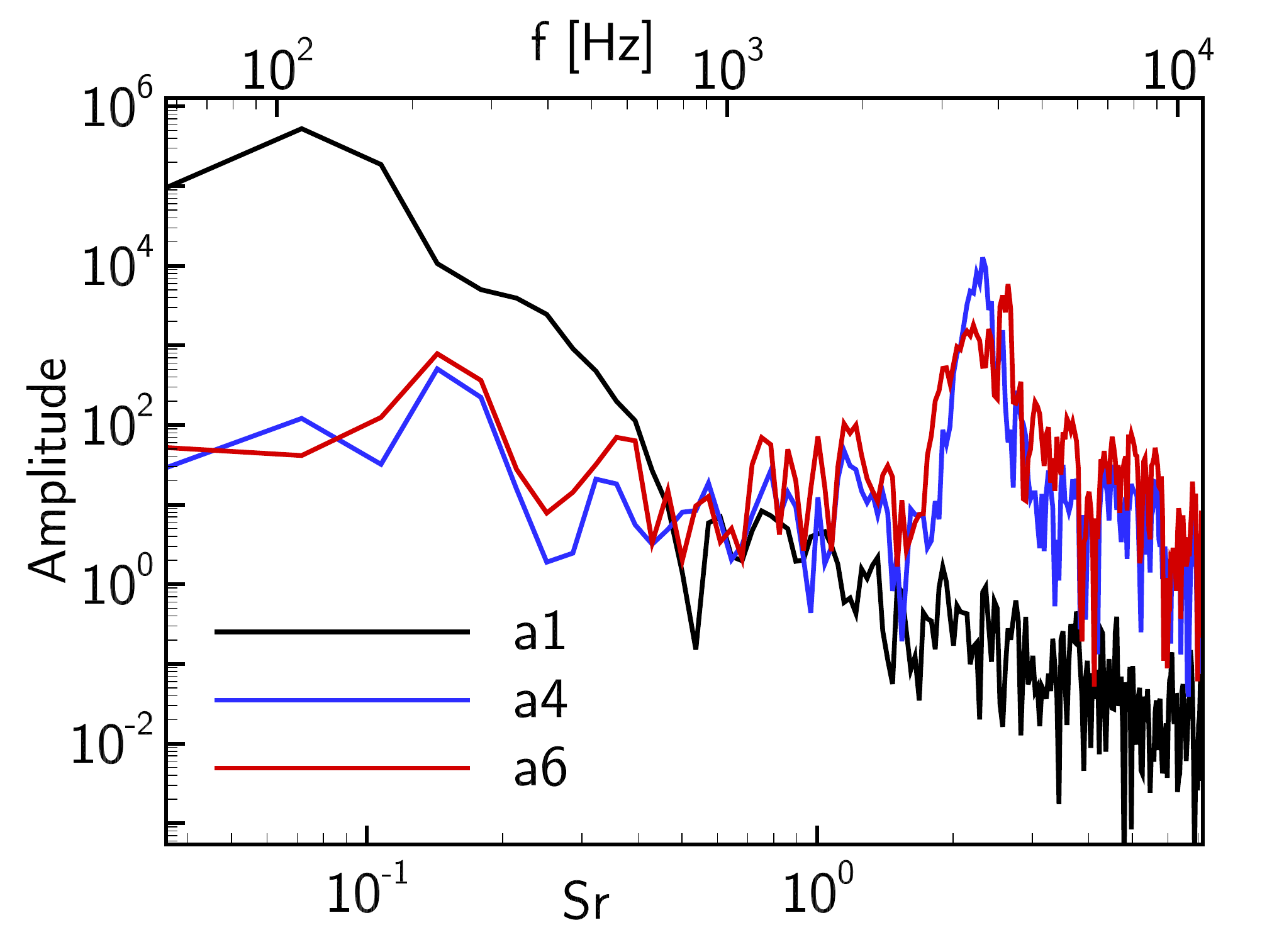}
      \caption{Spectral densities of selected POD modes}
      \label{fig:PODspec}
\end{subfigure}
 \caption{Time series and spectra of selected POD mode coefficients in the spanwise midplane of the front wing segment }\label{fig:PODmodes1}
 \end{figure}

Modal analyses of the flow field using Proper Orthogonal Decomposition (POD) make it possible to distinguish different flow phenomena and their respective temporal behavior.
The POD technique decomposes (the unsteady component of) a time-dependent flow field into different spatial modes with corresponding amplitude signals (or temporal coefficients), based on a singular value decomposition of the data set. 
Unlike a (spatial) Fourier transformation which breaks down the field into Fourier modes, a POD results in an optimal basis of orthogonal spatial functions for the given flow field data.
This decomposition is optimal in the least squares sense, e.g.~the modes are chosen optimal to capture the kinetic energy if the velocity field is used as input, for example.
The resulting modes then represent the major time-dependent and coherent features of the flow field, like vortex streets for example, depending on the respective case.
More details on POD theory and the involved algorithms can be found e.g.~in~\cite{taira2017}.
The time series of three POD coefficients of the horizontal velocity obtained in the midplane are shown in Fig.~\ref{fig:PODts}. The first mode pair exhibits a clearly periodic shape and can be associated with the shock motion of the transonic buffet present on the front wing segment.
Spectral densities of the time signals of the POD modes are depicted in Fig.~\ref{fig:PODspec}.  
As already mentioned earlier, buffet occurs at a distinct frequency of $\mathrm{Sr\approx0.075}$, which corresponds to $\mathrm{f\approx120Hz}$ and is the dominant frequency of the first POD mode. 
\begin{figure}[h]%
 \begin{subfigure}{.49\textwidth}
    \centering
    \includegraphics[width=1\textwidth]{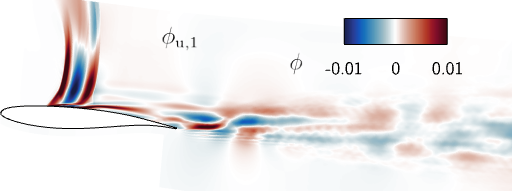}
    \caption{POD mode 1}\label{fig:POD1}
 \end{subfigure}
 \begin{subfigure}{.49\textwidth}
    \centering
    \includegraphics[width=1\textwidth]{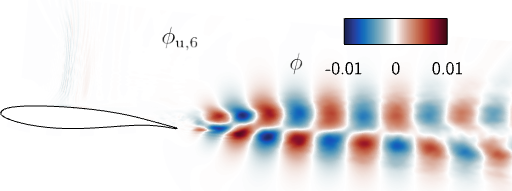}
    \caption{POD mode 6}\label{fig:POD6}
    \end{subfigure}
    \caption{POD modes in the spanwise midplane of the front wing segment}\label{fig:PODmodes2}
\end{figure}
The time series of the POD modes 4 and 6, as shown in Fig.~\ref{fig:PODts}, also yield some kind of periodicity with the buffet frequency. Here, oscillations of high frequency occur during the upstream motion of the shock during each buffet cycle. During the downstream motion, when the flow is mainly attached, the amplitudes are significantly lower, as described above. These modes can be associated with flow separation behind the shock and the resulting wake motion. Fig.~\ref{fig:PODspec} shows that these modes yield distinct spectral peaks at $\mathrm{Sr=2.0..2.6}$, which agrees well with the peaks of the velocity components in the wake shown in Fig.~\ref{fig:wake_point_PSD}.
Contours of the first and sixth POD mode of the horizontal velocity component are shown in Fig.~\ref{fig:PODmodes2}. The first mode exhibits clear maxima with changing sign in the shock region, which shows its main connection to the shock motion. The regularly spaced regions of positive and negative values of the sixth POD mode in the wake can be attributed to the shedding of wake vortices already noticed above that mainly takes place during the upstream motion of the shock. The corresponding frequency range falls in the region already discussed for the spectra of $w$ in Fig.~\ref{fig:wake_point_PSD}.

\subsection{Interaction of the Wake with the Rear Wing Segment}\label{subsec52}

The following section discusses the wake interactions in the configurations (A) and (B), for the first setting of the angle of incidence of the rear wing segment ($\alpha_I = -4^{\circ}$). Buffet occurs only on the front wing segment in these configurations, which strongly impacts the aerodynamics of the rear wing segment.
\begin{figure}[h]%
    \centering
    \includegraphics[clip,trim={80 80 200 210},width=0.5\textwidth]{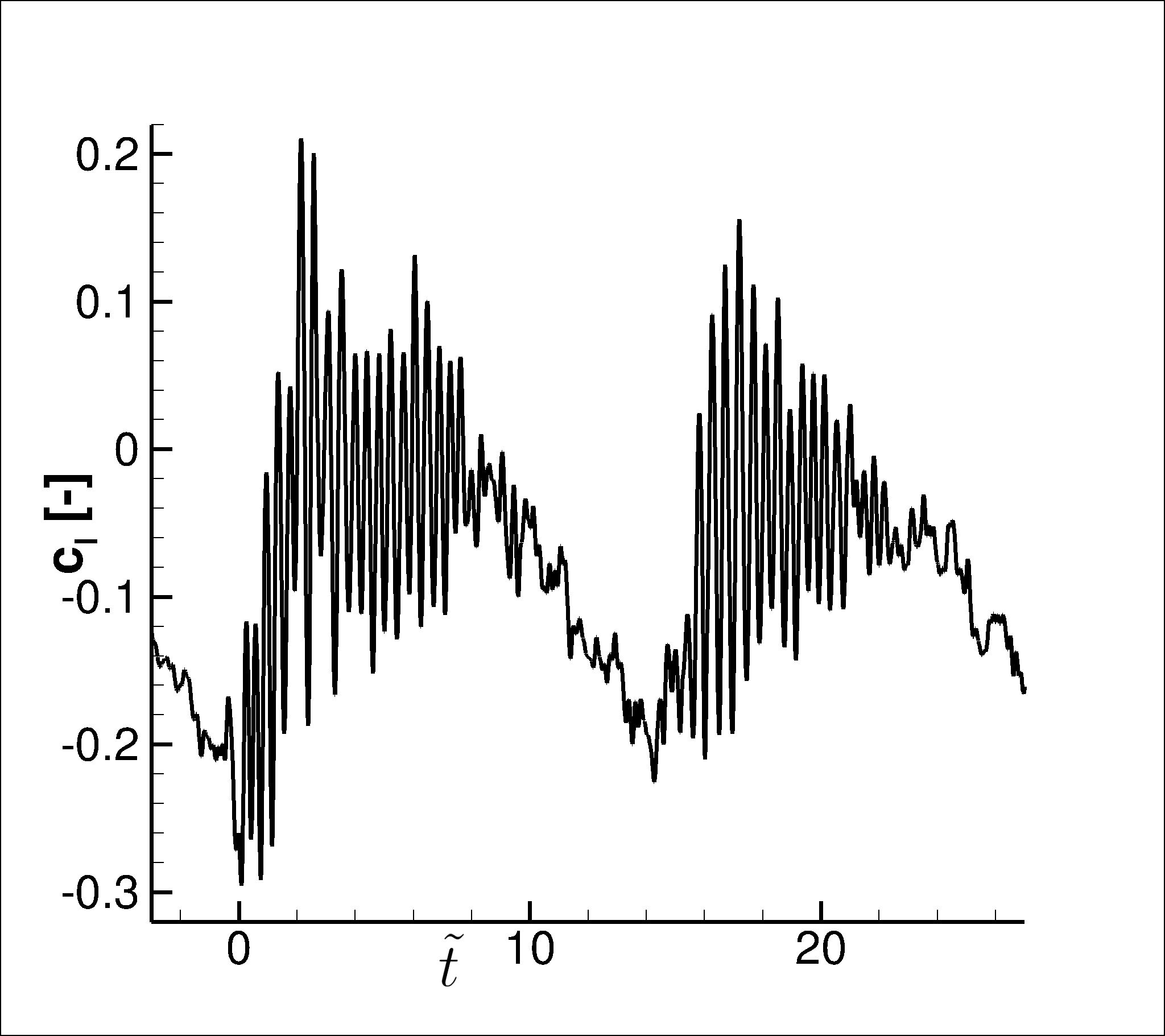}
    \caption{Lift coefficient of the rear wing segment over two buffet periods, configuration (A)}\label{fig:ca-hinten}
\end{figure}
The wake impingement causes a strong variation of the rear wing segment's loading, as shown in Fig.~\ref{fig:ca-hinten}. The figure displays the evolution of the lift coefficient of the rear wing segment over two buffet periods. At first look, a relatively low-frequent oscillation of the loading can be seen which shows a period that is equal to the buffet period of the front wing segment. 
This low-frequent oscillation dominates the load fluctuation with a variation of the lift coefficient between approximately $c_{l,min}=-0.3$ and $c_{l,max}=0.2$, corresponding to an amplitude of $\hat{c_l}=0.25$. This is caused by the temporal variation of the downwash intensity behind the front wing segment during the buffet cycle. With the oscillation of lift of and circulation around the front wing segment, the induced vertical velocity downstream from trailing edge oscillates too. This induces a variation of the effective angle of attack of the rear wing segment. Consequently, the lift coefficient of the rear wing segment reaches its minimum shortly after the lift coefficient of the front wing segment reaches its maximum value, considering the time delay of the propagation of the downwash in downstream direction. The same correlation holds between the time of the lift maximum of the rear and the lift minimum of the front wing segment. 
\begin{figure}[h]%
    \centering
    \includegraphics[clip,trim={40 30 240 510},width=0.6\textwidth]{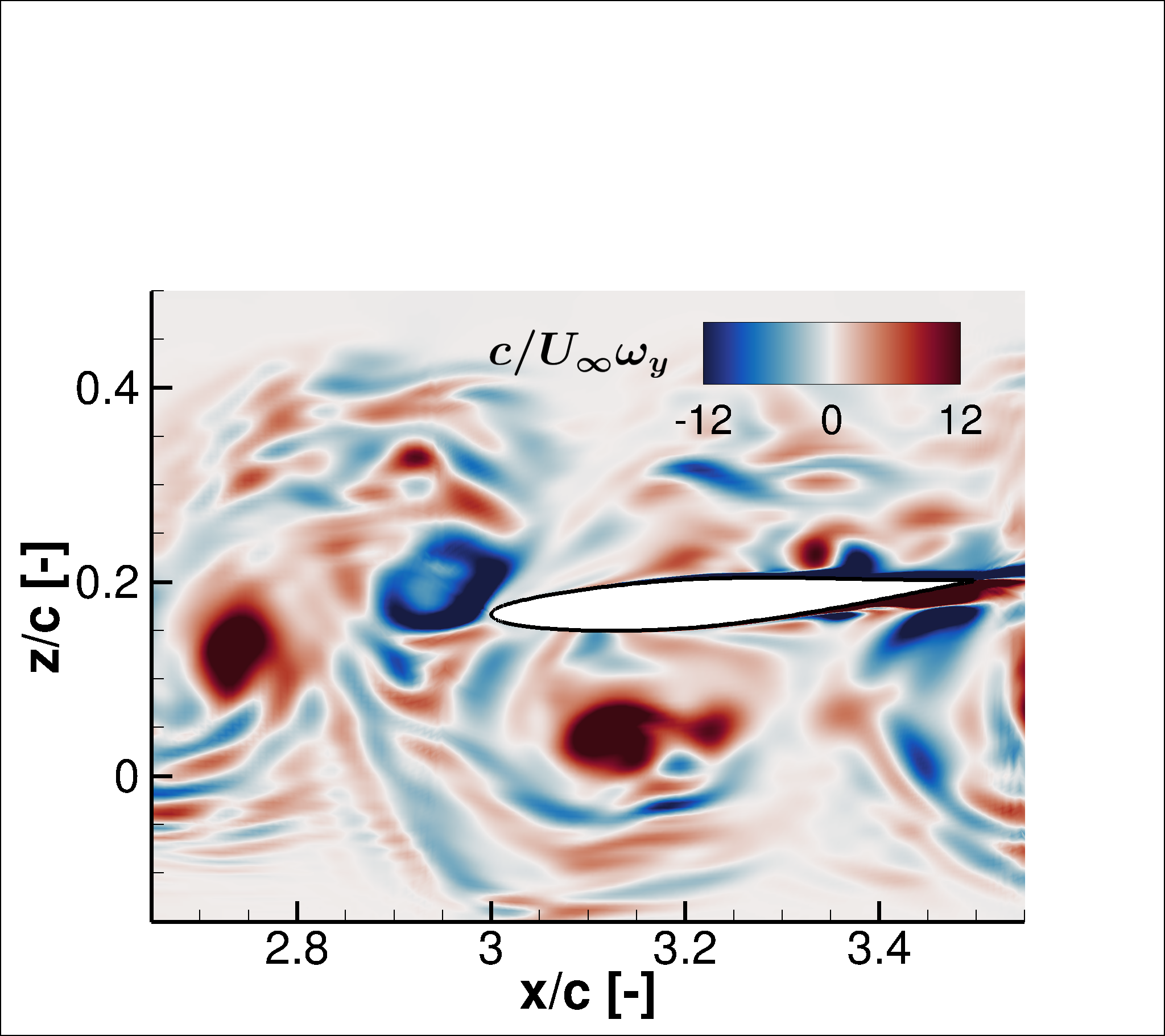}
    \caption{Instantaneous spanwise vorticity field during vortex impingement onto the rear wing segment, configuration (A)}\label{fig:ma-feld}
\end{figure}
The low-frequent load oscillation at the rear wing segment is superposed by relatively high-frequent oscillations of varying amplitude and frequency. 

These high-frequent oscillations are caused by the impingement of the turbulent structures or vortices in the wake discussed in section~\ref{subsec51} upon the rear wing segment. As an example, such a vortex impingement is depicted in Fig.~\ref{fig:ma-feld}, which displays the instantaneous dimensionless spanwise vorticity in the immediate vicinity of the rear wing segment. A turbulent vortex  pair, indicated by the region of strongly positive and negative vorticity, respectively, impinges on the leading edge of the rear wing segment at the depicted moment in time. Additionally, a second turbulent structure can be identified on the upper side close to the trailing edge, which appears notably stretched and distorted due to the interaction with the wing segment. 

The impact of the turbulent vortices leads to a change of both the effective angle of attack and the effective inflow velocity for the rear wing segment, resulting in changes to the acting forces. Furthermore, the smaller turbulent structures also distort the local pressure distribution on the surface as they pass, contributing to the load fluctuation. 

As evident from Fig.~\ref{fig:ca-hinten}, the load oscillations caused by the vortex impingement exhibit significant amplitudes reaching up to $\hat{c_l}=0.2$, which is comparable to the amplitude of the low-frequent oscillation. 
Moreover, the frequencies and especially the amplitudes of the high-frequent oscillations vary significantly within one buffet period. Phases with a low level of fluctuation alternate with phases of high fluctuation, which is explained by the alternation of comparatively weak fluctuations of velocity and pressure in the wake at times when only small vortices are present, and strong fluctuations, respectively, accompanied by the presence of large vortices. Consequently, the vortex-induced force oscillations are stronger when the vortices impinge on the rear segment that have been generated upstream at a moment earlier in time, when the amount of separation on the front wing segment has been large. It should be noted here that there is a time delay between the formation of the vortices and their impingement due to their propagation time through the wake, which can be roughly approximated by $D/U_{\infty}$, with $D$ being the distance between the wing segments and the free stream velocity $U_{\infty}$. As the mean convective velocity in the wake is smaller than $U_{\infty}$, however, the actual time delay proves to be slightly larger. 
\begin{figure}[h]%
    \centering
    \includegraphics[clip,trim={30 30 240 210},width=0.5\textwidth]{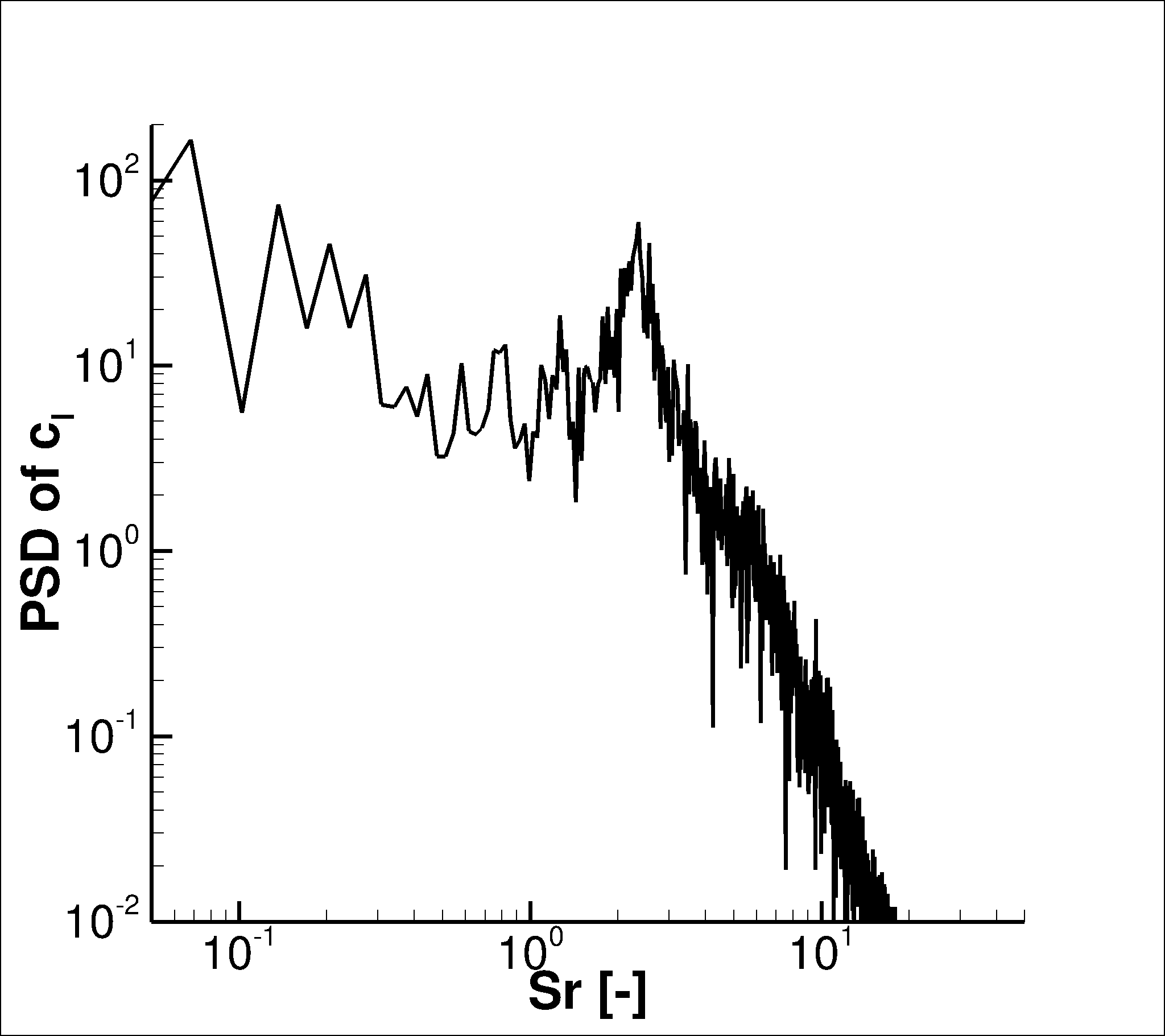}
    \caption{Power spectral density of the lift coefficient of the rear wing segment, configuration (A))}\label{fig:psd-lift-rear}
\end{figure}
The contribution of the impinging turbulent structures to the rear wing lift oscillation is also evident when looking at the spectral distribution of the latter, in form of the PSD of the lift coefficient shown in Fig.~\ref{fig:psd-lift-rear}. Beside the low-frequent oscillation caused by the change of the downwash magnitude around the buffet frequency, broadband spectral content contributions are visible at higher frequencies around $Sr \approx 1$ to  $Sr \approx 4$ due to the vortex impingement. The peak is around $Sr \approx 2.5$ which corresponds to the peak frequency identified in the wake, cf. Fig.~\ref{fig:wake_point_PSD}.

\begin{figure}[h]%
    \centering
    \includegraphics[clip,trim={80 60 240 220},width=0.5\textwidth]{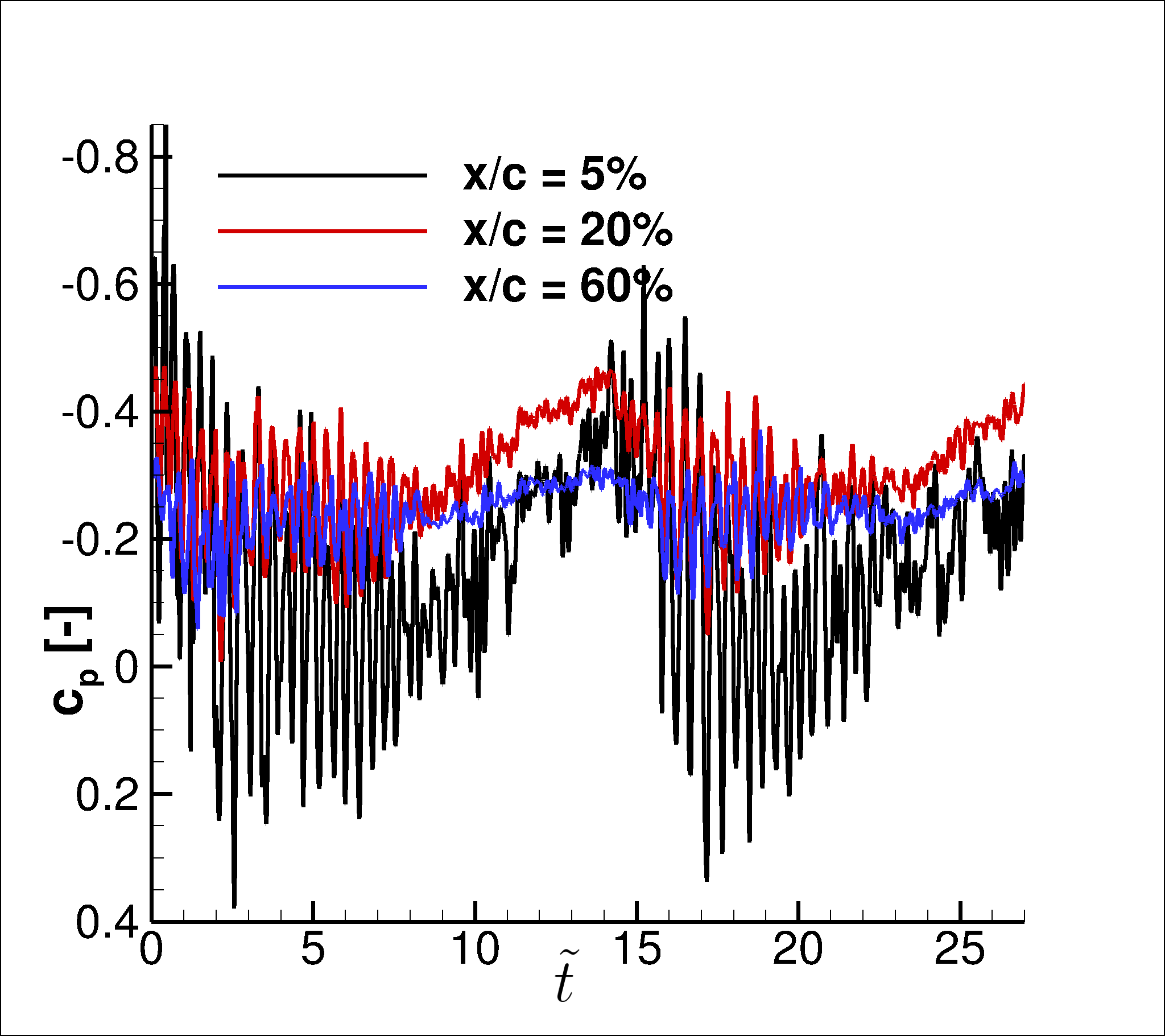}
    \caption{Time series of the pressure coefficient for different locations on the rear wing segment's surface, configuration (A)}\label{fig:cp-t-hinten}
\end{figure}

The interaction with the wake generates pressure fluctuations on the surface of the rear wing segment. Fig.~\ref{fig:cp-t-hinten} shows the evolution of the pressure coefficient for three locations on the lower surface of the rear wing segment, at $x/c=0.05$, $x/c=0.2$ and $x/c=0.6$. Both the low-frequent and the high-frequent components discussed above can be identified. 
Additionally, the intensity of the fluctuations decreases with increasing downstream position, as the turbulent structures are dissipated during the course of their interaction with the rear wing segment. 
\begin{figure}[h]%
    \centering
    \includegraphics[clip,trim={10 20 240 170},width=0.5\textwidth]{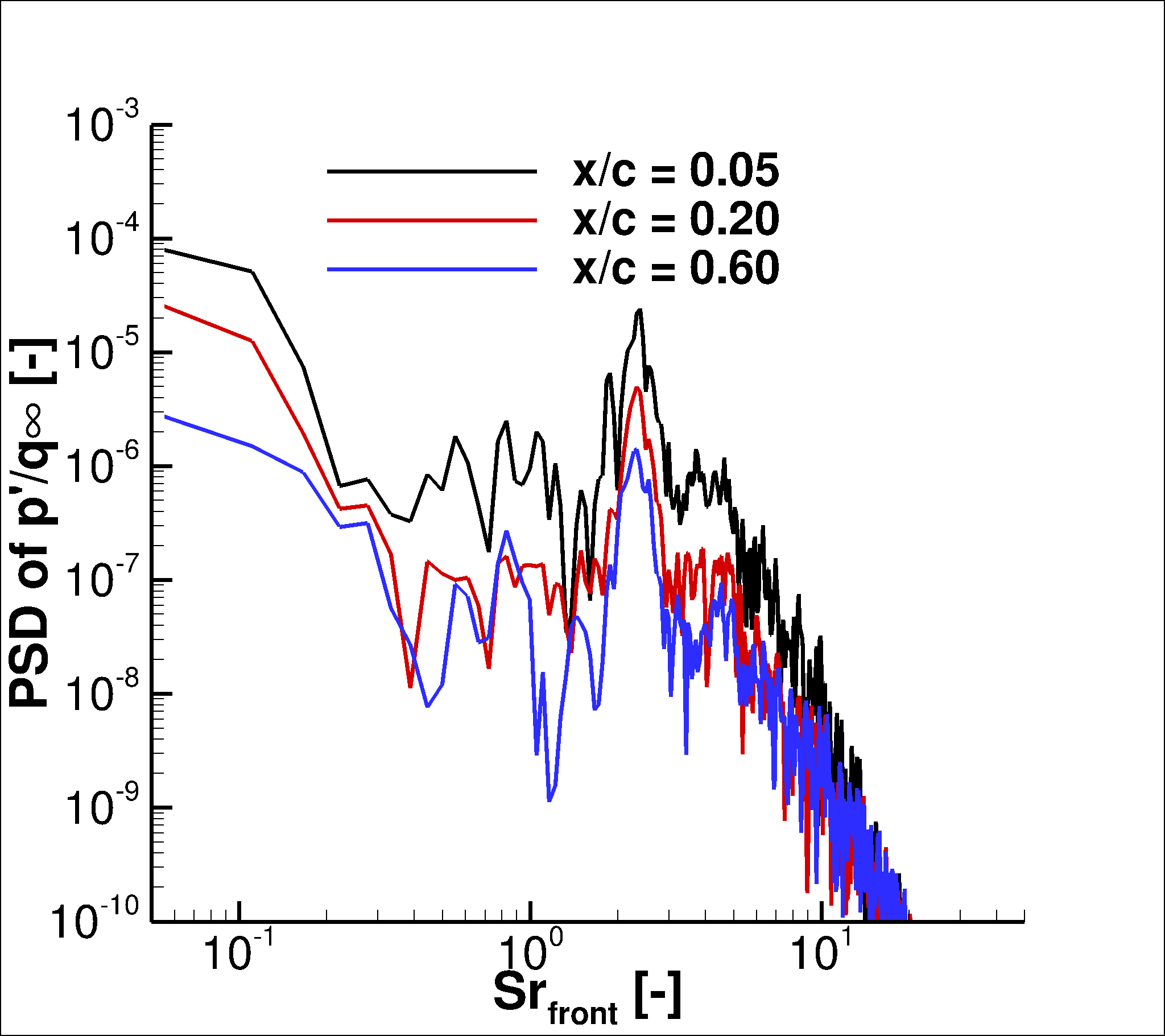}
    \caption{Power spectral density of the surface pressure for different locations on the rear wing segment's surface, configuration (A)}\label{fig:psd-hinten}
\end{figure}

The spectral distribution of the pressure fluctuations for the three locations on the surface is displayed in Fig.~\ref{fig:psd-hinten} in terms of the power spectral density (PSD) of the pressure coefficient. 
As for the lift coefficient, the fluctuations of the surface pressure are dominated by low-frequent oscillations at the front wing buffet frequency, which is related to the variation of the downwash as described above. Likewise, the increased level evident at high frequencies between $Sr \approx 1$ and $Sr \approx 5$ corresponds to the impingement of vortices of varying size and frequency. 

This correlation is also supported by the comparison of the surface pressure spectra with the corresponding spectra in the wake (cf. Figures \ref{fig:wake_point_PSD} and \ref{fig:PODspec}).
They appear very similar in shape and exhibit the same features,~i.e. the high amplitudes in the low frequency range at the buffet frequency and the broadband signal in the high frequency range with a dominant peak corresponding to the turbulent fluctuations. The turbulent structures themselves cause the pressure fluctuations on the rear wing segment. The comparison also reveals and underlines the direct correlation between the wake fluctuations and the variation of the rear wing segment loading. A similar relationship has been reported by M\"uller et al.~\cite{mueller2020} for the case of atmospheric turbulence.

\begin{figure}[h]%
    \centering
    \includegraphics[clip,trim={30 20 240 180},width=0.5\textwidth]{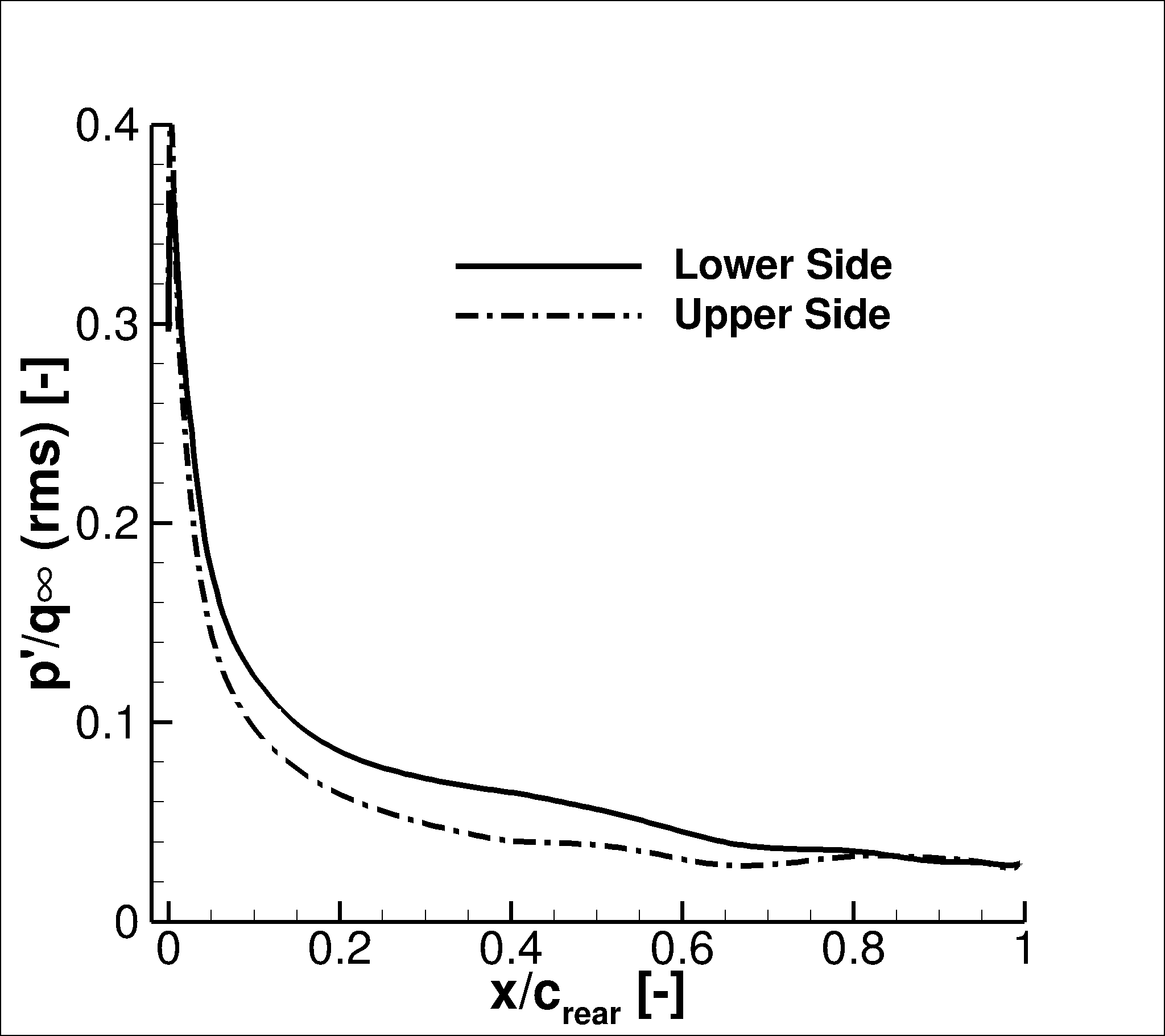}
    \caption{Root mean square (rms) of the pressure fluctuations on the rear wing segment, configuration (A))}\label{fig:cp-rms-hinten}
\end{figure}

Comparing the spectra at the different chordwise locations, a decrease of the spectral density over the whole spectrum can be seen for increasing chord position, which indicates the dissipation of the turbulent structures due to the interaction with the rear wing segment. This effect also becomes apparent in the decreasing rms of the pressure fluctuations with increasing downstream position, as shown in Fig.~\ref{fig:cp-rms-hinten}. 
As the vortices impinge on the leading edge of the rear wing segment, the highest fluctuation levels are found there, continuously decreasing towards the trailing edge. Furthermore, the effect of the wake on the upper (or pressure) side is a smaller than the impact on the lower (or suction) side, as both the flow velocity and the rate of change of the pressure with angle of attack are higher on the latter.

\subsubsection*{Influence of the Position of the Rear Wing Segment}

\begin{figure}[h]%
    \centering
    \includegraphics[clip,trim={80 70 200 220},width=0.5\textwidth]{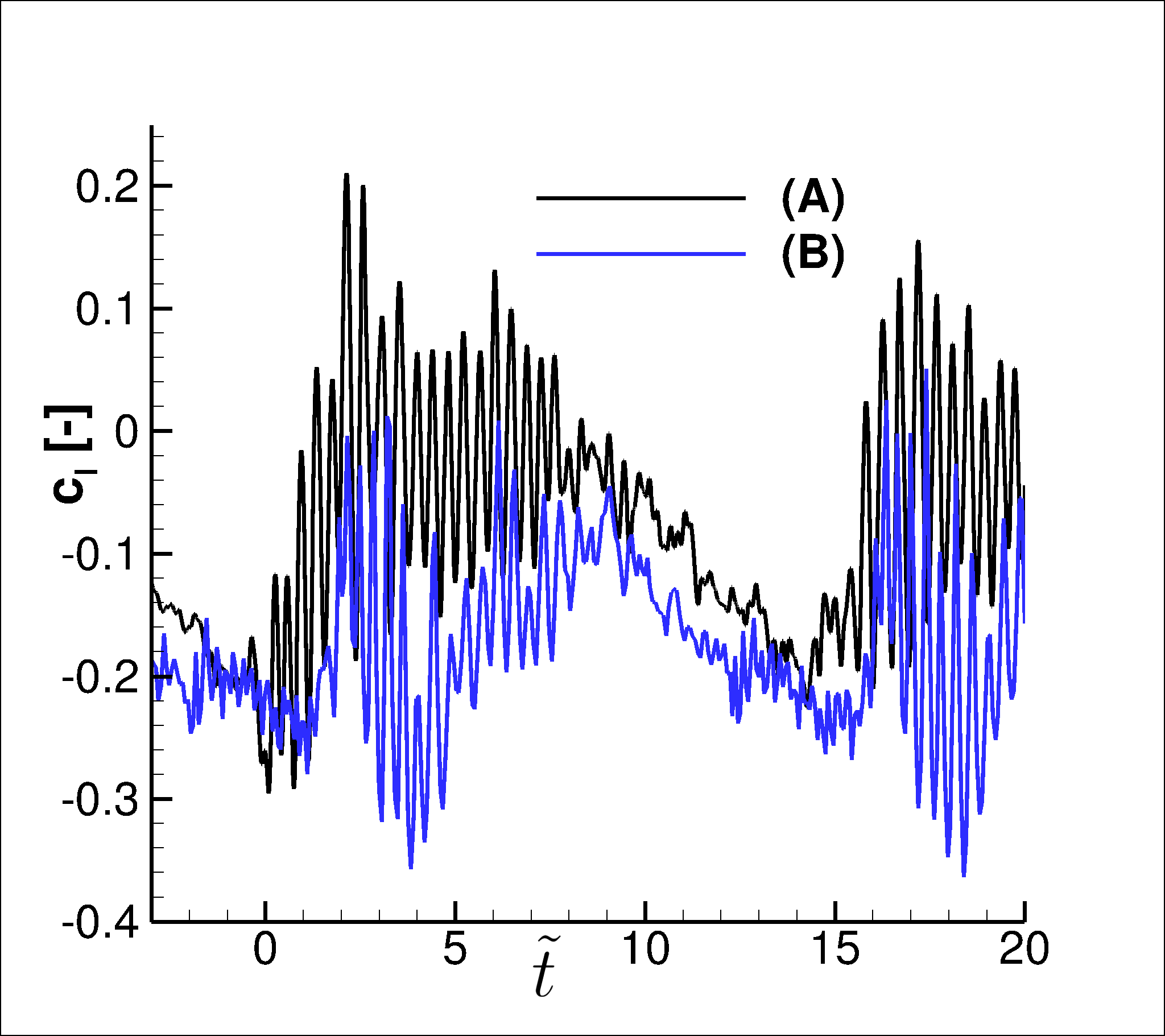}
    \caption{Lift coefficient of the rear wing segment over two buffet periods for configurations (A) and (B)}\label{fig:ca-hinten-vergleichpos}
\end{figure}

The influence of the position of the rear wing segment on the wake interaction phenomena is shown in Fig.~\ref{fig:ca-hinten-vergleichpos}, which displays the time series of the lift coefficient for both positions. Here, the $c_l$ for the higher vertical position ($z/c=1/6$) is marked in black, which corresponds to the configuration considered above (A), and the lower position ($z/c=1/20$) in blue, i.e.~configuration (B). Generally, it is evident that the effect of the wake on the load fluctuation is qualitatively the same for both positions of the rear wing. Both the low-frequent lift variation caused mainly by the downwash and the high-frequent oscillations due to the impact of the wake's vortices are visible, as discussed above, which is expected as the lower position of configuration (B) is still well inside the region covered by the wake.
\begin{figure}[h]%
    \centering
    \includegraphics[clip,trim={40 20 240 190},width=0.5\textwidth]{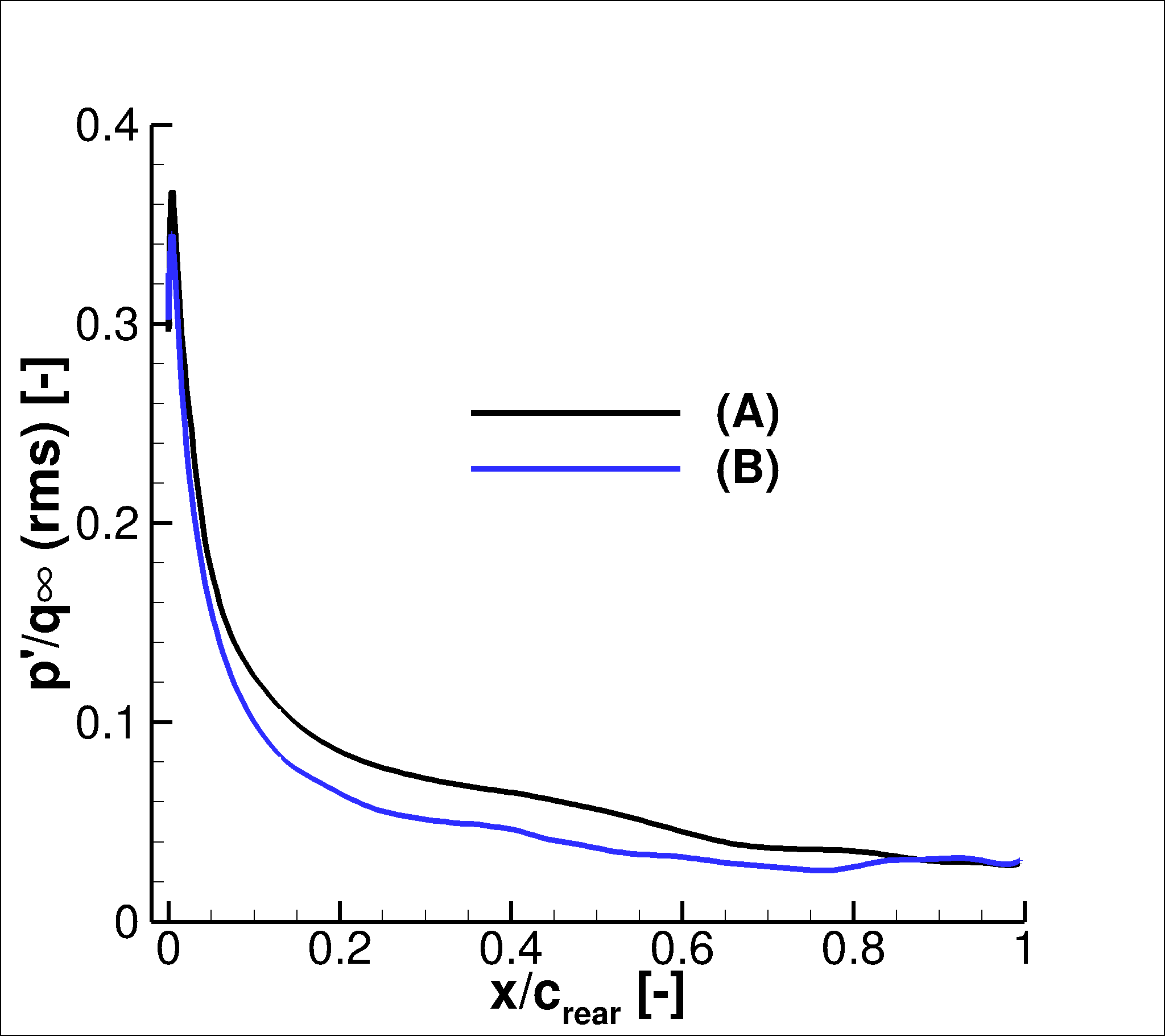}
    \caption{Root mean square (rms) of the pressure fluctuation on the rear wing segment for configurations (A) and (B)}\label{fig:cp-rms-hinten-vergleichpos}
\end{figure}
However, qualitative differences can be seen. Because the rear segment is positioned slightly below the center of the wake for configuration (B), the amplitudes of the load fluctuation are smaller for this case.
Notably, the average lift is also lower due to the smaller velocity deficit at this position, which leads to a higher effective dynamic pressure of the flow that increases the - negative - loading of the rear segment.
The impact on the surface pressure fluctuations is shown in Fig.~\ref{fig:cp-rms-hinten-vergleichpos}. Again, the characteristic remains qualitatively identical; however, the fluctuation levels are reduced for configuration (B).
Although the wake interactions are qualitatively similar, in general, the quantitative differences considering the small shift between both positions.

\subsection{Wake Interactions for Buffet Flow at the Rear Wing Segment}\label{subsec53}

In the following, the wake interaction phenomena of the tandem wing configuration with the adapted setting of the angle of incidence of the rear wing segment ($\alpha_I=-11^{\circ}$, configuration (C)) are presented.
To ensure that buffet is also present on the rear wing segment under the considered inflow conditions, unsteady RANS simulations of the isolated rear segment (configuration (D)) were performed in advance, increasing the angle of attack until buffet onset. A pronounced shock oscillation was found for $\alpha_{rear}=6^{\circ}$, which has consequently been analysed with hybrid simulations.
The buffet cycle and the corresponding variation of lift and flow separation of the rear segment with the NACA64A110 airfoil are similar to the case analyzed above in section~\ref{subsec51} for the front segment exhibiting the OAT15A airfoil. A lift amplitude of $\hat{c_l}= 0.16$ is found with a buffet frequency of 244\,Hz, which corresponds to a Strouhal number of 0.077 with regard to the rear segment's chord, or $Sr=0.154$ in relation to the front wing segment, respectively. Whereas the reduced frequency - in relation to the respective chord - is similar to that of the OAT15A airfoil, the absolute frequency is higher because of the smaller chord length.
\begin{figure}[h]%
    \centering
    \includegraphics[clip,trim={80 60 240 190},width=0.5\textwidth]{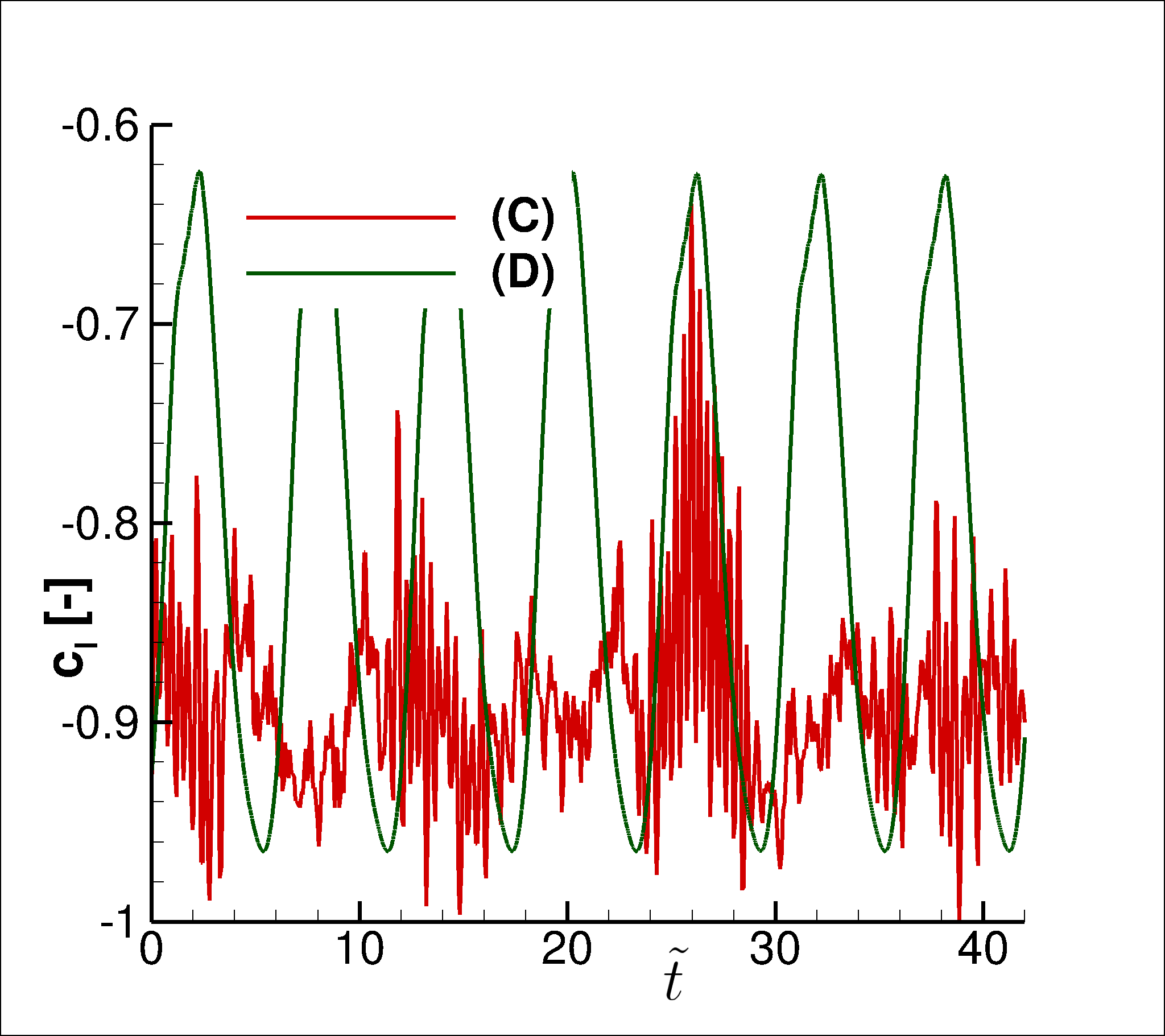}
    \caption{Lift coefficient of the rear wing segment over several buffet periods for configurations (C) and (D)}\label{fig:ca-hinten-htp-buffet}
\end{figure}
Changing the angle of incidence of the rear wing segment to $\alpha_I=-11^{\circ}$ for configuration (C) leads to the same angle of attack relative to the free inflow as in configuration (D), which allows for a comparison to analyze the wake impact under buffet conditions. However, the airfoil is mounted upside down in the tandem setup, thus creating a down force.
The time series of the lift coefficient for both configurations is depicted in Fig.~\ref{fig:ca-hinten-htp-buffet}. Here, the lift of the isolated segment (D) is shown pre-multiplied by -1 to allow for a better comparison. Notably, the harmonic oscillation of the lift of the uninfluenced buffet seen for configuration (D) is disrupted by the impact of the vortices of the wake. The characteristic is more closer to the one found for configurations (A) and (B), as the wake impact dominates the aerodynamics under the analyzed conditions. An unambiguous periodicity of the loading can not be determined. Yet, lift maxima reoccur after 9.6 to 14.3 dimensionless time units, corresponding to a frequency range of $f=110..165$\,Hz or a Strouhal number of $Sr=0.07..0.1$, which is close to the buffet frequency of the front wing segment.
\begin{figure}[h]%
    \centering
    \includegraphics[clip,trim={30 20 190 210},width=0.5\textwidth]{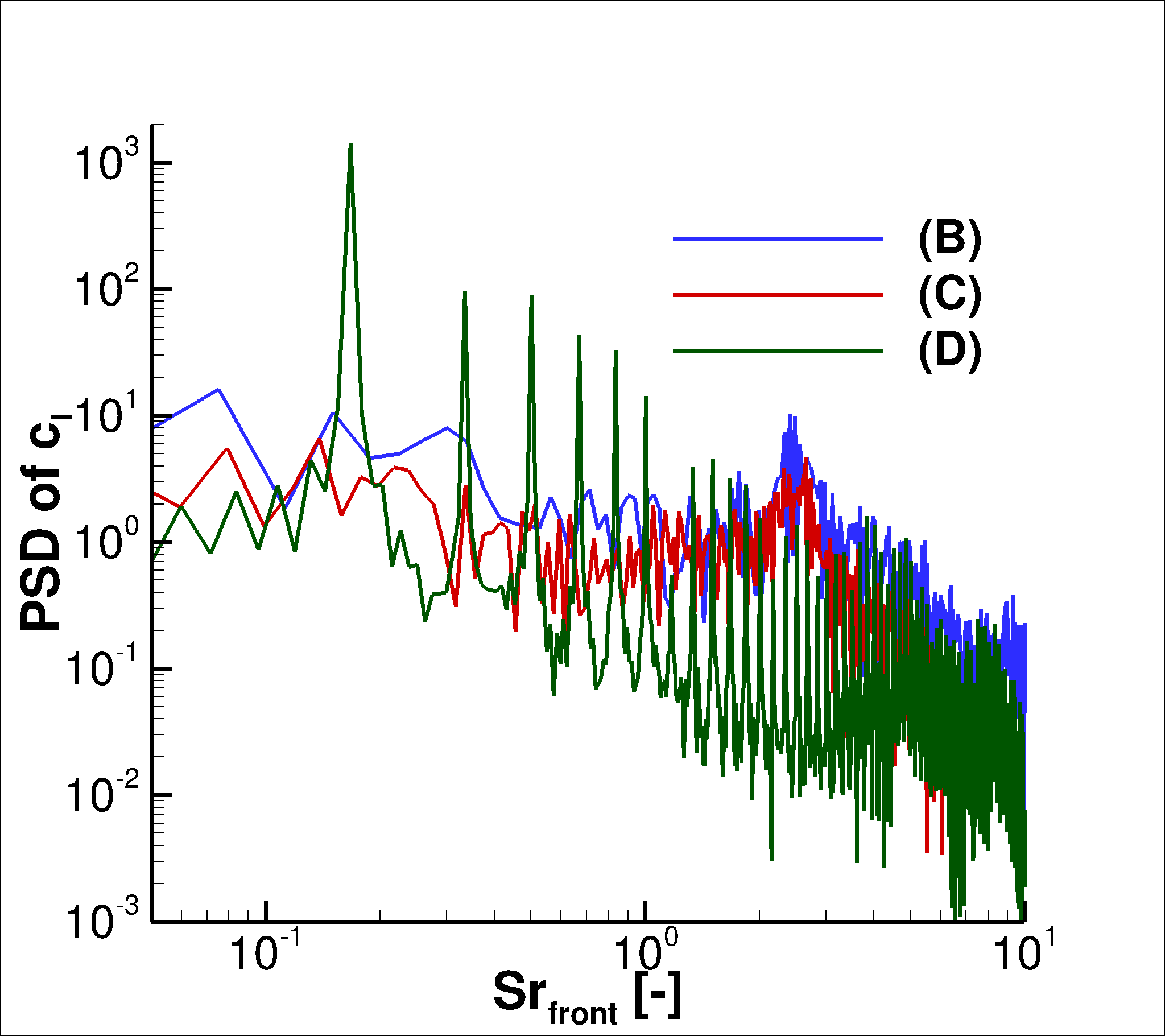}
    \caption{Power spectral density (PSD) of the lift coefficient for configurations (B), (C) and (D)}\label{fig:PSD-hinten-htp-buffet}
\end{figure}
A spectral analysis of the lift oscillations, depicted in Fig.~\ref{fig:PSD-hinten-htp-buffet}, confirms that the shock oscillation of the unhindered buffet at $Sr=0.154$ is suppressed and the lift oscillations are dominated by the wake fluctuations at $Sr \approx 1..4$ as for the case of $\alpha_{I}=-4^{\circ}$ in configuration (B).

\section{Conclusion}\label{sec6}

The characteristics of the turbulent wake downstream from a wing segment in buffet conditions, and the interaction phenomena occurring between the wake and a rear wing segment were investigated for a tandem wing configuration using hybrid RANS/LES simulations with the AZDES method. 
The configuration consists of two straight, untapered and unswept wing segments, creating flow phenomena representative of a wing-wake-tail plane configuration. 
A suitable airfoil for the rear wing was selected based on a preliminary study using the MSES potential solver toolbox.

For the considered flow conditions, a pronounced shock oscillation is present on the front wing segment, with a lift amplitude of $\hat{c_l}=0.145$ and a buffet frequency of $f=118.5$\,Hz, corresponding to a Strouhal number of $Sr=0.0745$, which is comparable to previous experimental investigations of 2D buffet \cite{jacquin2009} and recent measurements of the research unit \cite{schauerte2022}. 
The separated flow behind the shock forms a turbulent wake that impinges on the rear wing segment positioned downstream. Turbulent vortices develop behind the front wing segment and high levels of velocity and pressure fluctuations can be seen in the wake.
It is found that the characteristics of the wake vary strongly within the buffet cycle. During the downstream movement of the shock, the amount of shock-induced separation reaches its minimum, the wake is comparably thin, and small, regular alternating vortices are present in the wake. In this phase, the corresponding fluctuations of velocity and pressure also reach their minimum. In contrast, during the upstream movement of the shock, the amount of separation is at its maximum, the wake exhibits its largest extent, and large vortices are forming in the wake, which break into more irregular, chaotic turbulent structures. In this phase, the velocity and pressure fluctuations reach their maximum level, with the vertical fluctuations $w'$ being the largest reaching up to a half of the inflow velocity $u_{\infty}$.
All fluctuation quantities as well as the velocity deficit decrease with streamwise distance in the wake, reflecting momentum exchange and mixing of outer and inner flow, which eventually leads to the dissipation of the wake.
Apart from high amplitudes at low frequencies caused by the shock oscillation with $Sr=0.0745$, a high level of high-frequent fluctuations is found in the range of $Sr \approx 1.5$ to $Sr \approx 8$ that can be attributed to the vortices or turbulent structures present in the wake. 
A distinct peak at about $Sr \approx 2.4$ can be associated with the shedding of wake vortices, identified by means of a modal analysis of the flow field using Proper Orthogonal Decomposition (POD).

The aerodynamics and loading of the rear wing segment are strongly influenced by the interaction with the unsteady turbulent wake which results from the buffet occurring on the front wing segment. The impingement of the wake causes a strong variation of the loading, with a lift amplitude of $\hat{c_l}=0.25$ for configuration (A). 
A comparably low-frequent oscillation of the lift coefficient, attributed to the change of intensity of the downwash caused by the front segment, can be distinguished from fluctuations of high frequency. The latter are caused by the impingement of the vortices and turbulent structures in the wake and reach amplitudes comparable to those of the low-frequent oscillation.
As the amplitude of the velocity and pressure fluctuations varies strongly within the buffet cycle depending on the phase of the buffet, the same holds true for the loading fluctuations caused by the wake impingement. 
The same oscillating characteristic is also found for the pressure fluctuations on the rear segment's surface.
The similarity of the spectra in the wake and on the rear wing segment show that the turbulent fluctuations in the wake correlate with the pressure fluctuations on the rear wing segment, which is consistent with the findings reported by M\"uller et al.~\cite{mueller2020} for the case of atmospheric turbulence.
The highest levels of fluctuations are found on the leading edge, where the vortices directly impact on the surface. 

Comparing the different vertical positions (A) and (B) of the rear wing segment, it is found that the effect of the wake on the load fluctuation is qualitatively the same for both positions. As the rear segment is positioned slightly below the center of the wake for configuration (B), the amplitude of the load fluctuation is smaller for this case. Those quantitative differences are notable considering the the small vertical shift between both positions. 

Changing the angle of incidence of the rear wing segment so that a relative angle of attack of $\alpha_{rear}=6^{\circ}$ is reached, buffet occurs for the isolated rear wing segment (D) with a frequency of $Sr = 0.154$. In the tandem configuration (C), however, the shock oscillation on the rear segment is suppressed by the wake impact, and the lift oscillations are dominated by the wake fluctuations at $Sr \approx 1..4$ as for the case of $\alpha_I = -4^{\circ}$ in configuration (B).

\backmatter

\bmhead{Acknowledgments}

The authors gratefully acknowledge the Deutsche Forschungsgemeinschaft DFG (German Research Foundation) for funding this work in the framework of the research unit FOR\,2895. 
The authors also would like to thank ONERA for providing the geometry of the OAT15A airfoil,
as well as the H\"ochstleistungsrechenzentrum Stuttgart (HLRS) for providing computation time on the Hawk cluster.

\section*{Declarations}

\bmhead{Conflict of Interest}

Co-author Thorsten Lutz is Guest Editor of the special issue of the CEAS Aeronautical Journal in which this paper will be included. The authors have no other conflicts of interest nor competing interests to declare that are relevant to the content of this article.

\bmhead{Funding}

The research leading to these results received funding from the Deutsche Forschungsgemeinschaft DFG (German Research Foundation) in the framework of the research unit FOR\,2895 (Unsteady flow and interaction phenomena at high-speed stall conditions), subproject TP4, under the grant number LU809/11-1.
The computation time needed for performing the numerical simulations is provided by the H\"ochstleistungsrechenzentrum Stuttgart (HLRS) as part of SCBOPT project.

\bibliography{ceas_refs}

\end{document}